\documentclass[usenatbib]{aa}  

\usepackage{graphicx}
\usepackage{txfonts}
\def\Chan{{\sl Chandra}}
\def\XMM{XMM-{\sl Newton}}
\def\COS{HST-COS}

\def\NS{{\sl NuSTAR}}
\def\INT{INTEGRAL}
\def\SW{{\sl Swift}}

\def\CV{C\,{\sc v}}
\def\CVI{C\,{\sc vi}}
\def\CVII{C\,{\sc vii}}

\def\NVI{N\,{\sc vi}}
\def\NVII{N\,{\sc vii}}
\def\OIII{O\,{\sc iii}}

\def\OVII{O\,{\sc vii}}
\def\OVIII{O\,{\sc viii}}
\def\OIX{O\,{\sc ix}}

\def\NeIX{Ne\,{\sc ix}}
\def\NeX{Ne\,{\sc x}}
\def\NeXI{Ne\,{\sc xi}}

\begin{document}

   \title{Anatomy of the AGN in NGC 5548}

   \subtitle{VI. Long-term variability of the warm absorber}

   \author{J. Ebrero\inst{1}
          \and
          J. S. Kaastra\inst{2,3}
	  \and
	  G. A. Kriss\inst{4,5}
          \and
          L. Di Gesu\inst{2}
          \and
          E. Costantini\inst{2}
          \and
          M. Mehdipour\inst{2}
          \and
          S. Bianchi\inst{6}
          \and
          M. Cappi\inst{7}
          \and
          R. Boissay\inst{8}
          \and
          G. Branduardi-Raymont\inst{9}
          \and
          P.-O. Petrucci\inst{10,11}
          \and
          G. Ponti\inst{12}
          \and
          F. Pozo N\'u\~nez\inst{13}
          \and
          H. Seta\inst{14}
          \and
          K. C. Steenbrugge\inst{15}
          \and
          M. Whewell\inst{9}
          }

   \institute{XMM-Newton Science Operations Centre, ESAC,
              Camino Bajo del Castillo s/n, Urb. Villafranca del Castillo, 28692 Villanueva de la Ca\~nada, Madrid, Spain\\
              \email{jebrero@sciops.esa.int}
              \and
              SRON Netherlands Institute for Space Research,
	      Sorbonnelaan 2, 3584 CA, Utrecht, The Netherlands
	      \and
	      Leiden Observatory, Leiden University,
	      P.O. Box 9513, 2300 RA, Leiden, The Netherlands
              \and
              Space Telescope Science Institute,
              3700 San Martin Drive, Baltimore, MD 21218, USA
              \and
              Department of Physics and Astronomy, The Johns Hopkins University,
              Baltimore, MD 21218, USA
              \and
              Dipartimento di Matematica e Fisica, Universit\`a degli Studi Roma Tre,
              Via della Vasca Navale 84, 00146, Roma, Italy
              \and
              INAF-IASF Bologna,
              Via Gobetti 101, 40129 Bologna, Italy
              \and
              Department of Astronomy, University of Geneva,
              16 Ch. d'Ecogia, 1290 Versoix, Switzerland
              \and
              Mullard Space Science Laboratory, University College London,
              Holmbury St. Mary, Dorking, Surrey, RH5 6NT, UK
              \and
              University Grenoble Alpes, IPAG, 38000 Grenoble, France
              \and
              CNRS, IPAG, 38000 Grenoble, France
              \and
              Max-Planck-Institut f\"ur Extraterretrische Physik,
              Giessenbachstrasse, 85748 Garching, Germany
              \and
              Astronomisches Institut, Ruhr-Universit\"at Bochum,
              Universit\"atstrasse 150, 44801 Bochum, Germany
              \and
              Department of Physics, Tokyo Metropolitan University,
              1-1 Minami-Osawa, Hachioji, Tokyo, Japan
              \and
              Instituto de Astronom\'ia, Universidad Cat\'olica del Norte,
              Avenida Angamos 0610, Casilla 1280, Antofagasta, Chile
              }

   \date{Received <date>; accepted <date>}

 
  \abstract
   {We observed the archetypal Seyfert 1 galaxy NGC 5548 in 2013-2014 in the context of an extensive multiwavelength campaign involving several satellites, which revealed the source to be in an extraordinary state of persistent heavy obscuration.}
   {We re-analyzed the archival grating spectra obtained by \XMM{}~and \Chan{}~between 1999 and 2007 in order to characterize the classic warm absorber (WA) using consistent models and up-to-date photoionization codes and atomic physics databases and to construct a baseline model that can be used as a template for the physical state of the WA in the 2013 observations.}
   {We used the latest version of the photoionization code CLOUDY and the SPEX fitting package to model the X-ray grating spectra of the different archival observations of NGC 5548.}
   {We find that the WA in NGC 5548 is composed of six distinct ionization phases outflowing in four kinematic regimes. The components seem to be in the form of a stratified wind with several layers intersected by our line of sight. Assuming that the changes in the WA are solely due to ionization or recombination processes in response to variations in the ionizing flux among the different observations, we are able to estimate lower limits on the density of the absorbing gas, finding that the farthest components are less dense and have a lower ionization. These limits are used to put stringent upper limits on the distance of the WA components from the central ionizing source, with the lowest ionization phases at several pc distances ($<50$, $<20$, and $<5$~pc, respectively), while the intermediately ionized components lie at pc-scale distances from the center ($<3.6$ and $<2.2$~pc, respectively). The highest ionization component is located at $\sim 0.6$~pc or closer to the AGN central engine. The mass outflow rate summed over all WA components is $\sim 0.3$~$\rm M_{\odot}$~yr$^{-1}$, about six times the nominal accretion rate of the source. The total kinetic luminosity injected into the surrounding medium is a small fraction ($\sim 0.03$\%) of the bolometric luminosity of the source. After adding the contribution of the UV absorbers, this value augments to $\sim 0.2$\% of the bolometric luminosity, well below the minimum amount of energy required by current feedback models to regulate galaxy evolution.}
   {}


   \authorrunning{J. Ebrero et al.}
   \titlerunning{Anatomy of the AGN in NGC 5548. VI.}

   \maketitle
%

\section{Introduction}
\label{intro}

Active galactic nuclei (AGN) are powered by gravitational accretion of matter on to the central supermassive black hole (SMBH) that resides in their center (\citealt{Rees84}), emitting vast amounts of energy across the spectrum. X-ray emission is a characteristic feature of AGN spectra and finds its origin in the inverse Compton scattering of lower energy photons in a hot corona in the vicinity of the SMBH (e.g., \citealt{ST80}; \citealt{HM93}). With the advent of medium- and high-resolution X-ray spectroscopy, it was discovered that absorption lines of photoionized species blueshifted with respect to the systemic velocity of the host galaxies (\citealt{Kaa00}), were present in the soft X-ray spectra of more than 50\% of nearby Seyfert 1 galaxies (\citealt{Rey97}; \citealt{Geo98}). This so-called warm absorber (WA) typically shows multiple ionization phases with temperatures $\sim 10^4 - 10^6$~K, column densities ranging from $10^{20} - 10^{24}$~cm$^{-2}$, and outflow velocities of a few hundred to thousands of km s$^{-1}$ (e.g., \citealt{Blu05}).

While their ubiquity suggests that WA are an important structural feature in AGN, their contribution as sources of cosmic feedback (the relation between the SMBH growth and star formation in the host galaxy) has been a matter of debate for many years. Recent studies seem to indicate that the contribution of WA with low-to-moderate velocities is negligible in terms of feedback, with the exception of the so-called ultra-fast outflows (UFOs), which would be able to carry enough momentum to be a significant source of feedback (see \citealt{Tom13}, and references therein). Current models show that kinetic luminosities (kinetic energy carried by the outflow per unit time) ranging between 0.5\% and 5\% of the bolometric luminosity (\citealt{DM05}; \citealt{HE10}) must be fed back in order to significantly affect the interstellar medium (ISM) of the galaxy and reproduce the observed $M - \sigma$ relation (\citealt{FM00}; \citealt{Geb00}).

A key parameter for reliably estimating the mass outflow rate (hence the kinetic luminosity injected into the ISM) is the actual location of the WA. Current scenarios for the origin of these ionized outflows are: (i) disk-driven winds, which are launched within hundreds of gravitational radii (\citealt{MC95}; \citealt{Elvis00}); and (ii) thermal winds evaporating from an irradiated dusty torus that surrounds the AGN engine, which lie at pc-scale distances (\citealt{KK01}). Ideally, tight limits can be put on the distance by measuring variability on the ionization properties of the WA in response to changes in the incident ionizing flux (e.g., \citealt{Det08}; \citealt{Lon10}; \citealt{Kaa12}). High-density gas will ionize/recombine faster than low-density gas, allowing us to determine reliable lower limits to the density and hence strict upper limits on the distance where this gas crosses our line of sight. This is, however, very challenging because the observations of a given object are often sparse, washing out the possible effects of variability, and lacking the signal-to-noise ratio required to significantly measure the expected changes.

NGC 5548 is an archetypal Seyfert-1 galaxy ($M_{\rm BH} = 3.24 \times 10^7$~$\rm M_{\odot}$, \citealt{Pan14}), the first in which narrow absorption lines in the X-rays were detected (\citealt{Kaa00}). Since then, its ionized outflow has been widely studied, both in the X-rays (\citealt{Kaa02}; \citealt{Ste03}; \citealt{Ste05}; \citealt{Det08}; \citealt{Det09}; \citealt{Kro10}; \citealt{AV10}) and in the ultraviolet (\citealt{CK99}; \citealt{Bro02}; \citealt{Arav02}; \citealt{Cre03}; \citealt{Cre09}).

Following the success of the multiwavelength campaign on Mrk 509 (introduced in \citealt{Kaa11}), NGC 5548 was observed extensively in 2013-2014 in the context of a large dedicated multiwavelength campaign led by \XMM{}, comprising also observations with \Chan{}, \NS{}, \INT{}, \COS{}, and including \SW{}~monitoring, with the objective of characterizing the properties of the WA with unprecedented detail. An overview of the campaign can be found in \citet{Meh14a} (Paper I). Interestingly, the soft X-ray flux of NGC 5548 during the campaign was remarkably low, about 25 times lower than the median flux of the source in previous observations. Our analysis suggests that this is caused by a long-lasting stream of obscuring material situated at light-day distances, partially covering the X-ray source and the broad line region (BLR), as reported in \citet{Kaa14}.

This paper is organized as follows. In Sect.~\ref{revisit}~we briefly summarize the motivation for the re-analysis of the grating X-ray spectra of NGC 5548, and the observations used in this paper. In Sect.~\ref{analysis} we describe the analysis of the data, while in Sect.~\ref{discussion} we discuss the results. Our conclusions are reported in Sect.~\ref{conclusions}. Throughout this paper we assume a cosmological framework with $H_{\rm 0} = 70$~km s$^{-1}$~Mpc$^{-1}$, $\Omega_{\rm M} = 0.3$, and $\Omega_{\Lambda} = 0.7$. The quoted errors refer to 68.3\% confidence level ($\Delta C = 1$~for one parameter of interest) unless otherwise stated.


\section{NGC 5548 revisited}
\label{revisit}

\subsection{Motivation}
\label{motivation}

\begin{figure}
  \centering
  \includegraphics[width=6.5cm,angle=-90]{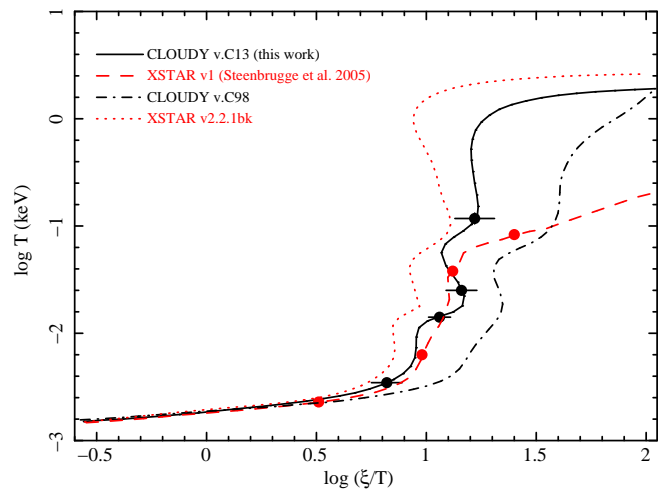}
  \caption{\label{scall_fig}Stability curves obtained with different versions of XSTAR and CLOUDY (black solid line, CLOUDY v13; black dot-dashed line, CLOUDY v98; red dashed line, XSTAR v1; red dotted line, XSTAR v2.2.1bk) for the same input spectral energy distribution. The solid dots represent the WA components of \citet{Ste05}.}
\end{figure}

The reasons to carry out a comprehensive re-analysis of the archival high-resolution grating spectra of NGC 5548 are two-fold. The main consequence of the extraordinary obscuration observed during the 2013 campaign (\citealt{Kaa14}) is the depression of the soft X-ray continuum, which makes the detection of WA features very challenging, although it gives us an excellent opportunity to study the X-ray emission of the narrow line region (NLR) of a Seyfert 1 galaxy with unprecedented detail (\citealt{Whe15}). The depressed X-ray flux is not intrinsically low, since the hard X-ray continuum is comparable to the historical fluxes of NGC 5548 in that energy band, but it is caused by the presence of obscuring material crossing our line of sight. Moreover, according to \SW{}~monitoring, this obscuration has lasted since at least February 2012 (\citealt{Meh14b}). An analysis of the data from our multiwavelength campaign reveals that the obscurer, possibly in the form of a clumpy stream of gas, is located within a few light days of the central SMBH, extending to outside of the BLR but closer in than the expected location of the WA (\citealt{Kaa14}). The ionizing photons that reach the WA are therefore severely diminished.

By comparing the properties of the WA with those measured earlier, we can take advantage of this unexpected situation to measure the response of the ionized gas, giving us an opportunity to constrain the location of the WA components with unprecedented accuracy. This, however, poses an important problem as very few counts are left in the soft X-ray spectrum (where the bulk of the WA features lie) in spite of the large exposure time ($\sim 660$~ks with \XMM{}~RGS at the core of the campaign). Our strategy to model the WA during 2013 relied on the assumption that most of the parameters of the classic persistent WA outflow will remain the same with respect to the archival observations, at least the number of components, their outflow velocity, broadening of the lines and, possibly, their column density. If one gets an accurate characterization of the WA from past observations, this baseline model can then be applied to the obscured spectrum of NGC 5548, dramatically decreasing that number of free parameters that would prevent a spectral analysis otherwise.

\begin{table*}
  \centering
  \caption[]{NGC 5548 archival observations log.}
  \label{obslog}
  \begin{tabular}{l c c c c c}
    \hline\hline
    \noalign{\smallskip}
    Observatory  &  Grating  &  ID  &  Start Time (UTC)  & Exp. Time  &  Label\tablefootmark{a} \\
                 &           &      &  (yyyy-mm-dd \, hh:mm:ss) & (ks) &  \\
    \noalign{\smallskip}
    \hline
    \noalign{\smallskip}
    \XMM{}      &  RGS  & 0109960101 & 2000-12-24 \, 22:12:11 & 26.1 & R00x \\
    \XMM{}      &  RGS  & 0089960301 & 2001-07-09 \, 15:45:59 & 95.8 & R01y \\
    \XMM{}      &  RGS  & 0089960401 & 2001-07-12 \, 07:34:56 & 39.1 & R01z \\
    \noalign{\smallskip}
    \hline
    \noalign{\smallskip}
    \Chan{}     & LETG  & 330  &  1999-12-11 \, 22:51:20 & 86.0 & L99 \\
    \Chan{}     & HETG  & 837  &  2000-02-05 \, 15:37:29 & 82.3 & H00 \\
    \Chan{}     & HETG  & 3046 &  2002-01-16 \, 06:12:34 & 153.4 & H02 \\
    \Chan{}     & LETG  & 3045 &  2002-01-18 \, 15:57:02 & 169.4 & L02\tablefootmark{b} \\
    \Chan{}     & LETG  & 3383 &  2002-01-21 \, 07:33:57 & 171.0 & L02\tablefootmark{b} \\
    \Chan{}     & LETG  & 5598 &  2005-04-15 \, 05:18:18 & 116.4 & L05\tablefootmark{b} \\
    \Chan{}     & LETG  & 6268 &  2005-04-18 \, 00:31:12 & 25.2 & L05\tablefootmark{b} \\
    \Chan{}     & LETG  & 7722 &  2007-08-14 \, 20:57:57 & 99.6 & L07\tablefootmark{b} \\
    \Chan{}     & LETG  & 8600 &  2007-08-17 \, 03:54:51 & 37.2 & L07\tablefootmark{b} \\
    \noalign{\smallskip}
    \hline
  \end{tabular}
  \tablefoot{
    \tablefoottext{a}{Label used to identify the spectra throughout the paper; }
    \tablefoottext{b}{Observations with the same label are co-added to produce a single spectrum for analysis.}
  }
\end{table*}

Our second motivation stems from the fact that different atomic physics codes were used to derive the WA properties reported in the literature. The ionization state of the WA components is usually computed using a photoionization code that calculates the transmission through a slab of material where all ionic column densities are linked through a photoionization balance model. Previous modeling of the NGC 5548 WA made use of XSTAR\footnote{\tt http://heasarc.gsfc.nasa.gov/docs/software/xstar/xstar.html} (\citealt{KK99}) or CLOUDY\footnote{\tt http://www.nublado.org} (\citealt{Fer13}), which use slightly different atomic physics databases. Furthermore, since past observations of NGC 5548 span many years, the published WA parameters are computed using different versions of these codes, which evolve and are updated with time.

Differences between photoionization codes and versions of codes are not negligible. To illustrate this, in Fig.~\ref{scall_fig} we show four stability curves obtained from the output ionization balance provided by different versions of XSTAR and CLOUDY. In all cases the input spectral energy distribution (SED) was that of NGC 5548 reported by \citet{Ste05}. While the four lines agree reasonably well at low ionization values, the differences between them are dramatic for higher values of the ionization parameter $\xi$. In particular, the ionization balance used by \citet{Ste05}, computed with XSTAR v1, lacks several unstable areas (the branches with negative slope) and the Compton shoulder at high temperatures. Even the more recent versions of CLOUDY and XSTAR (v13 and v2.2.1, respectively), although they reproduce the same overall shape reasonably well, show strong deviations with respect to each other. The solid dots overplotted on top of the curves represent the positions where the WA components detected by \citet{Ste05} fall, showing that a fit to the same spectrum with such different ionization balances provide significantly different ionization parameter $\xi$ values.

Therefore, in order to obtain a comprehensive picture of the WA properties that can be used as a template for the 2013-2014 campaign as well as consistent parameter values, it is essential to re-analyze the archival grating spectra of NGC 5548 using the same, and most up-to-date, photoionization code for all of them.

\subsection{Archival data}
\label{data}

We used a total of 12 \Chan{}~and \XMM{}~individual archival observations in this paper. In some cases, the observations were taken in consecutive orbits and with the same instrumental configuration, allowing us to co-add the spectra in order to obtain a better signal-to-noise ratio. Below we briefly describe each of the observations. The observations log is shown in Table~\ref{obslog}.

Prior to the 2013 campaign, NGC 5548 was observed three times by \XMM{}, once in 2000 (ObsID 0109960101) for 26.1 ks, and twice in 2001 (ObsIDs 0089960301 and 0089960401) for $\sim96$~ and 36 ks, respectively. In this paper we used solely the Reflection Grating Spectrometer (RGS, \citealt{Her01}) data, that were processed using the \XMM{}~Science Analysis System (SAS) v13.5 (\citealt{Gab04}). The data were rebinned by a factor of 3 to avoid oversampling. In what follows, the RGS spectra will be identified in the paper as R00x, R01y, and R01z, respectively.

There are nine archival observations of NGC 5548 taken with the gratings onboard \Chan{}, seven with the Low Energy Transmission Grating Spectrometer (LETGS, \citealt{Bri00}) and two with the High Energy Grating Spectrometer (HETGS, \citealt{Can05}). The HETGS/ACIS-S spectra were extracted from the TGCat archive\footnote{\tt http://tgcat.mit.edu}. The data were binned by a factor of two, and the Gehrels-statistic used by default in HETGS data was replaced by the true Poissonian error. In order to obtain consistent results for the MEG and HEG gratings, the MEG fluxes were scaled by a factor of 0.954 relative to the HEG fluxes. The observations were performed in 2000, with an exposure of 82 ks, and in 2002 as part of an UV/X-ray campaign with an exposure of 153~ks (\citealt{Ste05}). Throughout this paper these observations will be identified as H00 and H02, respectively.

The LETGS/HRC-S spectra were reduced using the \Chan{}~Interactive Analysis of Observations software (CIAO, \citealt{Fru06}) v4.5, until the level 1.5 event files were created. The rest of the procedure, until the creation of the final level 2 event files, was performed following an independent method first reported in \citet{Kaa02}. This results in an improved wavelength accuracy determination and effective area generation. NGC 5548 was observed with \Chan{}~LETGS once in 1999 for 86 ks (labeled as L99 hereafter), and twice in 2002, 2005, and 2007. In the latter three cases the observations were taken in consecutive orbits, allowing us to co-add the two spectra to obtain a higher signal-to-noise. Therefore, we ended up with a $\sim340$~ks spectrum in 2002, $\sim141$~ks in 2005, and $\sim138$~ks in 2007, that are labeled throughout the paper as L02, L05, and L07, respectively.


\section{Data analysis}
\label{analysis}

\subsection{Strategy}
\label{strategy}

The archival observations of NGC 5548 sum up to $\sim1.1$~Ms of exposure time of grating spectra, although with great disparity in data quality among the different datasets. A modeling strategy that makes the most of this large amount of data is therefore required in order to maximize the scientific return. On one hand, fits to the individual observations in which all the parameters are left free would lead to too large error bars for most of the relevant parameters, or in those observations with a low S/N in completely unconstrained values. On the other hand, a joint fit of all the spectra together (3 RGS, 2 HETGS, and 4 LETGS) is not feasible, as it requires a prohibitive amount of computing time due to the large number of data channels and models that need to be evaluated together.

We therefore adopted a mixed approach for modeling these data. Since the 2002 data constitute the bulk of the archival data ($\sim500$~ks taking the L02 and H02 observations together), and they are dominant in terms of S/N contribution, we used them to create a model template for the rest of the archival observations. We first obtained a best-fit model for the 2002 data, including the continuum, broad and narrow emission lines, radiative recombination continua (RRCs), and absorption troughs produced by the WA (see the detailed modeling in Sect.~\ref{modeling}). This best-fit model was then used as a starting point to analyze the rest of the archival spectra, leaving as free parameters only those of the continuum, the flux of the \OVII{}~forbidden line, and the ionization parameter $\log \xi$ of each WA component. The \OVII{}~forbidden line flux was set free since it is known to vary (\citealt{Det09}), whereas the rest of the lines are too weak to measure significant changes or are not even significantly detected (partially due to the wavenlength coverage, partially due to the lower S/N). Therefore, all other parameters describing the emission lines as well as the absorption lines ($N_{\rm H}$, outflow velocity, velocity broadening) are kept frozen.

With this strategy we aim at determining the most relevant parameters that are known to vary between observations, such as the spectral shape of the continuum and the WA ionization parameter. The dynamics of the persistent outflows in NGC 5548 have barely changed over $\sim20$~years, i.e. the different kinematic components can be easily cross-identified between the different observations, but their ionization state can be substantially different. We therefore assume that the observed variations in the WA are mainly due to ionization changes rather than being caused by the dissapearance of any of these outflows, or the appearance of newer ones.

\begin{figure}
  \centering
  \includegraphics[width=6.5cm,angle=-90]{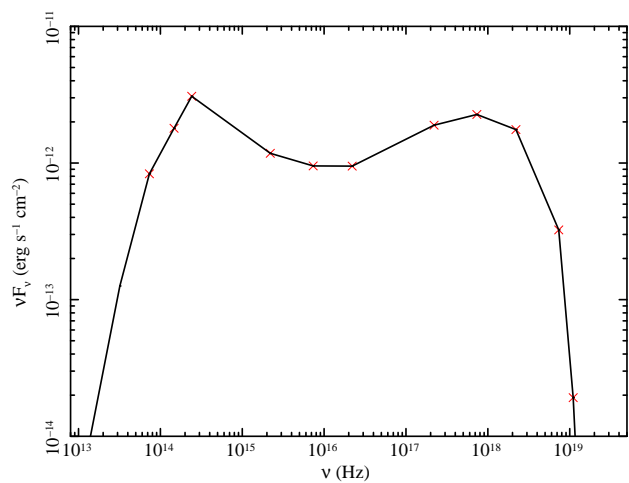}
  \caption{\label{sed_fig}Spectral energy distribution of NGC 5548. Data points are those of \citet{Ste05}.}
\end{figure}

\subsection{Modeling}
\label{modeling}

The archival grating spectra of NGC 5548 were analyzed using the SPEX fitting package\footnote{\tt http://www.sron.nl/spex} version 2.05 (\citealt{Kaa96}). The fitting method was C-statistics (\citealt{Cash79}). The adopted cosmological redshift for NGC 5548 was $z = 0.017175$~(\citealt{Vac91}). The foreground Galactic column density was set to $N_{\rm H} = 1.45 \times 10^{20}$~cm$^{-2}$~(\citealt{Wak11}) using the {\it hot} model in SPEX, with the temperature fixed to 0.5 eV to mimic a neutral gas. Throughout the analysis we assumed proto-Solar abundances of \citet{LP09}.

\subsubsection{The \Chan{}~2002 spectra}
\label{2002}

As mentioned in Sect.~\ref{strategy}, we modeled the L02 and H02 datasets in order to obtain a baseline model template that could be used as a starting point to analyze the remaining archival spectra. The methodology reported below was firstly described in the supplementary material of \citet{Kaa14}. The fits were carried out in the $2-60$~\AA~range for the LETGS spectrum, and in the $1.55-15.5$~\AA~and $2.5-26$~\AA~range for the HEG and MEG spectra, respectively. The continuum was modeled with a power law plus a modified black body ({\it mbb} model in SPEX), which considers modifications of a simple black body model by coherent Compton scattering based on the calculations of \citet{KB89}.

\begin{figure*}
  \centering
  \hbox{
  \includegraphics[width=6.5cm,angle=-90]{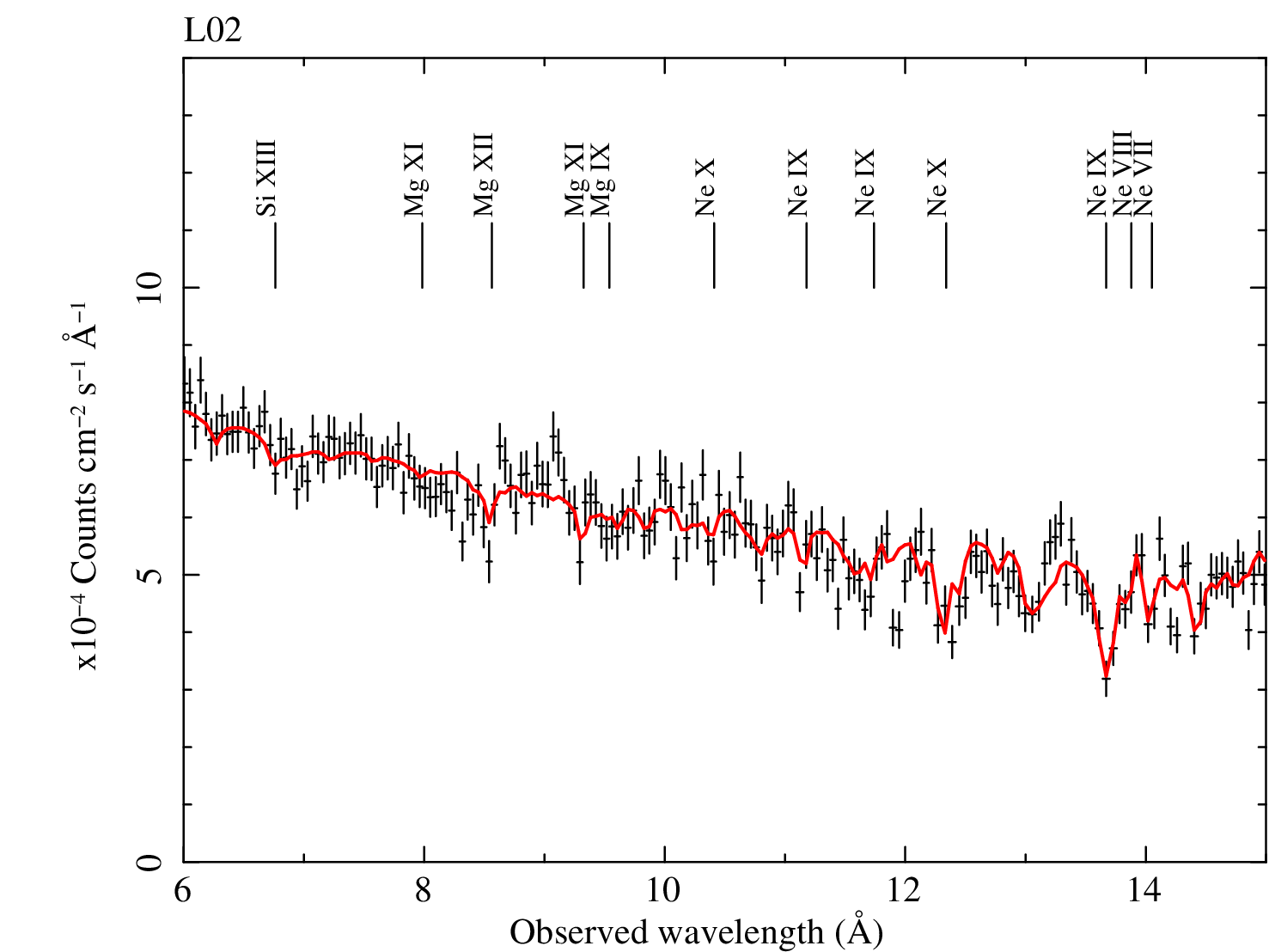}
  \includegraphics[width=6.5cm,angle=-90]{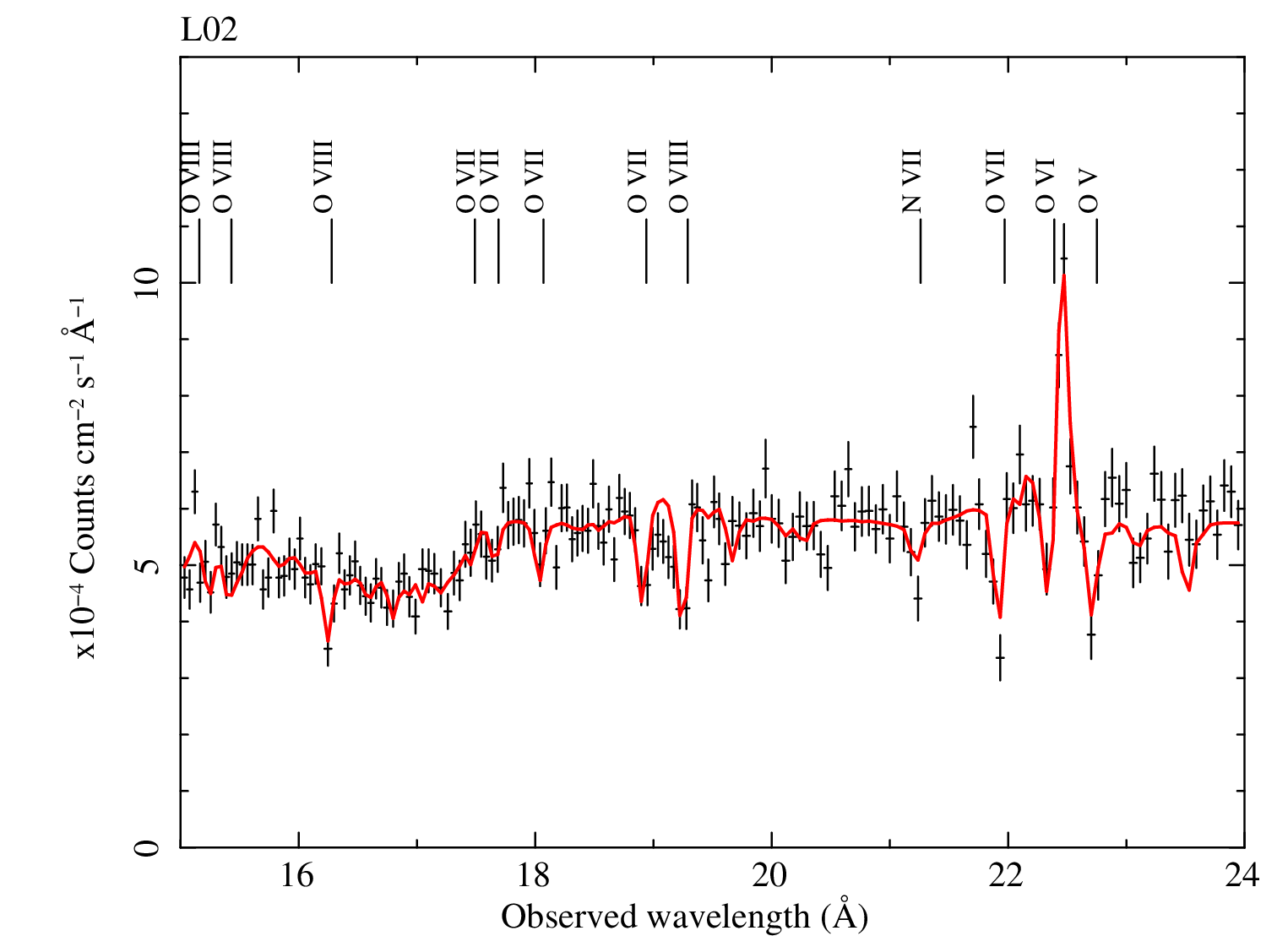}
  }
  \hbox{
  \includegraphics[width=6.5cm,angle=-90]{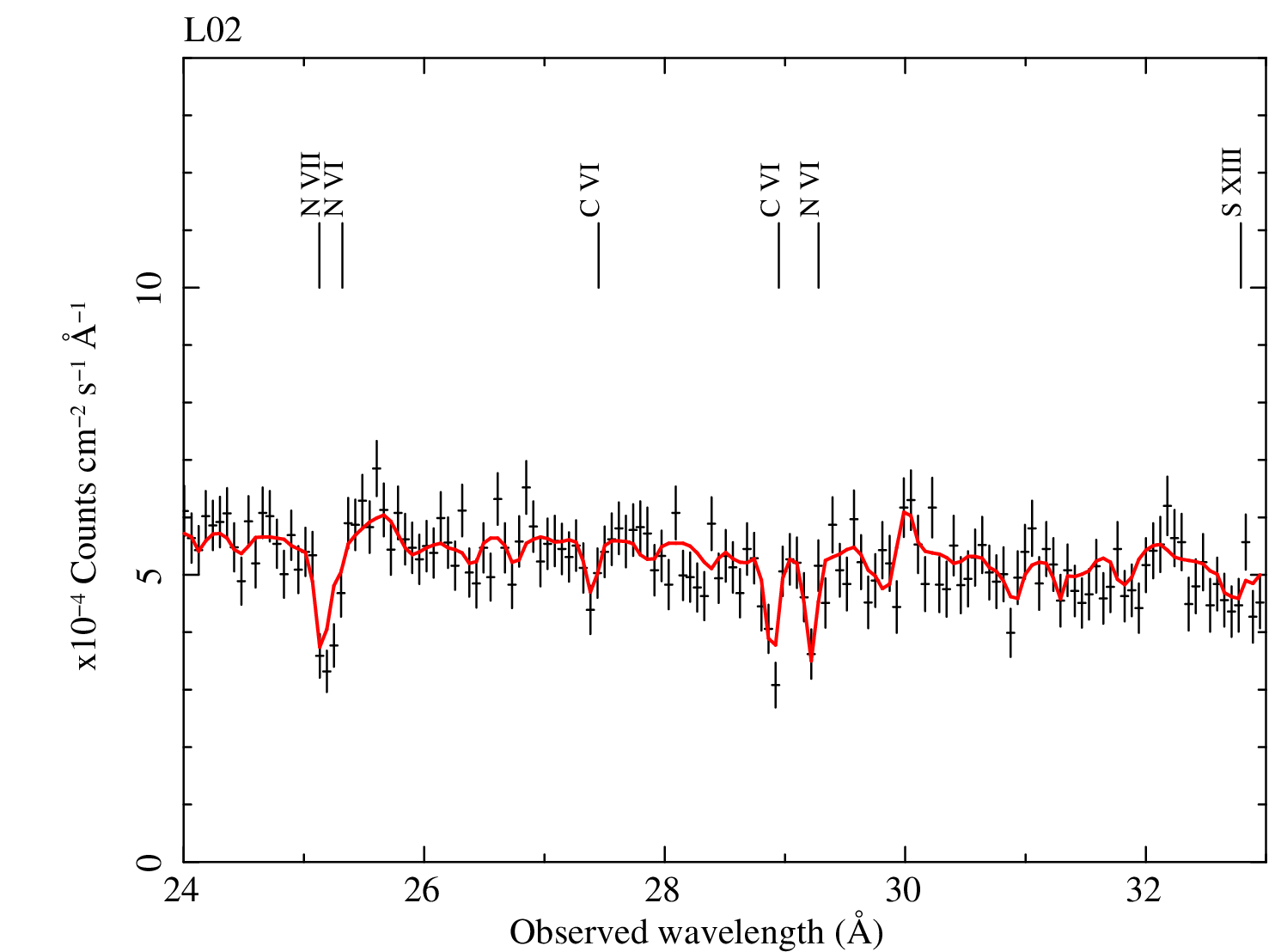}
  \includegraphics[width=6.5cm,angle=-90]{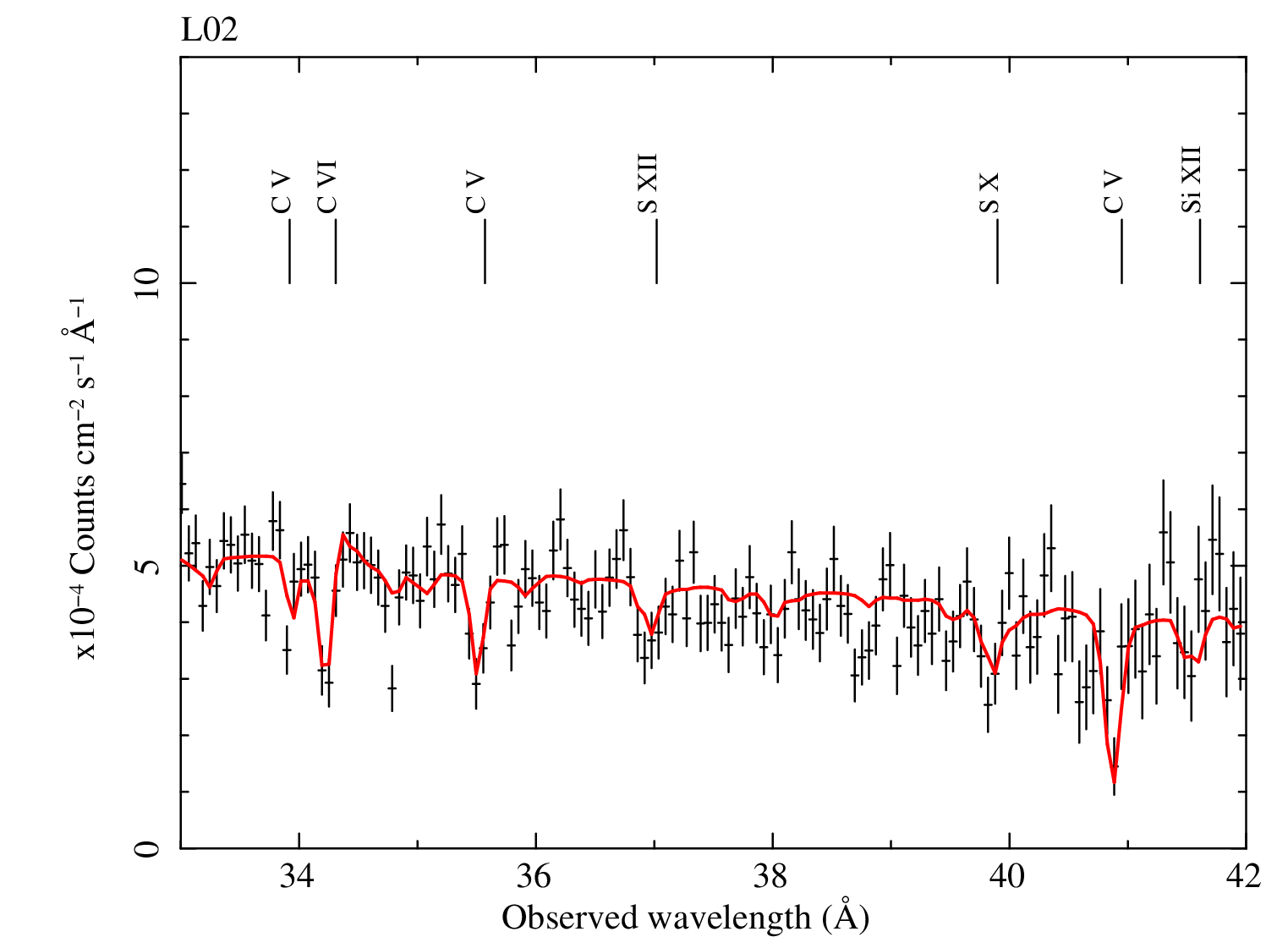}
  }
  \hbox{
  \includegraphics[width=6.5cm,angle=-90]{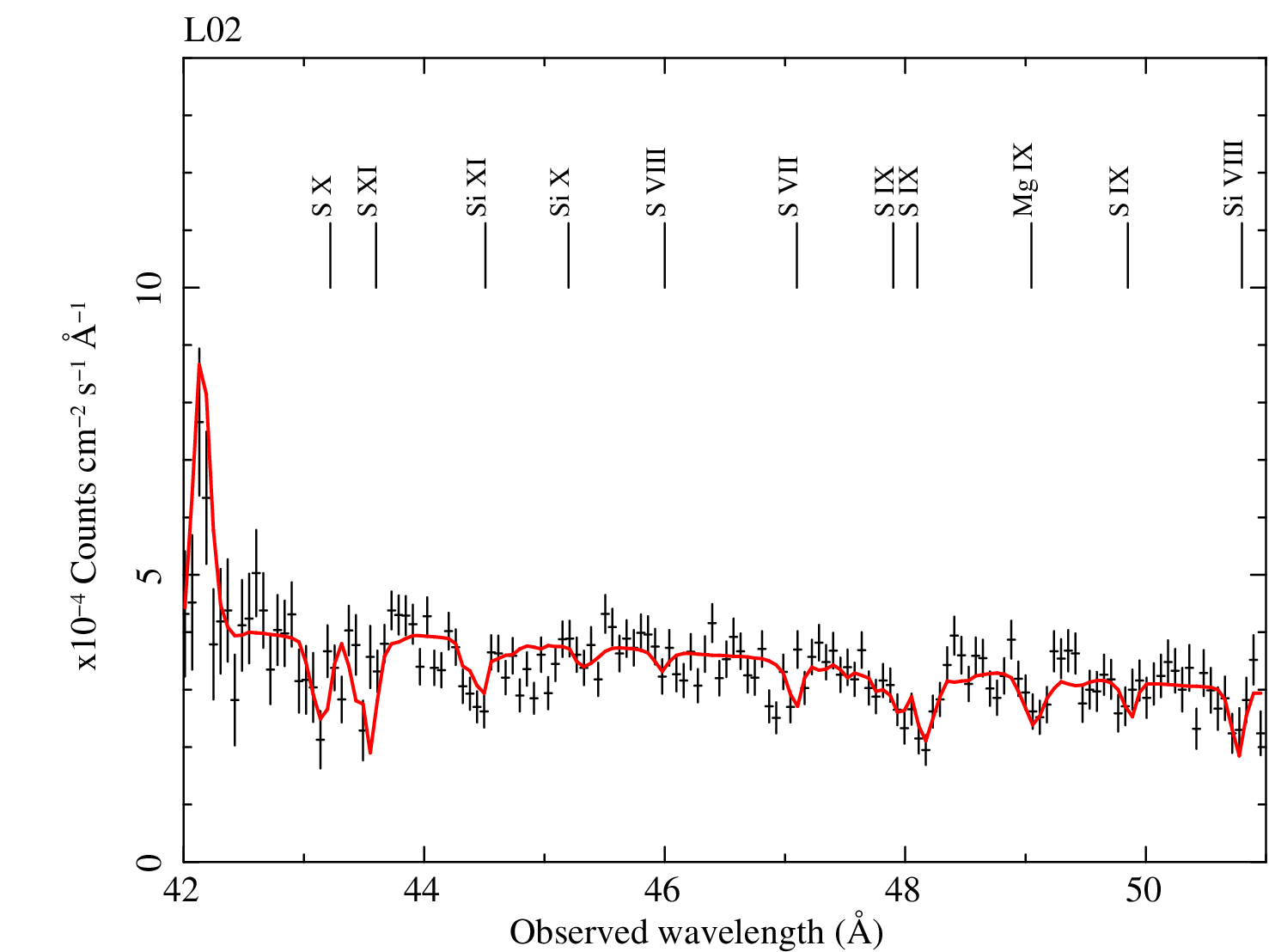}
  \includegraphics[width=6.5cm,angle=-90]{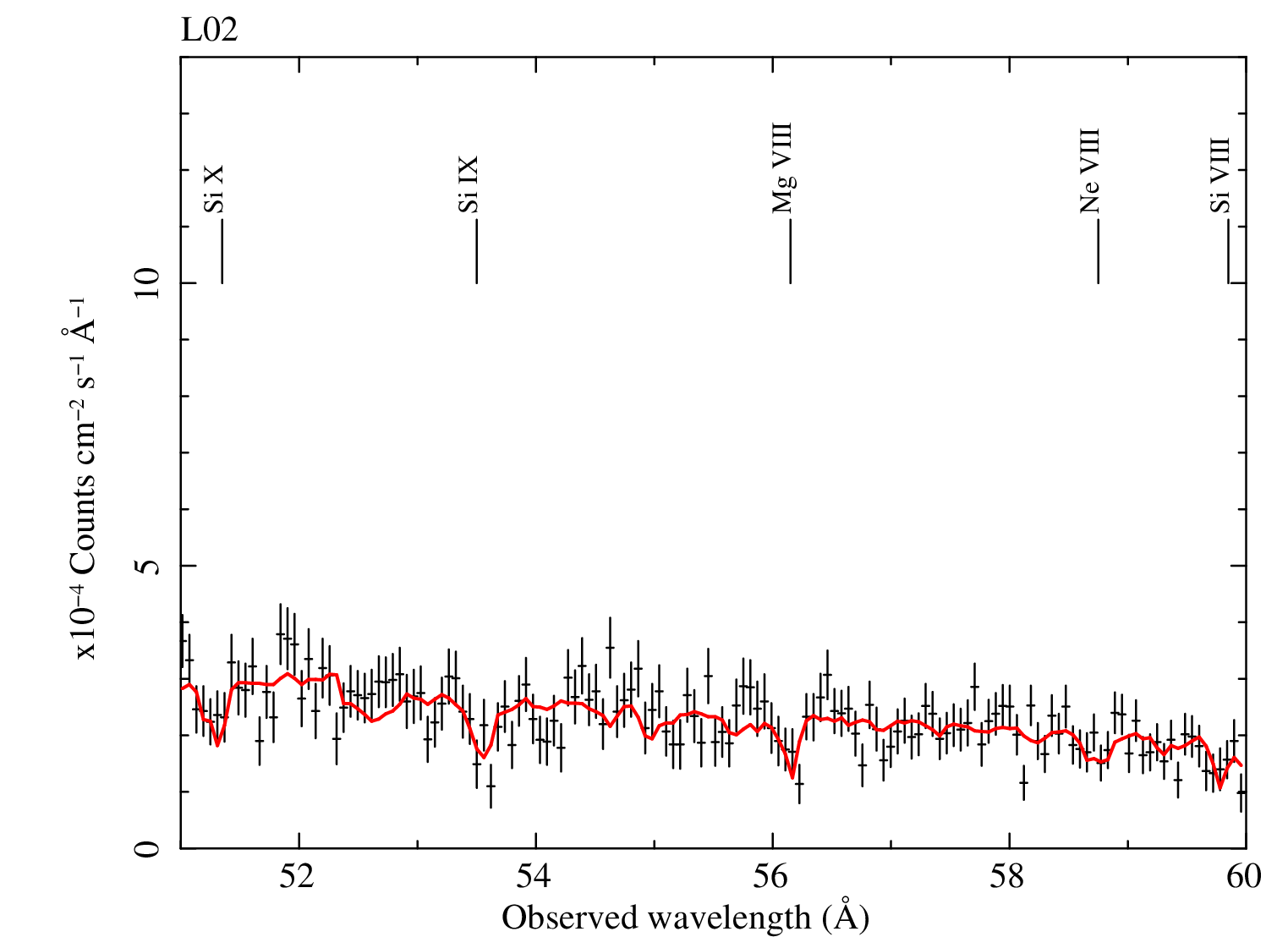}
  }
  \caption{\label{L02fluxed}\Chan{}~LETGS spectrum of NGC 5548 in 2002 (L02). The solid red line represents our best-fit model. Some relevant WA absorption features have been labeled.}
\end{figure*}

\begin{table*}
  \centering
  \caption[]{Continuum best-fit parameters of the 2002 observations.}
  \label{cont2002best}
  \begin{tabular}{l l c c}
    \hline\hline
    \noalign{\smallskip}
    Component      &  Parameter                                   &  LETGS 2002 (L02)  &  HETGS 2002 (H02)  \\
    \noalign{\smallskip}
    \hline
    \noalign{\smallskip}
    Power law      &  Photon index                                & $1.70 \pm 0.01$ & $1.65 \pm 0.01$ \\
                   &  Flux ($0.2-10$~keV)\tablefootmark{a}        & $5.40 \pm 0.06 \times 10^{-11}$ & $4.85 \pm 0.03 \times 10^{-11}$ \\
                   &  Luminosity ($0.2-10$~keV)\tablefootmark{b}  & $4.21 \pm 0.05 \times 10^{43}$ & $3.66 \pm 0.05 \times 10^{43}$ \\
    \noalign{\smallskip}
    \hline
    \noalign{\smallskip}
    Mod. black body &  Temperature\tablefootmark{c}                & $0.140 \pm 0.004$ & $0.115 \pm 0.010$ \\
                    &  Flux ($0.2-10$~keV)\tablefootmark{a}        & $2.8 \pm 0.2 \times 10^{-12}$ & $5.4 \pm 2.3 \times 10^{-12}$ \\
                    &  Luminosity ($0.2-10$~keV)\tablefootmark{b}  & $3.2 \pm 0.2 \times 10^{42}$ & $6.1 \pm 2.6 \times 10^{42}$ \\
    \noalign{\smallskip}
    \hline
  \end{tabular}
  \tablefoot{
    \tablefoottext{a}{In units of erg cm$^{-2}$~s$^{-1}$; }
    \tablefoottext{b}{in units of erg s$^{-1}$; }
    \tablefoottext{c}{in units of keV.}
  }
\end{table*}

Superimposed to the continuum we added four broad emission lines, modeled with Gaussians, representing the $1s-2p$~and $1s-3p$~transitions of \OVII{}, and the $1s-2p$~transition of \OVIII{}~and \CVI{}. Their parameter values were taken from the previous analysis of these datasets by \citet{Ste05}. In addition to them we included a set of narrow emission lines, namely the He-like triplets of \NeIX{}, \OVII{}, and \NVI{}, the forbidden line of the He-like \CV{}, and the $1s-2p$ transitions of \NeX{}~and \NVII{}, as well as radiative recombination continua (RRCs). With the exception of the forbidden lines of \OVII{}~and \CV{}, these lines are difficult to detect in the L02 and H02 spectra. Therefore we fixed their normalization to the values obtained in the 2013 campaign (\citealt{Whe15}), where the lines are clearly visible against the strongly obscured continuum. The narrow lines and RRCs are convolved with a velocity dispersion of $\sigma_{\rm v} = 463$~km s$^{-1}$ (see the supplementary material in \citealt{Kaa14}). For further information we point the reader to \citet{Whe15}, who present a detailed analysis of the narrow lines and RRCs in NGC 5548.

\begin{table*}
  \centering
  \caption[]{Emission features best-fit parameters of the 2002 observations. Values without uncertainties are kept frozen.}
  \label{em2002best}
  \begin{tabular}{l l c c c}
    \hline\hline
    \noalign{\smallskip}
    Component      &  Line            &  Rest wavelength\tablefootmark{b}  &  FWHM\tablefootmark{b} &  Luminosity\tablefootmark{c} \\
    \noalign{\smallskip}
    \hline
    \noalign{\smallskip}
    Broad emission          &  \CVI{}~$1s-2p$  & $33.74$ & $0.90$ & $1.9$ \\
    lines\tablefootmark{a}  &  \OVII{}~$1s-2p$  & $21.60$ & $0.58$ & $3.4$ \\
                            &  \OVII{}~$1s-3p$  & $18.63$ & $0.50$ & $1.4$ \\
                            &  \OVIII{}~$1s-2p$  & $18.97$ & $0.51$ & $2.8$ \\
    \noalign{\smallskip}
    \hline\hline
    \noalign{\smallskip}
    Component      &  Line  &  Rest wavelength\tablefootmark{b}  &  $\sigma_{\rm v}$\tablefootmark{d} &  Luminosity\tablefootmark{c} \\
    \noalign{\smallskip}
    \hline
    \noalign{\smallskip}
    Narrow emission   &  \OVII{}~$f$  & $22.101$ & $382$ & $4.7 \pm 0.4$ \\
    lines                   &  \CV{}~$f$  & $41.472$ & $382$ & $2.3 \pm 0.1$ \\
    \noalign{\smallskip}
    \hline\hline
    \noalign{\smallskip}
    Component      &  Line  &  Rest wavelength\tablefootmark{b}  &  $\sigma_{\rm v}$\tablefootmark{d} &  Luminosity\tablefootmark{c} \\
    \noalign{\smallskip}
    \hline
    \noalign{\smallskip}
    Narrow emission         &  \NeX{}~$1s-2p$  & $12.134$ & $463$ & $0.09$ \\
    lines\tablefootmark{e}  &  \NeIX{}~$r$  & $13.447$ & $''$ & $0.22$ \\
                            &  \NeIX{}~$i$  & $13.553$ & $''$ & $0.14$ \\
                            &  \NeIX{}~$f$  & $13.699$ & $''$ & $0.78$ \\
                            &  \OVIII{}~$1s-3p$  & $16.006$ & $''$ & $0.27$ \\
                            &  \OVII{}~$r$  & $21.602$ & $''$ & $1.09$ \\
                            &  \OVII{}~$i$  & $21.807$ & $''$ & $1.07$ \\
                            &  \NVII{}~$1s-2p$  & $24.781$ & $''$ & $0.33$ \\
                            &  \NVI{}~$r$  & $28.787$ & $''$ & $0.26$ \\
                            &  \NVI{}~$i$  & $29.084$ & $''$ & $0.25$ \\
                            &  \NVI{}~$f$  & $29.534$ & $''$ & $0.60$ \\
    \noalign{\smallskip}
    \hline\hline
    \noalign{\smallskip}
    Component      &  Ion  &  Emission measure\tablefootmark{f}  &  $\sigma_{\rm v}$\tablefootmark{d} &  Temperature\tablefootmark{g} \\
    \noalign{\smallskip}
    \hline
    \noalign{\smallskip}
    Radiative recombination          &  \CVI{} & $90$ & $463$ & $5.6$ \\
    continua (RRC)\tablefootmark{e}  &  \CVII{} & $61$ & $''$ & $''$ \\
                                     &  \NVII{} & $11$ & $''$ & $''$ \\
                                     &  \OVIII{} & $29$ & $''$ & $''$ \\
                                     &  \OIX{} & $7.3$ & $''$ & $''$ \\
                                     &  \NeX{} & $4.3$ & $''$ & $''$ \\
                                     &  \NeXI{} & $1.6$ & $''$ & $''$ \\
    \noalign{\smallskip}
    \hline
  \end{tabular}
  \tablefoot{
    \tablefoottext{a}{Values from \citet{Ste05}; }
    \tablefoottext{b}{in units of \AA; }
    \tablefoottext{c}{in units of $10^{40}$~erg s$^{-1}$; }
    \tablefoottext{d}{convolved velocity dispersion, in units of km s$^{-1}$; }
    \tablefoottext{e}{values from \citet{Whe15}; }
    \tablefoottext{f}{in units of $10^{59}$~cm$^{-3}$; }
    \tablefoottext{g}{in units of eV.}
  }
\end{table*}

The WA was modeled using the {\it xabs} model in SPEX, which calculates the transmission of a slab of material where all the ionic column densities are linked through a photoionization balance model. As shown in Sect.~\ref{motivation}, the choice of the photoionization code that computes this model is critical. Here we used CLOUDY version 13.01, and we provided as input the spectral energy distribution (SED) of \citet{Ste05} (see Fig.~\ref{sed_fig}). The output photoionization balance was provided to SPEX to fit the absorption troughs caused by an intervening WA. Since \citet{Ste05} used an older version of their photoionization code, we looked for further WA components that might have remained undetected in their analysis.

We started from the four WA components reported in \citet{Ste05}, and added additional {\it xabs} components on a one-by-one basis until the fit no longer improved. The {\it xabs} free parameters were the ionization parameter $\log \xi$, the hydrogen column density $N_{\rm H}$, the outflow velocity $v_{\rm out}$, and the Gaussian turbulent velocity $\sigma_{\rm v}$, while the covering factor was fixed to unity. The ionization parameter $\xi$ is a measure of the ionization state of the gas, and it is defined as

\begin{equation}
  \label{xidef}
  \xi = L/nR^2,
\end{equation}

\noindent where $L$ is the ionizing luminosity in the $1-1000$~Ryd range, $n$ is the hydrogen density of the gas, and $R$ is the distance of the gas to the ionizing source.

The L02 and H02 spectra were fit simultaneously, but we allowed the power law and modified black body parameters to be different for each dataset (to account for continuum variability), as well as the ionization parameters $\log \xi$ of the WA (to account for possible short-term variations, as the observations were taken 3.5 days apart). The rest of the parameters were coupled so that they had the same values for both datasets. We assumed that the WA components absorb the continuum and the broad emission lines, but not the narrow emission lines and RRCs.

We found that the WA in NGC 5548 is composed of six distinct ionization phases, labeled from $A$ to $F$ in order of increasing ionization. The final C-stat/d.o.f. was 2402/2090 for L02, and 2890/2560 for H02. The best-fit values of the joint L02 and H02 analysis are reported in Tables~\ref{cont2002best}, \ref{em2002best}, and \ref{WA2002best}. The $\log \xi$ values of these observations presented here differ from those shown in Table S2 in the supplementary material of \citealt{Kaa14} by $\sim 0.25$~dex due to a bug in CLOUDY, which caused the ionizing flux to be calculated in the $1-\infty$ range instead of the conventional range of $1-1000$~Ryd. The L02 spectrum together with the best-fit model are shown in Fig.~\ref{L02fluxed}.

\begin{table*}
  \centering
  \caption[]{Warm absorber best-fit parameters of the 2002 observations.}
  \label{WA2002best}
  \begin{tabular}{l c c c c c}
    \hline\hline
    \noalign{\smallskip}
    Component      &  $\log \xi$\tablefootmark{a} (L02) & $\log \xi$\tablefootmark{a} (H02) & $N_{\rm H}$\tablefootmark{b} & $\sigma_{\rm v}$\tablefootmark{c} & $v_{\rm out}$\tablefootmark{c}  \\
    \noalign{\smallskip}
    \hline
    \noalign{\smallskip}
    $A$  &  $0.51 \pm 0.08$  & $0.60 \pm 0.16$ & $2.0 \pm 0.6$ & $210 \pm 40$ & $-588 \pm 34$ \\
    $B$  &  $1.26 \pm 0.05$  & $1.42 \pm 0.06$ & $7.0 \pm 0.9$ & $61 \pm 15$ & $-547 \pm 31$ \\
    $C$  &  $1.90 \pm 0.03$  & $2.18 \pm 0.10$ & $15.3 \pm 3.3$ & $19 \pm 6$ & $-1148 \pm 20$ \\
    $D$  &  $2.10 \pm 0.03$  & $2.26 \pm 0.05$ & $10.7 \pm 1.6$ & $68 \pm 14$ & $-254 \pm 25$ \\
    $E$  &  $2.68 \pm 0.08$  & $2.64 \pm 0.05$ & $28.0 \pm 8.1$ & $24 \pm 12$ & $-792 \pm 25$ \\
    $F$  &  $2.87 \pm 0.05$  & $3.10 \pm 0.04$ & $57.1 \pm 17.0$ & $34 \pm 13$ & $-1221 \pm 25$ \\
    \noalign{\smallskip}
    \hline
  \end{tabular}
  \tablefoot{
    \tablefoottext{a}{In units of erg cm s$^{-1}$; }
    \tablefoottext{b}{in units of $10^{20}$~cm$^{-2}$; }
    \tablefoottext{c}{in units of km s$^{-1}$.}
  }
\end{table*}

\subsubsection{The remaining archival spectra}
\label{otherarchive}

After obtaining a best-fit model for the 2002 \Chan{}~spectra, we used it as a template for modeling the remaining archival datasets. As we mentioned in Sect.~\ref{strategy}, we left as free parameters in the fits only the continuum parameters (power law and modified black body emission), the flux of the narrow \OVII{}~forbidden emission line, and the ionization parameter of each WA component. The remaining parameters were kept fixed to those obtained in the joint fit of the L02 and H02 datasets. In modeling the WA, the ionization balance used as input by {\it xabs} was based on the SED of \citet{Ste05}, because it is the one with a wider contemporaneous frequency coverage. We therefore adopted it as a template SED for the rest of the archival observations. We note that, although ideally one would need to derive a SED for each of the individual observations, the assumption of a template SED will not significantly affect the {\it xabs} parameters determination. In the absence of huge spectral changes (such as those originated by obscuration events), the different SEDs may change their normalization due to the intrinsic variability of the source but not their shape, which is the key factor in computing the {\it xabs} parameters. The best-fit results are shown in Table~\ref{archivalbest}.

The ionization structure of the WA in NGC 5548 depicted in the 2002 \Chan{}~observations, showing six distinct ionization components, is also seen in the rest of the archival datasets. In some cases, some of the components are either not detected or are blended with other components due to a poor signal-to-noise ratio of the spectrum. This is the case for R00x (very short exposure, $\sim26$~ks) or L07 (very low intrinsic flux). A more detailed view of the parameter variability between the different archival observations is discussed in Sect.~\ref{discussion}.

\begin{table*}
  \centering
  \caption[]{Best-fit parameters of the remaining archival observations.}
  \label{archivalbest}
  \begin{tabular}{l c c c c c c c}
    \hline\hline
    \noalign{\smallskip}
    Dataset  &  L99  &  H00  &  R00x  &  R01y  &  R01z  &  L05  &  L07  \\
    \noalign{\smallskip}
    \hline
    \noalign{\smallskip}
    PL $\Gamma$ & $1.81 \pm 0.02$ & $1.56 \pm 0.01$ & $2.25 \pm 0.05$ & $1.83 \pm 0.05$ & $1.91 \pm 0.07$ & $1.53 \pm 0.04$ & $1.70 \pm 0.06$ \\
    PL $F_{\rm 0.2-10}$\tablefootmark{a} & $7.00 \pm 0.12$ & $3.90 \pm 0.06$ & $4.24 \pm 0.15$ & $6.75 \pm 0.15$ & $8.48 \pm 0.25$ & $1.90 \pm 0.06$ & $1.05 \pm 0.06$ \\
    PL $L_{\rm 0.2-10}$\tablefootmark{b} & $5.46 \pm 0.10$ & $2.84 \pm 0.04$ & $4.36 \pm 0.16$ & $5.40 \pm 0.12$ & $6.97 \pm 0.21$ & $0.57 \pm 0.02$ & $0.95 \pm 0.05$ \\
    \noalign{\smallskip}
    \hline
    \noalign{\smallskip}
    mbb Temp.\tablefootmark{c} & $0.138 \pm 0.004$ & $0.134 \pm 0.012$ & $0.090 \pm 0.014$ & $0.145 \pm 0.006$ & $0.144 \pm 0.010$ & $0.070 \pm 0.027$ & $\dots$ \\
    mbb $F_{\rm 0.2-10}$\tablefootmark{a} & $0.64 \pm 0.05$ & $0.47 \pm 0.19$ & $0.16 \pm 0.10$ & $0.47 \pm 0.05$ & $0.54 \pm 0.11$ & $< 0.04$ & $\dots$ \\
    mbb $L_{\rm 0.2-10}$\tablefootmark{b} & $0.68 \pm 0.06$ & $0.48 \pm 0.20$ & $0.35 \pm 0.22$ & $0.52 \pm 0.06$ & $0.61 \pm 0.12$ & $< 0.15$ & $\dots$ \\
    \noalign{\smallskip}
    \hline
    \noalign{\smallskip}
    \OVII{}~$f$ Lum.\tablefootmark{d} & $5.3 \pm 1.0$ & $8.4 \pm 1.7$ & $7.2 \pm 1.3$ & $7.7 \pm 0.7$ & $4.9 \pm 1.3$ & $2.4 \pm 0.4$ & $1.9 \pm 0.4$ \\
    \noalign{\smallskip}
    \hline
    \noalign{\smallskip}
    WA $A$~$\log \xi$\tablefootmark{e}  & $1.01 \pm 0.08$  & $1.19 \pm 0.15$  & $\dots$          & $0.52 \pm 0.09$  & $0.58 \pm 0.15$  & $0.07 \pm 0.42$  & $1.01 \pm 0.35$  \\
    WA $B$~$\log \xi$\tablefootmark{e}  & $1.71 \pm 0.06$  & $1.84 \pm 0.09$  & $0.80 \pm 0.20$  & $1.31 \pm 0.06$  & $1.32 \pm 0.08$  & $0.01 \pm 0.24$  & $-2.08 \pm 0.34$  \\
    WA $C$~$\log \xi$\tablefootmark{e}  & $1.93 \pm 0.06$  & $2.43 \pm 0.10$  & $1.74 \pm 0.11$  & $1.86 \pm 0.02$  & $1.78 \pm 0.03$  & $0.75 \pm 0.20$  & $0.22 \pm 0.35$  \\
    WA $D$~$\log \xi$\tablefootmark{e}  & $2.26 \pm 0.08$  & $2.52 \pm 0.08$  & $2.14 \pm 0.09$  & $2.09 \pm 0.05$  & $2.08 \pm 0.08$  & $1.80 \pm 0.14$  & $1.98 \pm 0.30$  \\
    WA $E$~$\log \xi$\tablefootmark{e}  & $2.69 \pm 0.07$  & $3.06 \pm 0.08$  & $2.37 \pm 0.11$  & $2.62 \pm 0.10$  & $2.57 \pm 0.11$  & $1.96 \pm 0.10$  & $2.33 \pm 0.33$  \\
    WA $F$~$\log \xi$\tablefootmark{e}  & $3.09 \pm 0.10$  & $3.19 \pm 0.08$  & $2.93 \pm 0.14$  & $2.83 \pm 0.07$  & $2.81 \pm 0.06$  & $2.59 \pm 0.13$  & $2.30 \pm 0.21$  \\
    \noalign{\smallskip}
    \hline
    \noalign{\smallskip}
    C-stat/d.o.f  & 2299/2090 & 2764/2560 & 1517/1102 & 1445/1069 & 1461/1067 & 2697/2214 & 1731/1659 \\
    \noalign{\smallskip}
    \hline
  \end{tabular}
  \tablefoot{
    \tablefoottext{a}{Observed flux, in units of $10^{-11}$~erg cm s$^{-1}$; }
    \tablefoottext{b}{in units of $10^{43}$~erg s$^{-1}$; }
    \tablefoottext{c}{in units of keV; }
    \tablefoottext{d}{in units of $10^{40}$~erg s$^{-1}$; }
    \tablefoottext{e}{In units of erg cm s$^{-1}$.}
  }
\end{table*}


\section{Discussion}
\label{discussion}

\subsection{Spectral properties}
\label{specprop}

\begin{figure*}
  \centering
  \includegraphics[width=12cm,angle=-90]{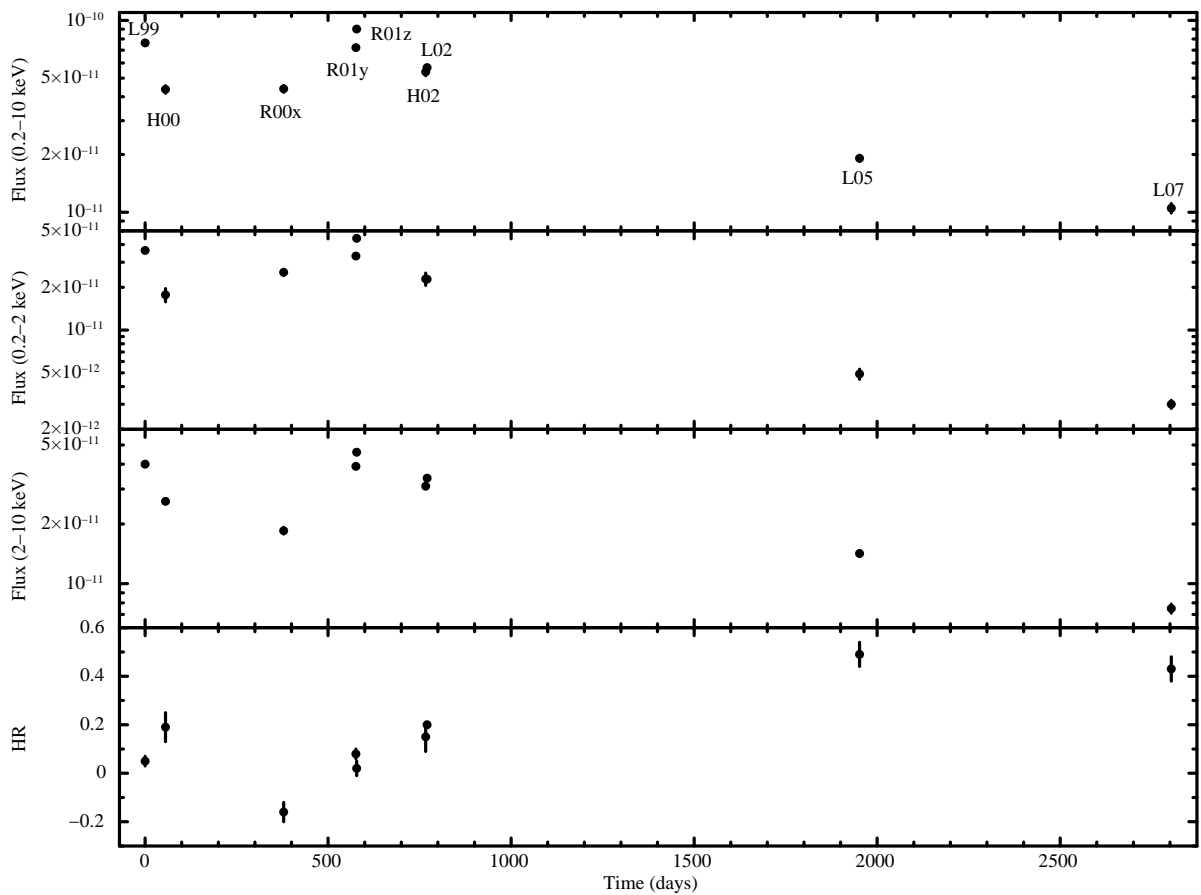}
  \caption{\label{vstime} Fluxes in the $0.2-10$~keV, $0.2-2$~keV, and $2-10$~keV bands, and hardness ratios ($HR = (H-S)/(H+S)$) as a function of time in days since the first observation (L99) for all of the observations used in this paper.}
\end{figure*}

The archival observations of NGC 5548 span almost 8 years, with the bulk of the datasets taken in the period $1999-2002$. In Fig.~\ref{vstime} we show the evolution in time of the overall $0.2-10$~keV flux of the source to provide a reference on the flux state of the source at the epoch of each observation, along with the soft ($0.2-2$~keV) and hard ($2-10$~keV) fluxes, and their hardness ratios ($HR = (H-S)/(H+S)$).

In the $1999-2002$ period, NGC 5548 experienced a maximum flux variability of a factor of $\sim 2$, with a median flux of $\sim 5 \times 10^{-11}$~erg cm$^{-2}$~s$^{-1}$. In 2005, the overall flux decreased by almost a factor of 3 with respect to the L02 observation, and even further in 2007 when it reached a minimum of $\sim 10^{-11}$~erg cm$^{-2}$~s$^{-1}$. The behavior of the soft ($0.2-2$~keV) and hard ($2-10$~keV) X-ray fluxes is very similar, following the same trends as the overall X-ray flux over the years. This suggests that the source variability is due to intrinsic changes rather than obscuration processes such as the one reported in \citet{Kaa14}. During the obscuration event the hard X-ray flux stayed at the historical values while the soft X-ray flux decreased dramatically. Furthermore, this supports our approach of using the 2002 SED for the WA analysis of all the archival observations, as the variations only involved the normalization but did not significantly changed the shape of the SED. More detailed studies of the broadband continuum of NGC 5548 is reported in \citet{Meh14a}~(Paper I) and \citet{Urs14}~(Paper III). An in-depth discussion on the origin and variability of the soft excess in NGC 5548 is also reported in Papers I and VI (\citealt{Meh14a, Meh14b}).

\subsection{Thermal stability of the warm absorber}
\label{scurves}

\begin{figure*}
  \centering
  \hbox{
  \includegraphics[width=6.5cm,angle=-90]{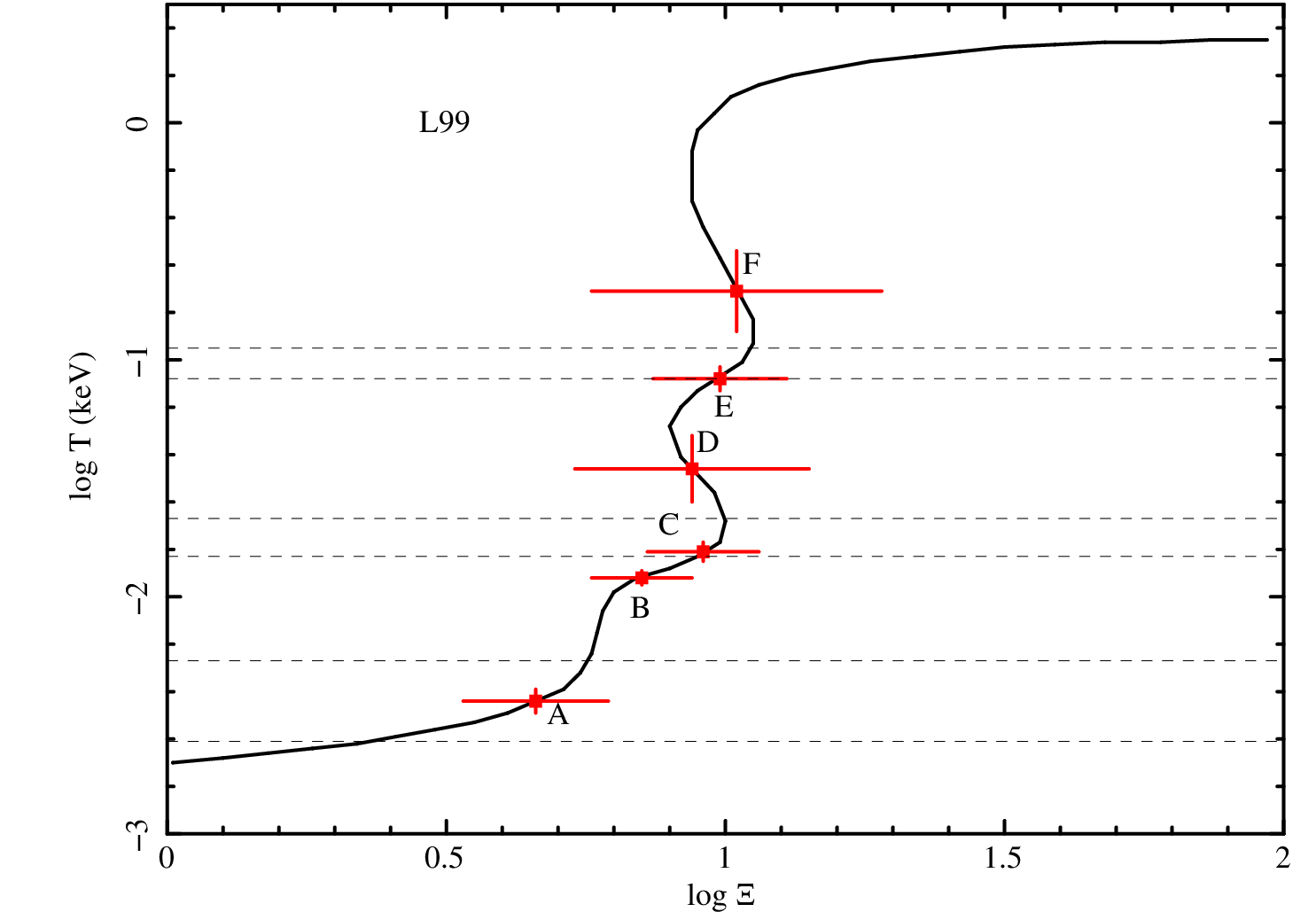}
  \includegraphics[width=6.5cm,angle=-90]{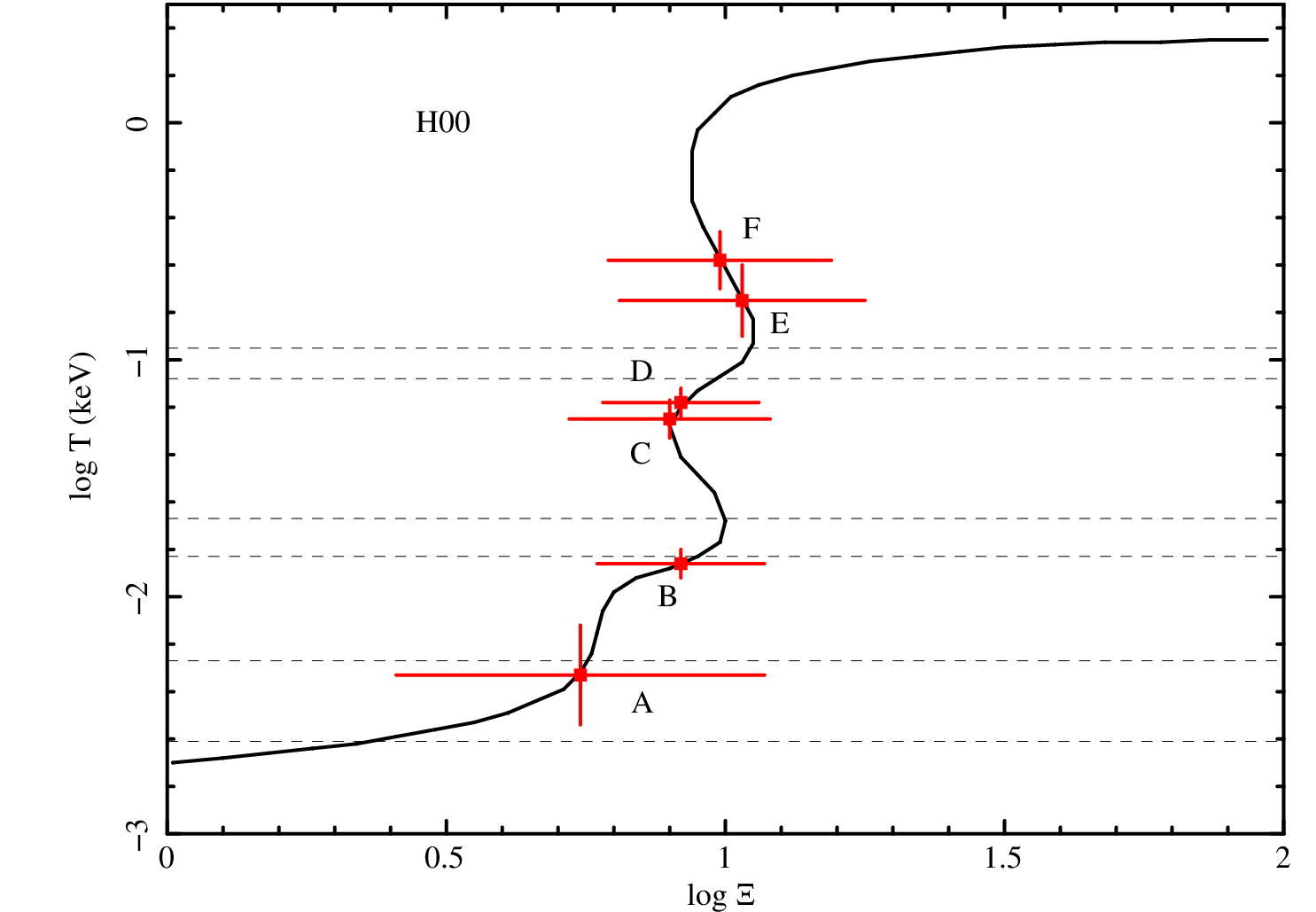}
  }
  \hbox{
  \includegraphics[width=6.5cm,angle=-90]{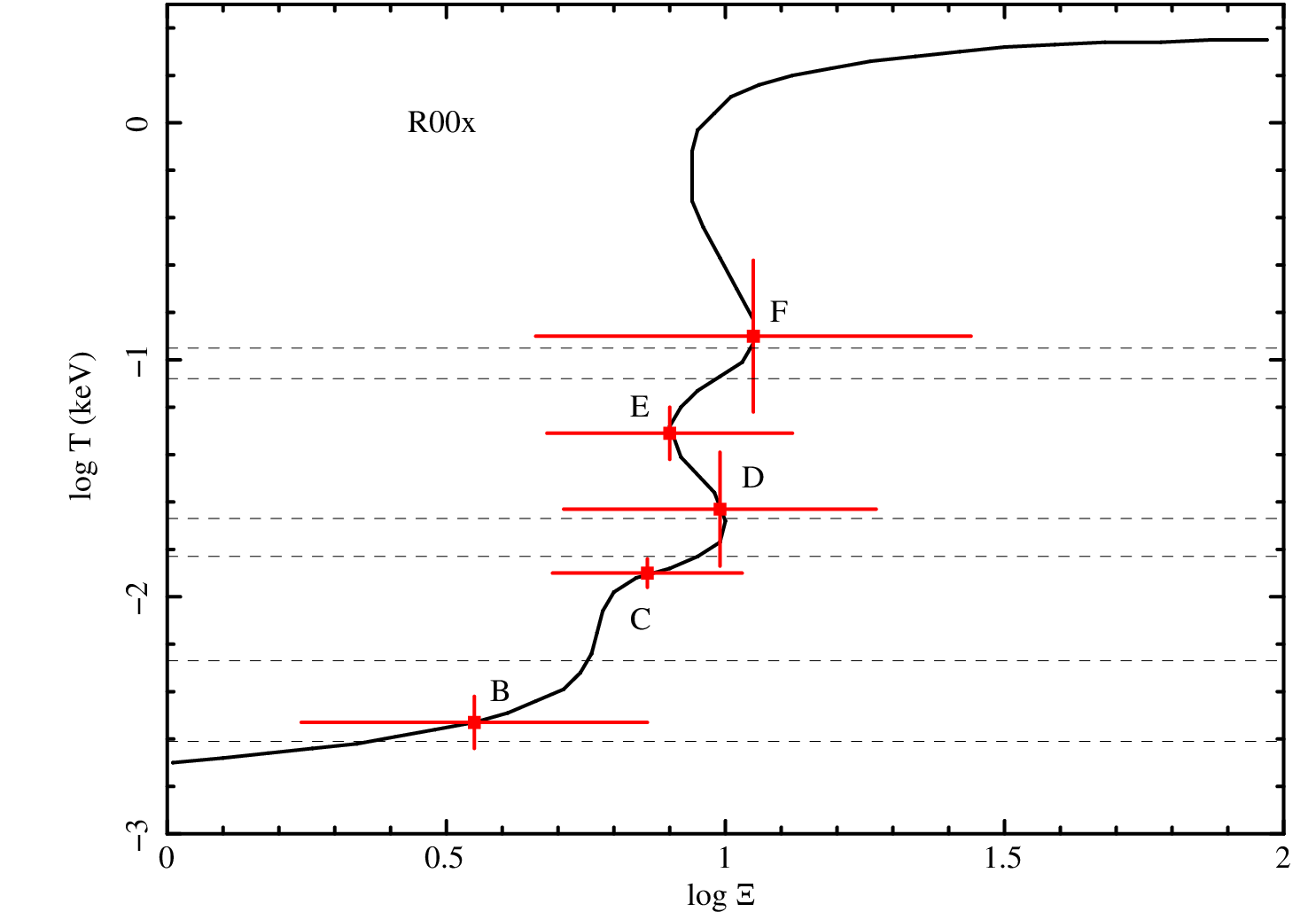}
  \includegraphics[width=6.5cm,angle=-90]{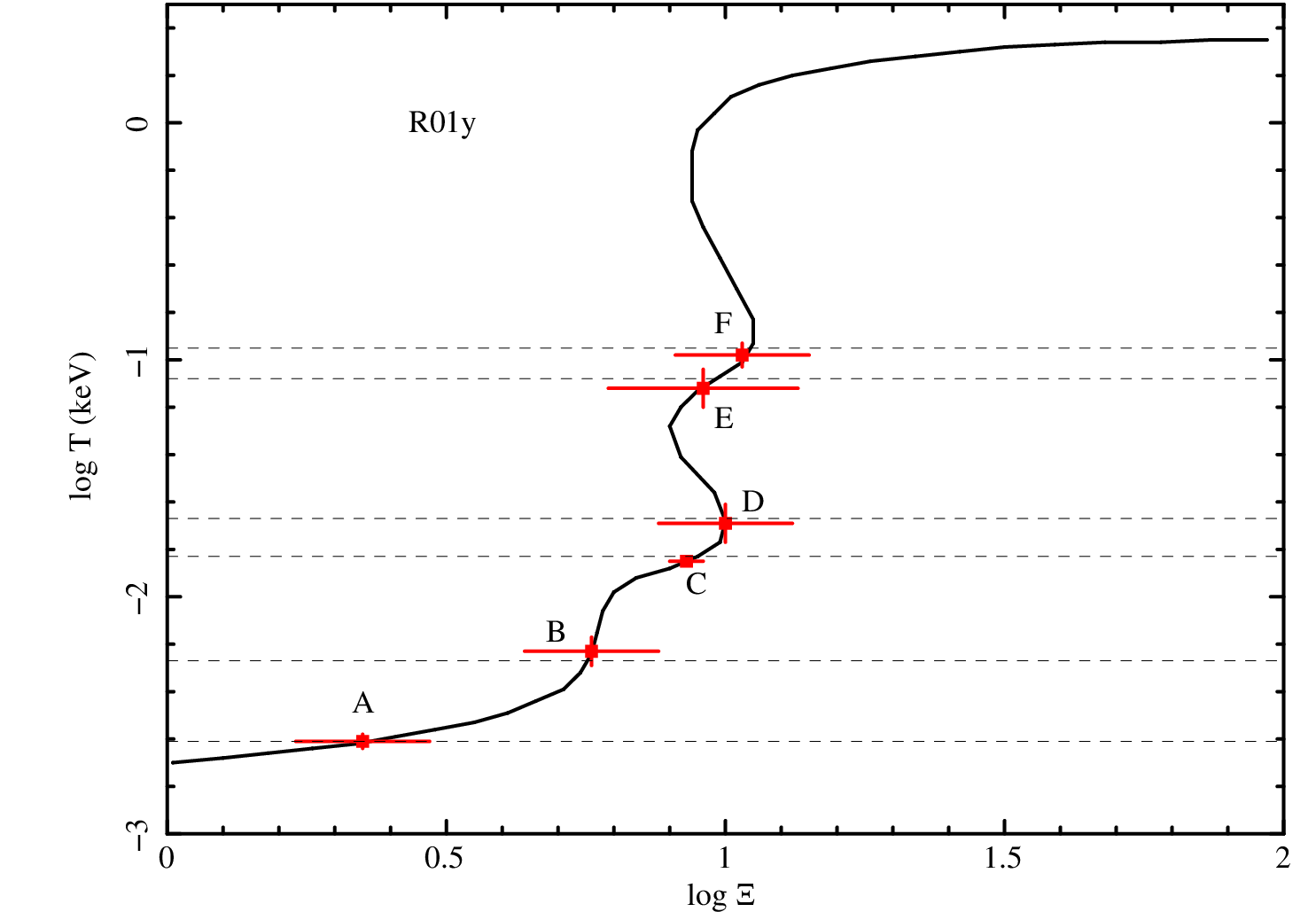}
  }
  \hbox{
  \includegraphics[width=6.5cm,angle=-90]{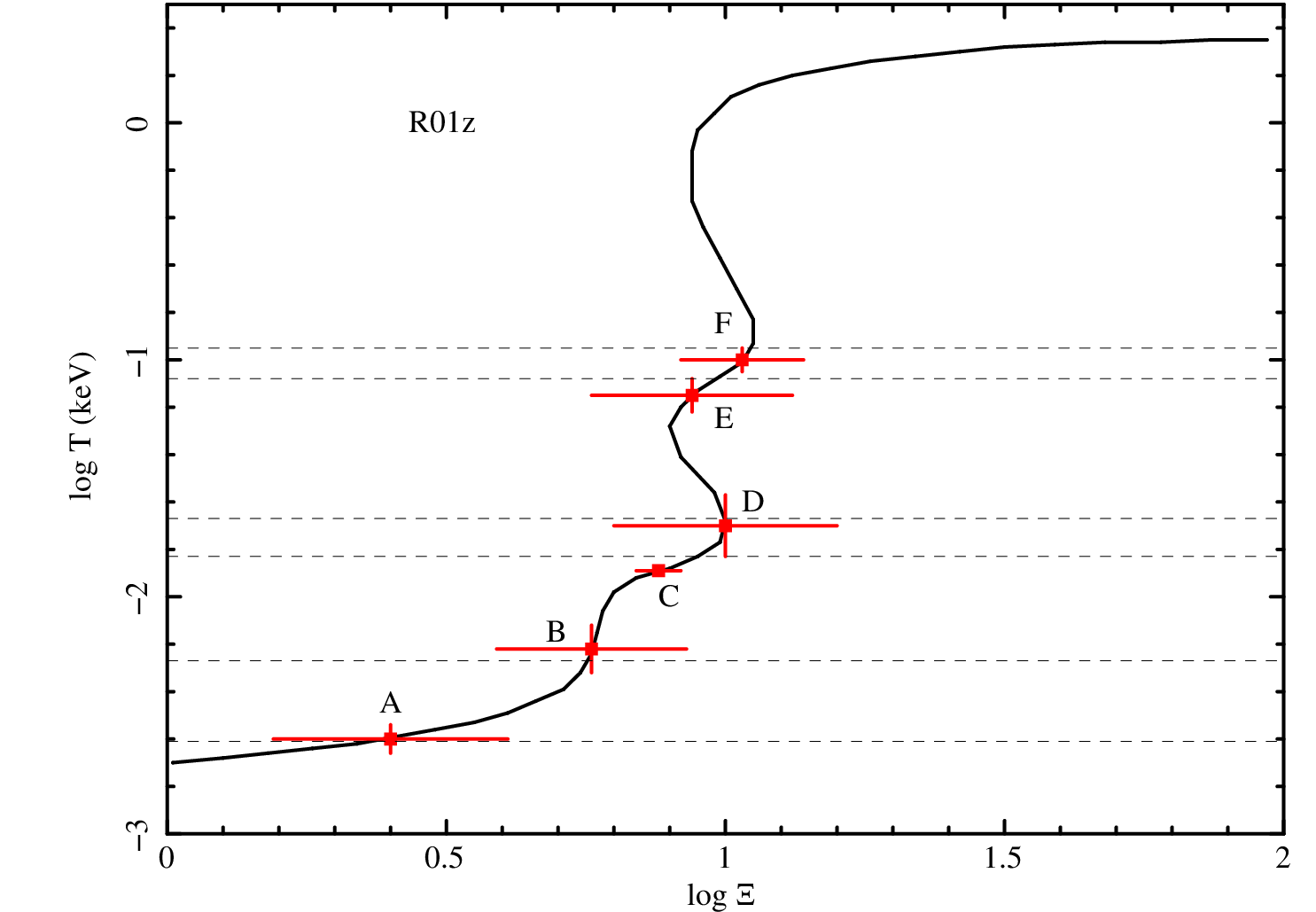}
  \includegraphics[width=6.5cm,angle=-90]{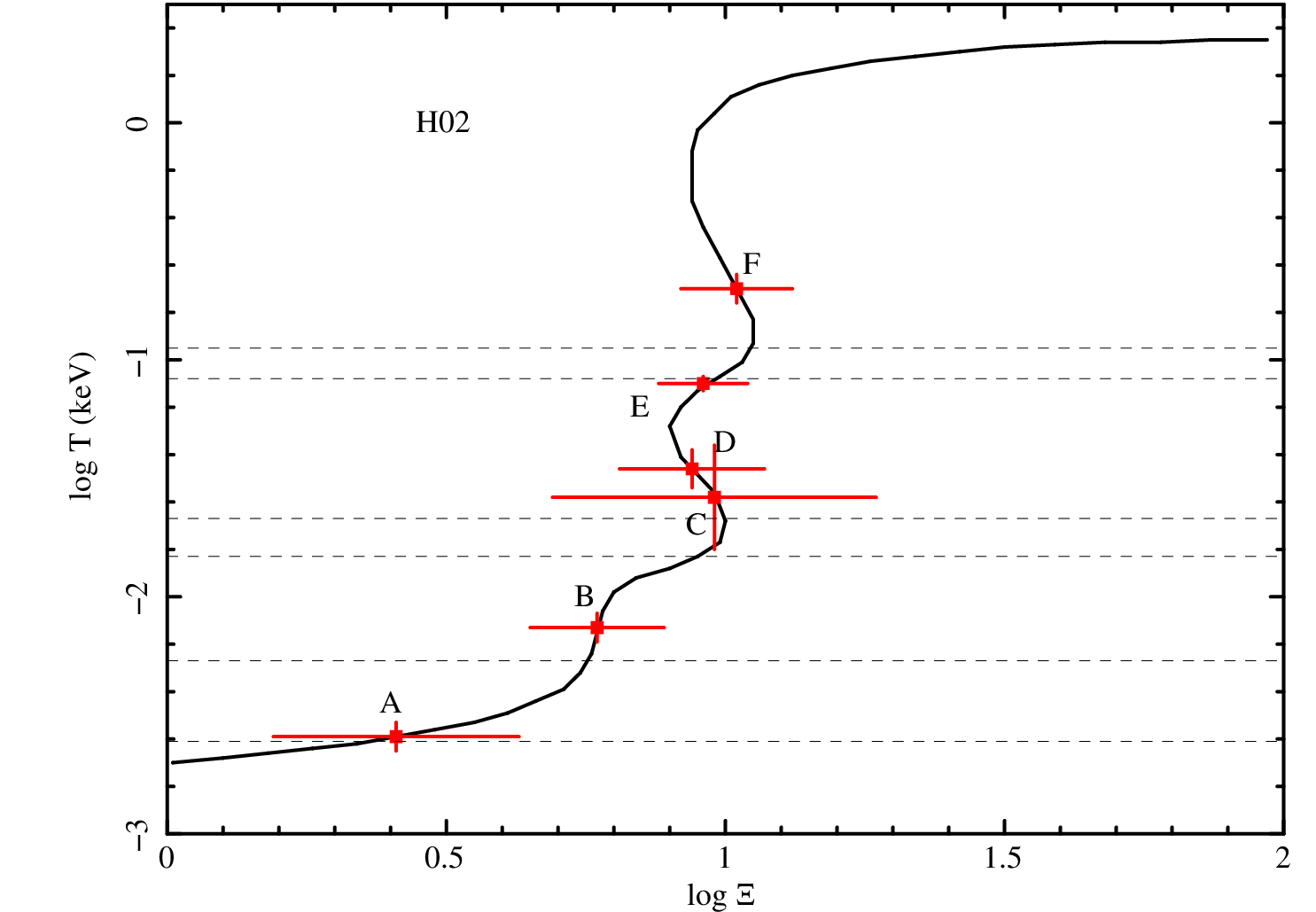}
  }
  \caption{\label{scurves1}Thermal stability curves of the NGC 5548 archival observations (L99 and H00 in the top panels; R00x and R01y in the middle panels; R01z and H02 in the bottom panels) showing the pressure ionization parameter as a function of the electron temperature (solid line). The curves have been computed from the SED of \citet{Ste05}. The X-ray WA components in each observation are represented as solid squares ($A$ to $F$). For comparison, the dashed horizontal lines represent in each panel the location of the WA components in the L02 observation.}
\end{figure*}

\begin{figure*}
  \centering
  \hbox{
  \includegraphics[width=6.5cm,angle=-90]{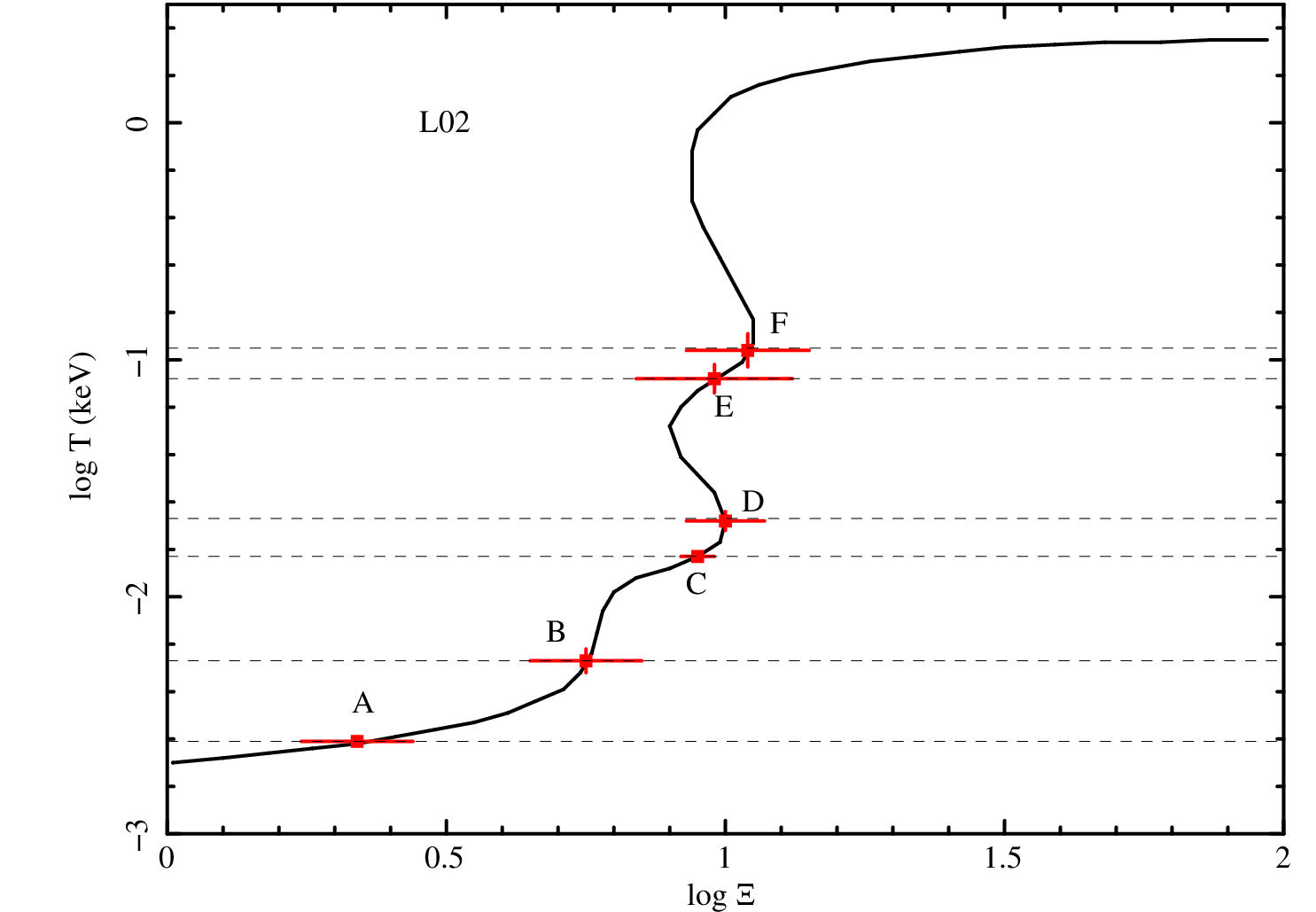}
  \includegraphics[width=6.5cm,angle=-90]{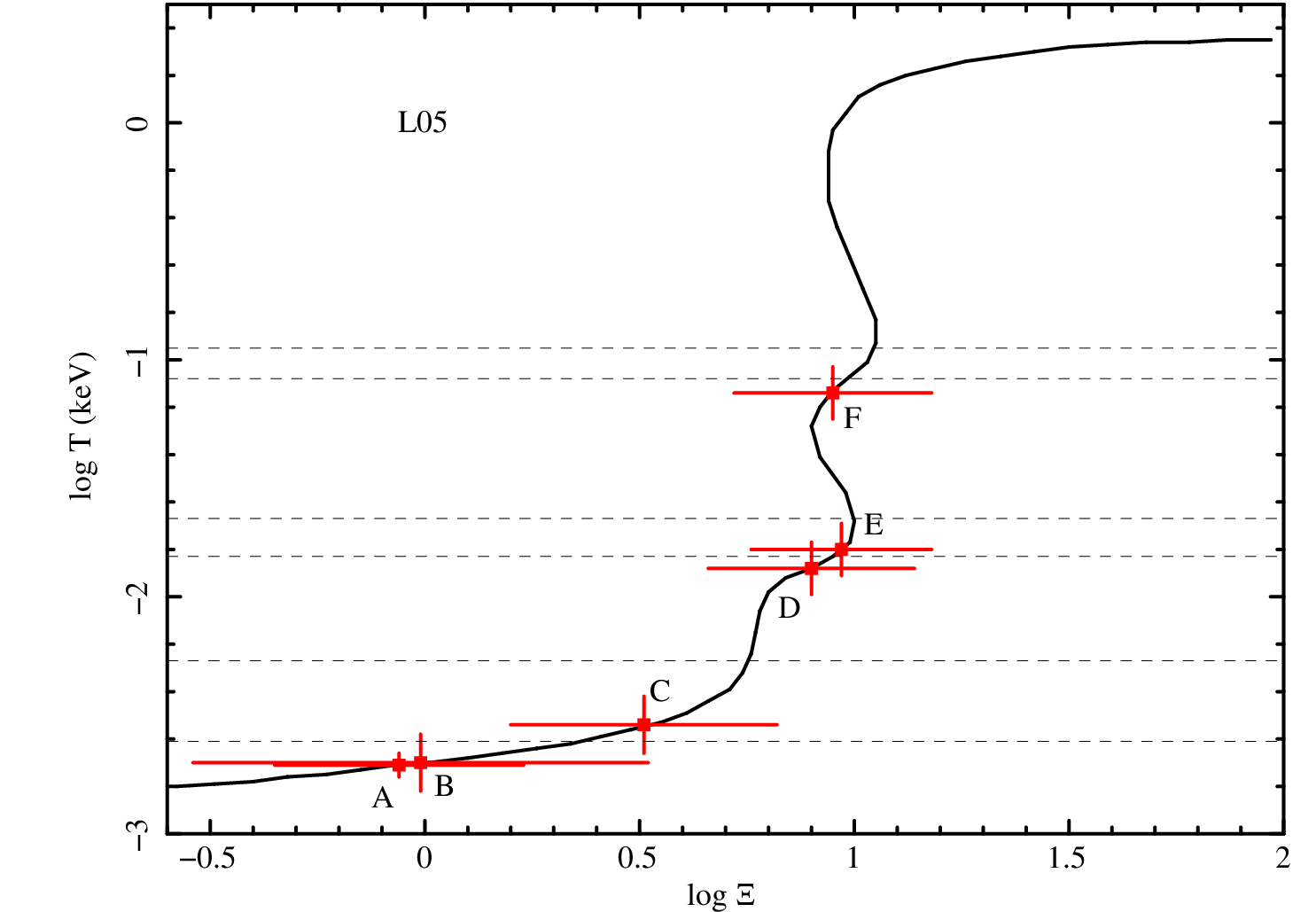}
  }
  \includegraphics[width=6.5cm,angle=-90]{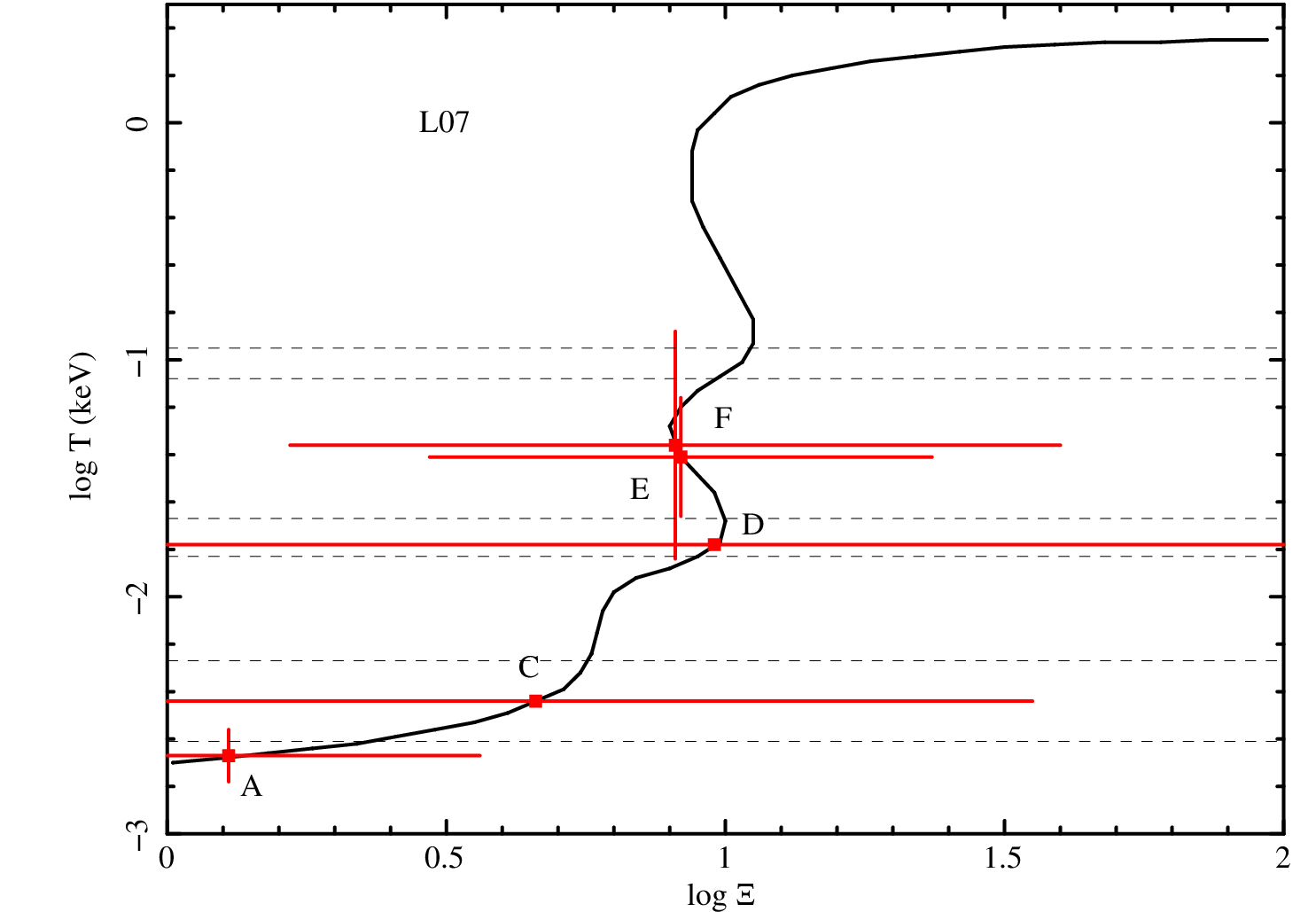}
  \caption{\label{scurves2}Thermal stability curves of the NGC 5548 archival observations (L02 and L05 in the top panels; L07 in the bottom panel) showing the pressure ionization parameter as a function of the electron temperature (solid line). The curves have been computed from the SED of \citet{Ste05}. The X-ray WA components in each observation are represented as solid squares ($A$ to $F$). For comparison, the dashed horizontal lines represent in each panel the location of the WA components in the L02 observation.}
\end{figure*}

Some clues on the structure of the X-ray WA can be drawn from the thermal stability curves of the photoionized gas. These curves, also known as cooling curves, are determined by the ionizing SED that illuminates the gas. They are typically represented as plots of the pressure ionization parameter $\Xi$ as a function of the electron temperature $T$. The pressure ionization parameter $\Xi$ (\citealt{Kro81}) is defined as

\begin{equation}
\label{xicapdef}
\Xi = L/4 \pi r^2 cp = \xi / 4\pi ckT,
\end{equation}

\noindent where $\xi$ is the ionization parameter defined in Eq.~\ref{xidef}, $c$ is the speed of light, $k$ is the constant of Boltzmann, and $T$ is the electron temperature. The different values of $\Xi$ were computed using CLOUDY version 13.01, which provided a grid of ionization parameters $\xi$ and their corresponding electron temperatures $T$ for a thin layer of gas irradiated by the ionizing continuum. For consistency with the analysis we performed in Sect.~\ref{modeling}, we used as input SED that of \citet{Ste05} for all of the archival observations, and assumed the proto-Solar abundances of \citet{LP09}.

The curve thus generated divides the $\Xi - T$ plane in two regions: below the curve heating dominates cooling, whereas above the curve cooling dominates heating. In the points on the curve, the heating rate equals the cooling rate so that the gas is in thermal equilibrium. The branches of the curve with negative derivative $dT/d\Xi < 0$ (those where the curve turns backwards) are unstable against isobaric thermal perturbations (e.g., vertical displacements from the curve). This means that for a photoionized gas in an unstable branch, if it is subject to a small positive temperature perturbation its temperature will keep increasing until it reaches a stable branch of the curve (those with positive derivative $dT/d\Xi > 0$). Conversely, a small negative temperature perturbation will cause the gas temperature to decrease until it falls in a stable region of the $\Xi - T$ plane. Moreover, gas components sharing the same $\Xi$ value are nominally in pressure equilibrium with each other, and therefore they are likely part of the same long-lived structure.

\begin{figure*}
  \centering
  \hbox{
  \includegraphics[width=6.5cm,angle=-90]{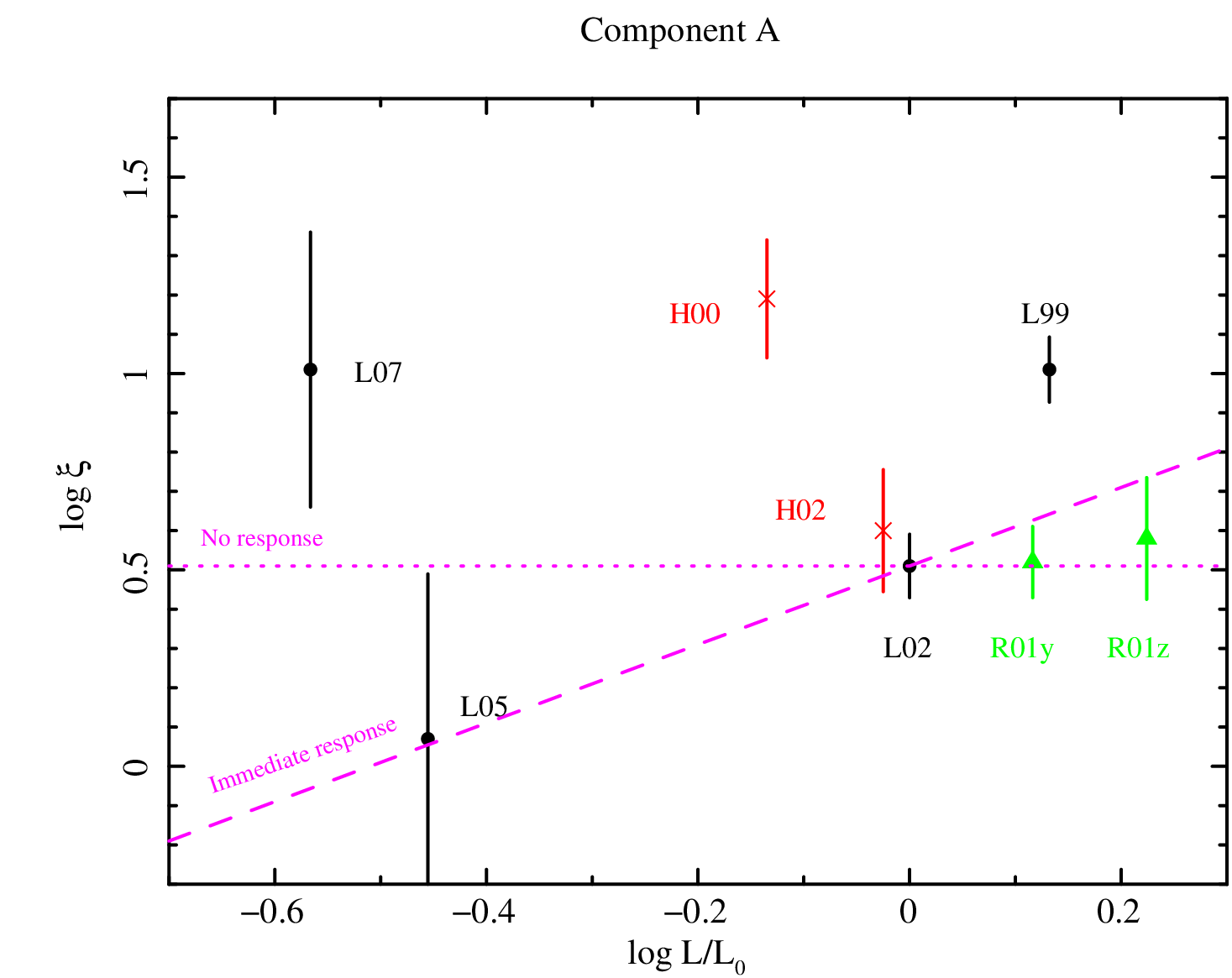}
  \includegraphics[width=6.5cm,angle=-90]{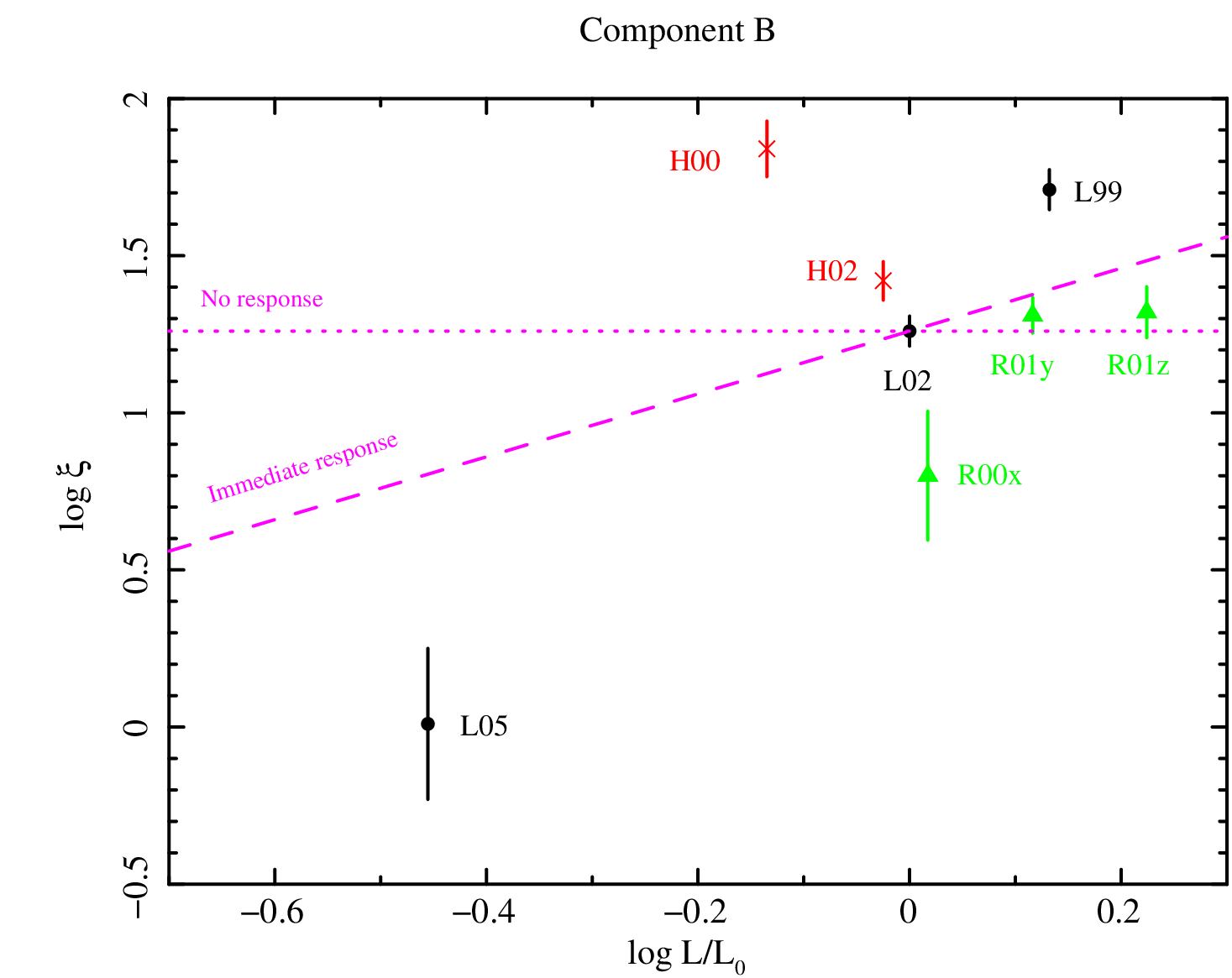}
  }
  \hbox{
  \includegraphics[width=6.5cm,angle=-90]{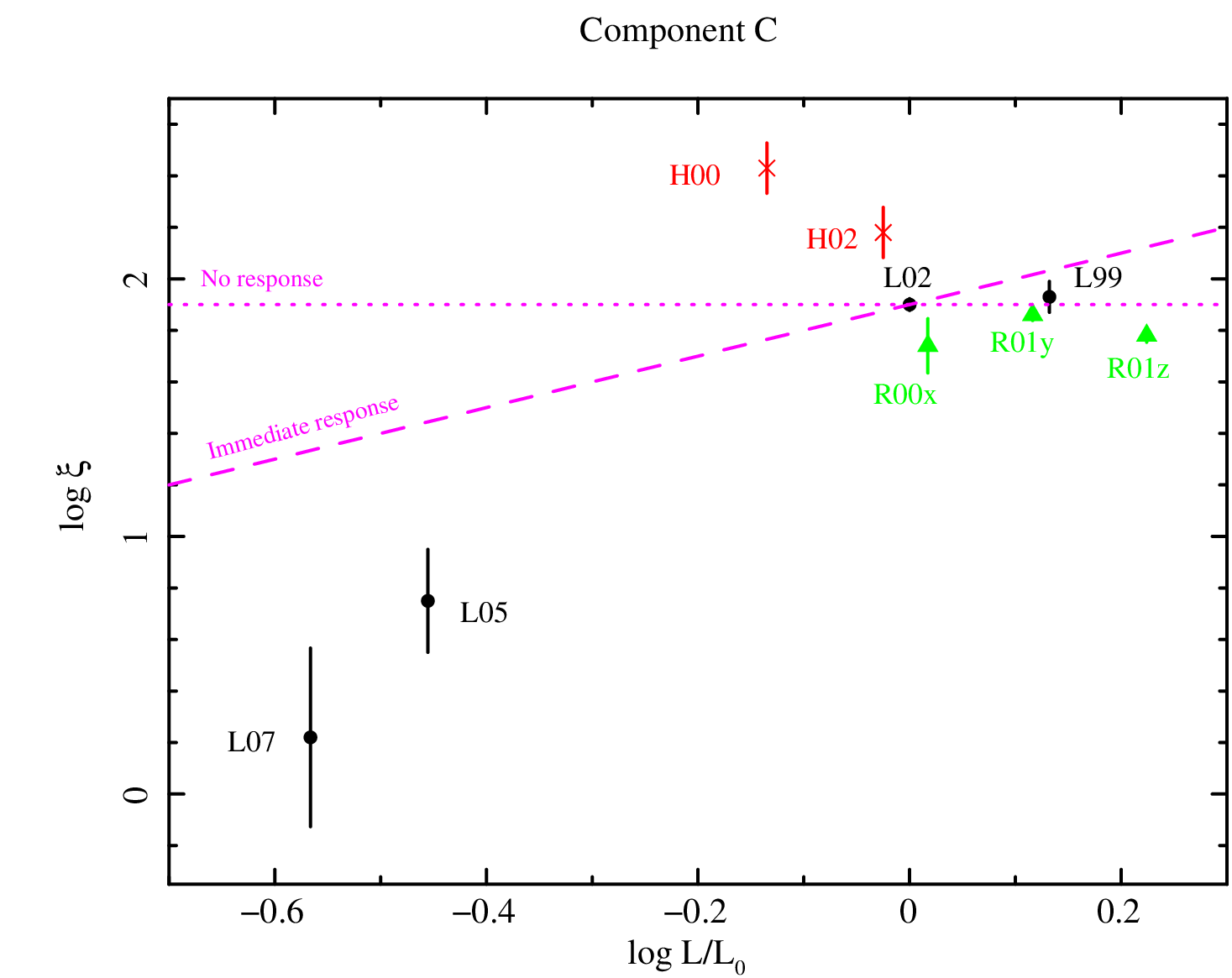}
  \includegraphics[width=6.5cm,angle=-90]{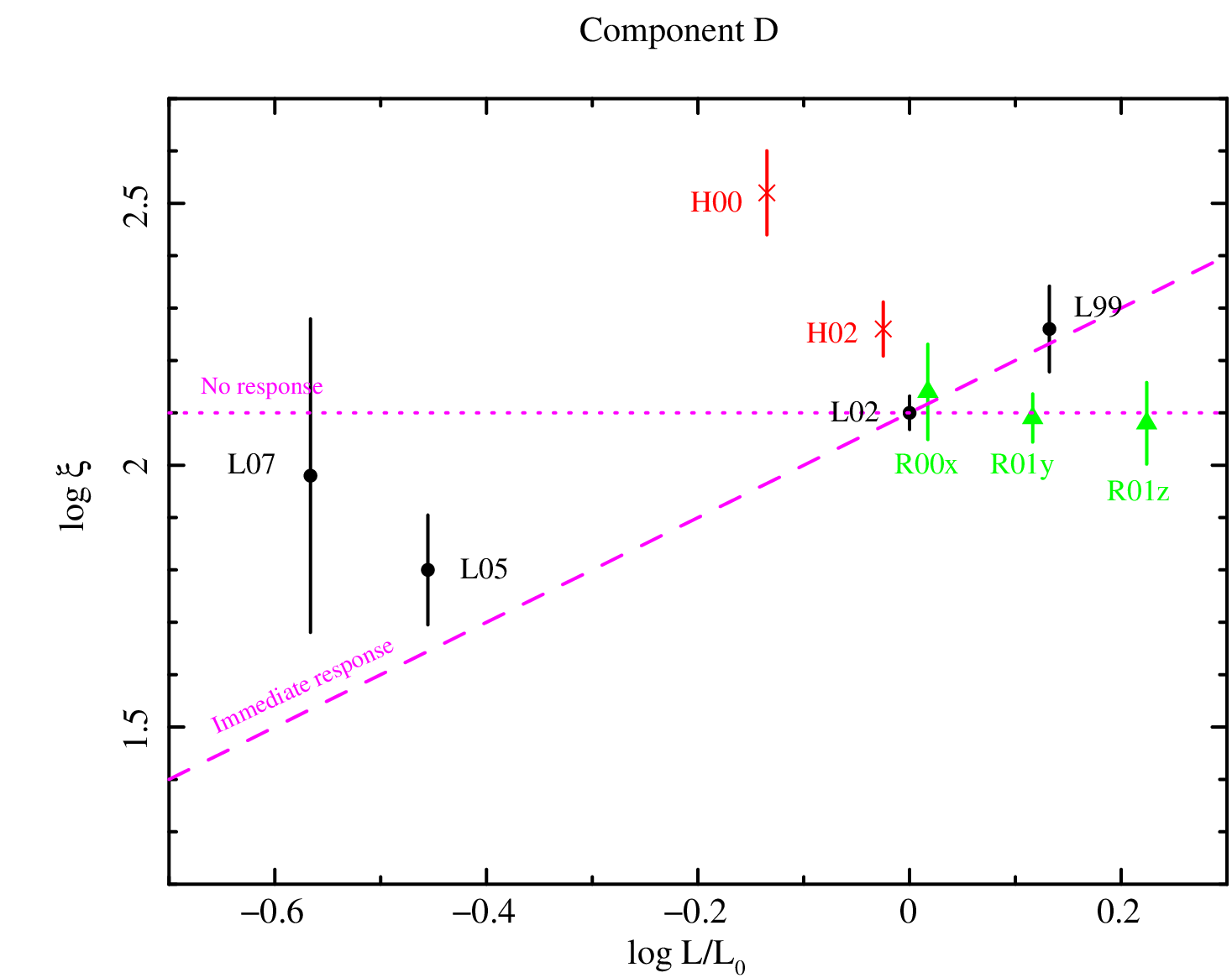}
  }
  \hbox{
  \includegraphics[width=6.5cm,angle=-90]{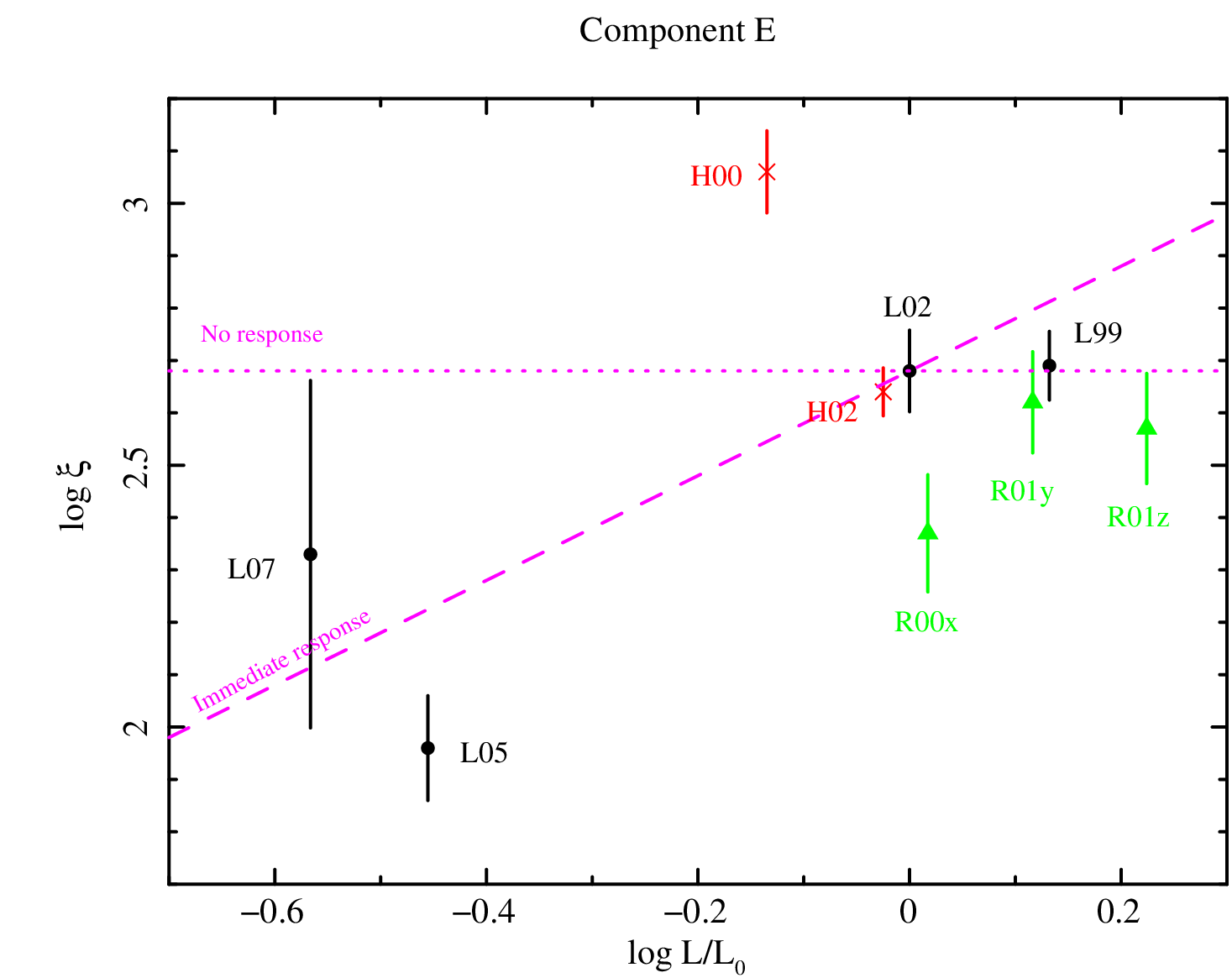}
  \includegraphics[width=6.5cm,angle=-90]{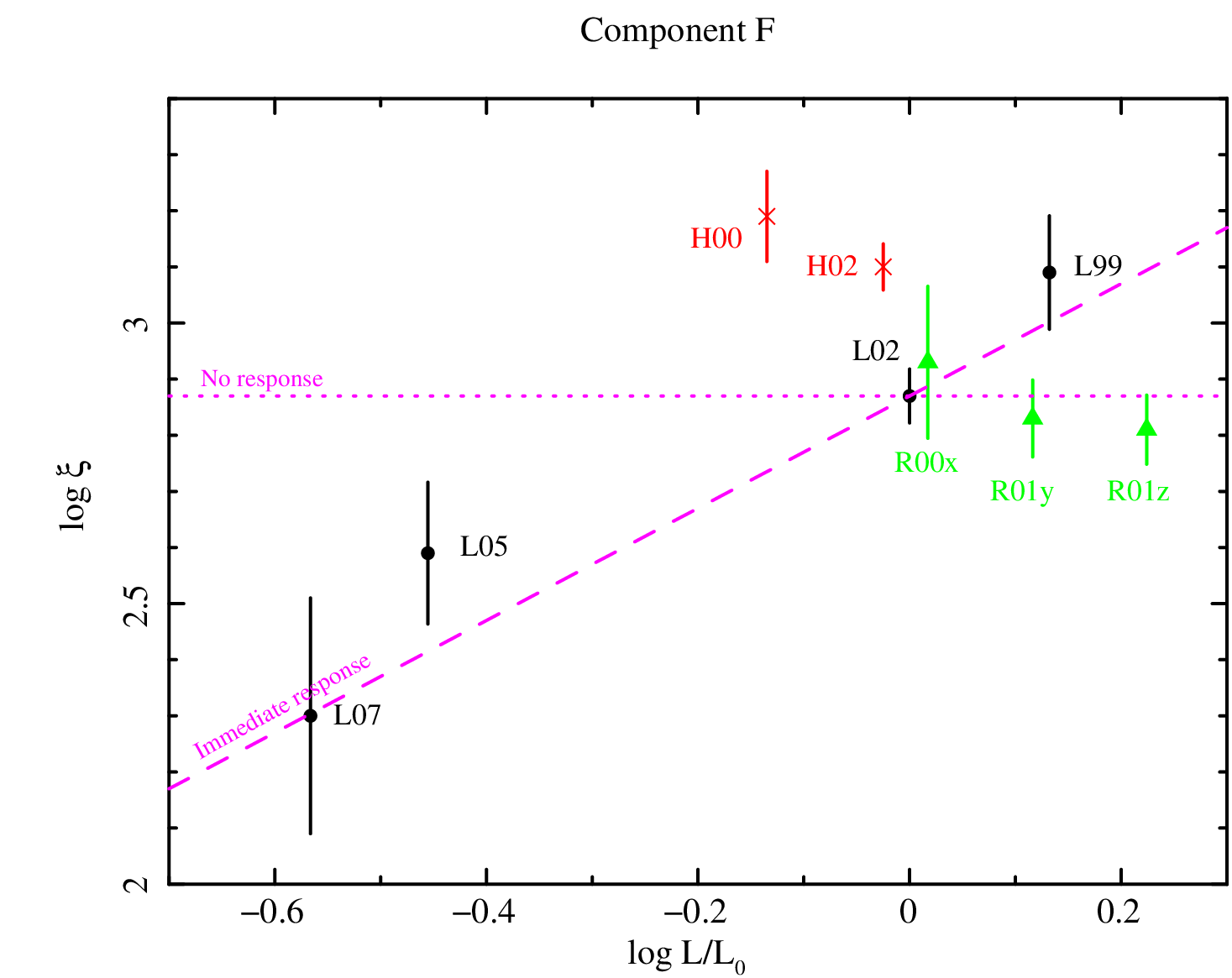}
  }
  \caption{\label{xilum}Ionization parameter $\xi$ against the ionizing luminosity $L$ in units of the luminosity of the L02 observation $L_0$ for components $A$ and $B$ (top panels), $C$ and $D$ (middle panels), and $E$ and $F$ (bottom panels). Black filled circles represent the LETGS observations, red crosses the HETGS observations, and green filled triangles the RGS observations. The dotted horizontal line indicates no response to continuum changes, while the dashed diagonal shows $\xi$ proportional to $L$ and indicates instantaneous response to variations in the ionizing flux.}
\end{figure*}

In Figs.~\ref{scurves1}~and \ref{scurves2}~we show the thermal stability curves of the archival observations of NGC 5548 analyzed in this paper. The curves are the same in all of the observations since the same input SED was used to generate the curve. Overplotted on the curves are the different WA components detected in each archival observation. While in most of the cases the WA components $B$ to $F$ seem to be in pressure equilibrium, the extent of the uncertainties makes the opposite equally likely, and therefore there are no grounds to claim pressure equilibrium. The large error bars associated with most of the observations are mainly due to the uncertainties in the determination of $\xi$. It is possible that some of the components might be truly in pressure equilibrium, but their different outflow velocities and turbulence suggest that they likely originate from different spatial regions.

On the other hand, we can examine the evolution of the WA components as they move along the curve between the different epochs of the archival observations. Since the uncertainty of the WA components in the vertical axis is much smaller than in the horizontal axis, it is possible to examine vertical displacements along the curve. For this reason, for comparison among the different datasets, we also show as horizontal dashed lines in Figs.~\ref{scurves1}~and \ref{scurves2} the location of the WA components in observation L02, the one with the best determined WA parameters. In the first grating observation (L99) the flux of the source was slightly higher than in L02. The WA components are higher on the curve than at the L02 epoch except for components $C$ and $E$, which were in a similar ionization state. In H00, an observation taken almost 2 months later, the overall flux decreased by a factor of 2 but all the WA components were in a higher ionization state. Component $C$ "jumped" the unstable branch that was just above its position in the L99 curve, almost blending with component $D$. Similar jumps happen for components $E$ and $F$. In this case, the drop in flux between L99 and H00 could have happened shortly before the observation, not giving the gas enough time to respond to this change. In support of this argument, R00x was observed almost 11 months after H00, maintaining a similar flux level, but with all components significantly less ionized. The uncertainties in this observation, however, are quite large due to the short exposure (only 26~ks). R01y, observed $\sim200$~days later, has a flux higher by a factor of $\sim 2$ compared to R00x, and it has all six components in an ionization state fully consistent with L02. Similarly for R01z, observed two days later. H02, taken six months later with a flux half of that in R01z, shows little change in the WA components, except for $D$ and $F$ which climbed in the curve toward the unstable branches above them. Again, we lack here information about the flux history of the source in the days or weeks prior to this observation, which could provide a hint on what initiated this change. Interestingly, when the L02 observation was taken 4 days later these components had returned to their positions in the stable branches of the curve, which tells us something about the time scales at which they respond (but see the discussion in Sect.~\ref{WAvar}). Finally, on much longer time scales, L05 was observed $\sim 3$~years later when the source had a flux level $\sim 3$~times lower than in the L02 epoch. Even though the error bars are rather large, one can see a dramatic change in the location of the WA components on the curve, all of them moving toward lower positions in the curve. The last observation in 2007 has the lowest overall flux, several times lower than that of L02, and the associated error bars are so large that no useful conclusion can be extracted from it other than the WA components seem to be located at similar positions as in 2005.

It is worth noting again that each observation is just a snapshot of a dynamical picture in which each gas component is reacting differently to changes in the ionizing flux over the years, a picture that is incomplete since we do not have a continuous monitoring of the flux history of the source between observations. However, this provides support for the fact that the response to flux variations on different time scales is motivated by the different densities of the absorbing gas components. If their densities could be constrained through their variability, we would have a powerful tool to also constrain their location with respect to the central ionizing source.

\subsection{Warm absorber variability}
\label{WAvar}

\begin{figure*}
  \centering
  \hbox{
  \includegraphics[width=6.5cm,angle=-90]{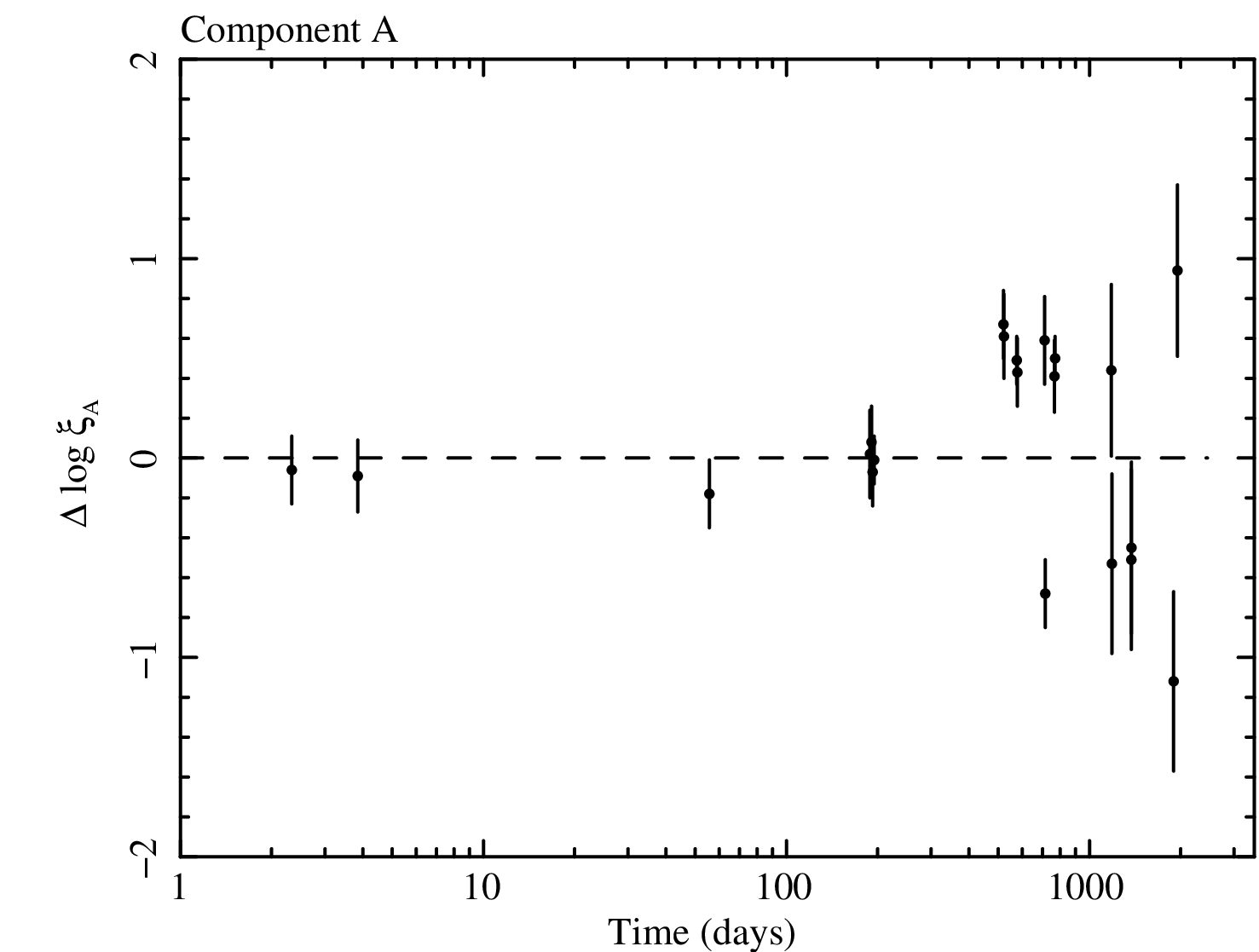}
  \includegraphics[width=6.5cm,angle=-90]{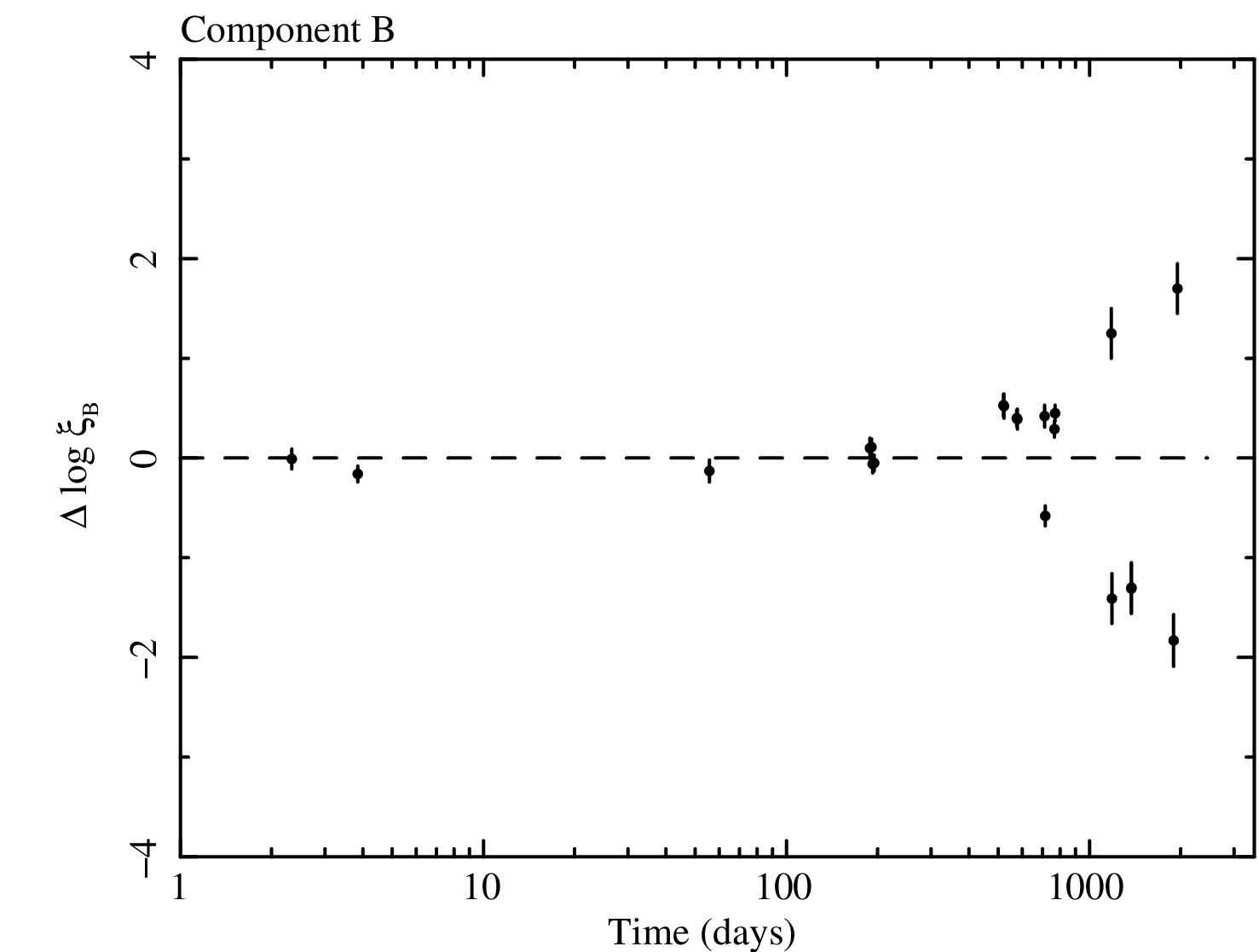}
  }
  \hbox{
  \includegraphics[width=6.5cm,angle=-90]{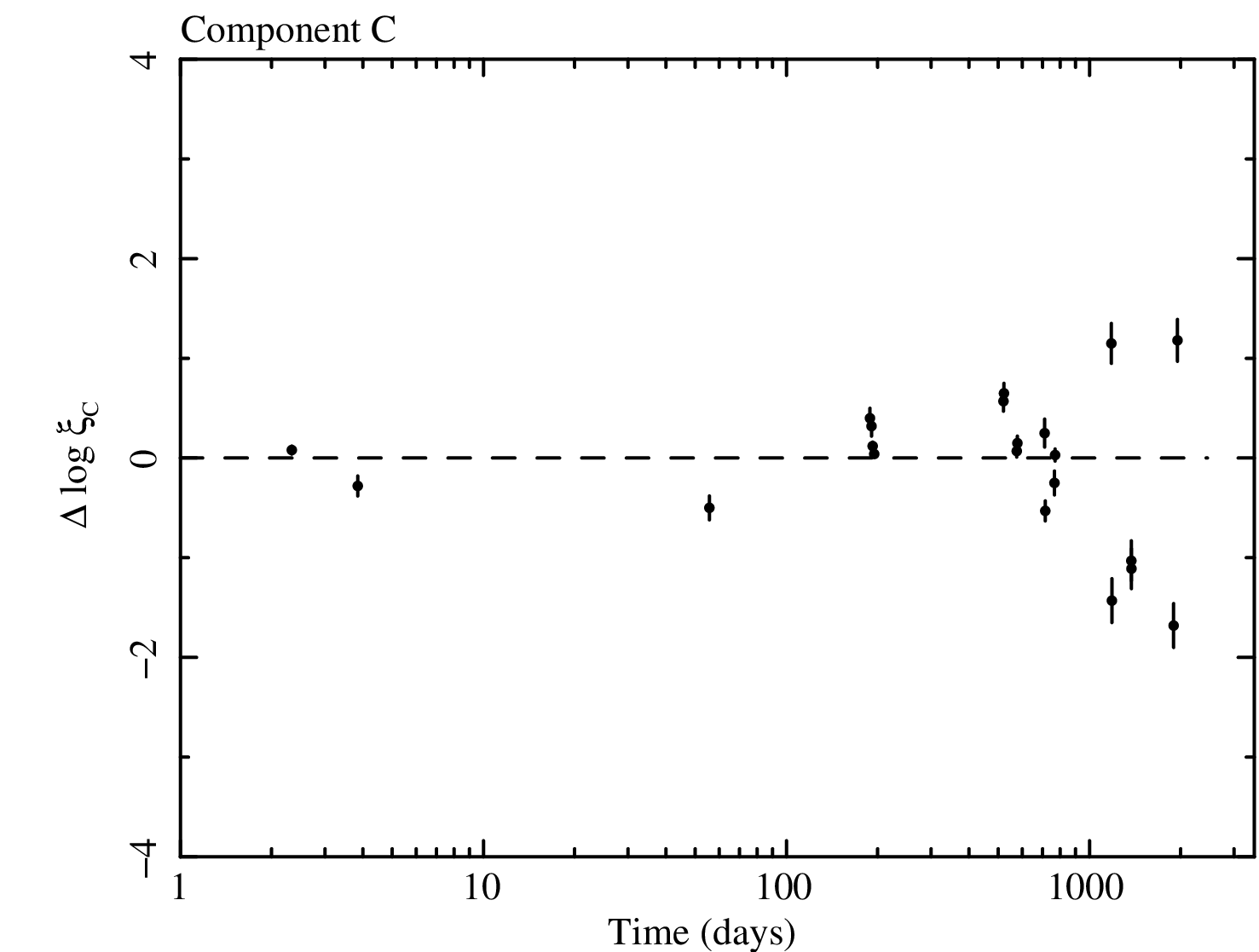}
  \includegraphics[width=6.5cm,angle=-90]{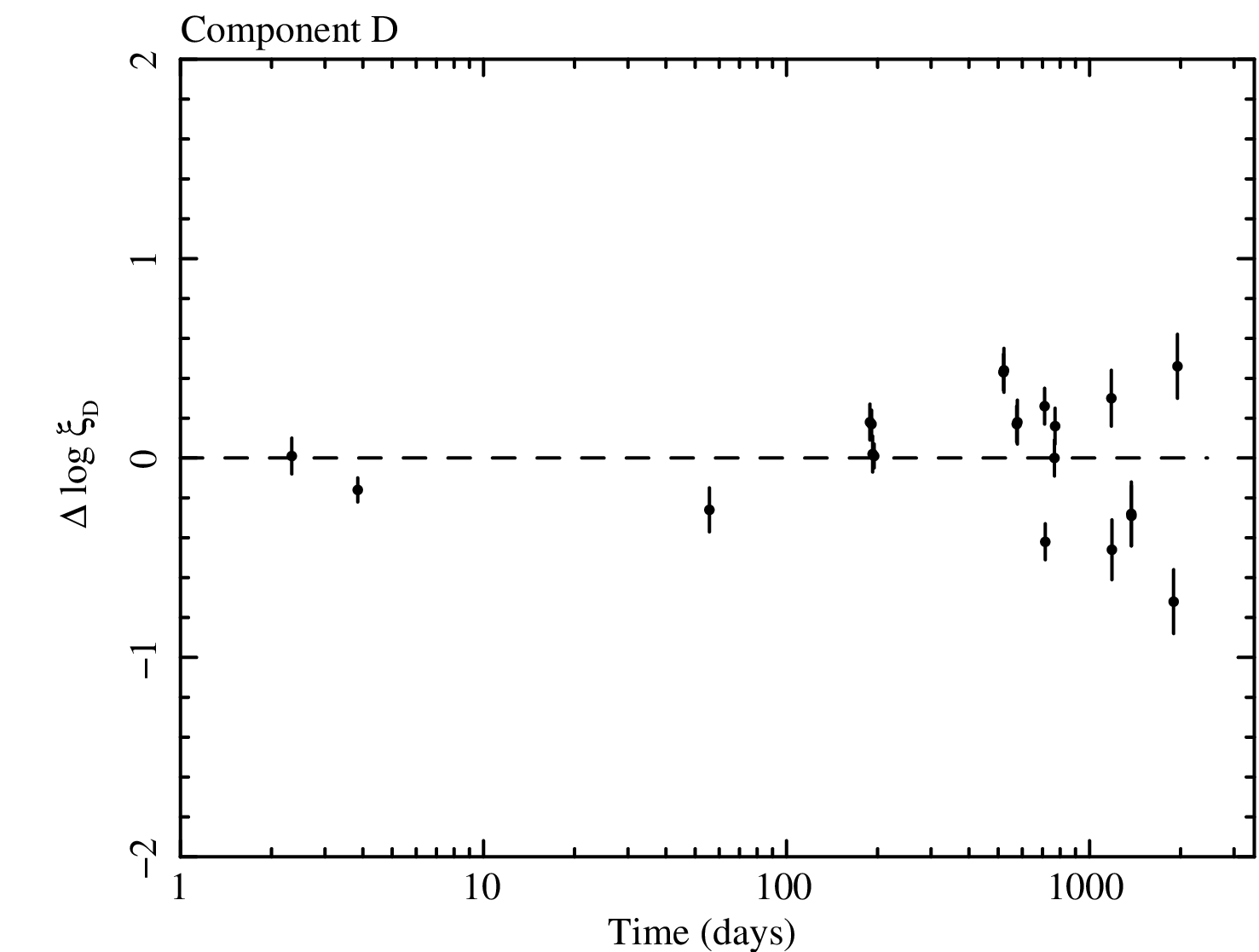}
  }
  \hbox{
  \includegraphics[width=6.5cm,angle=-90]{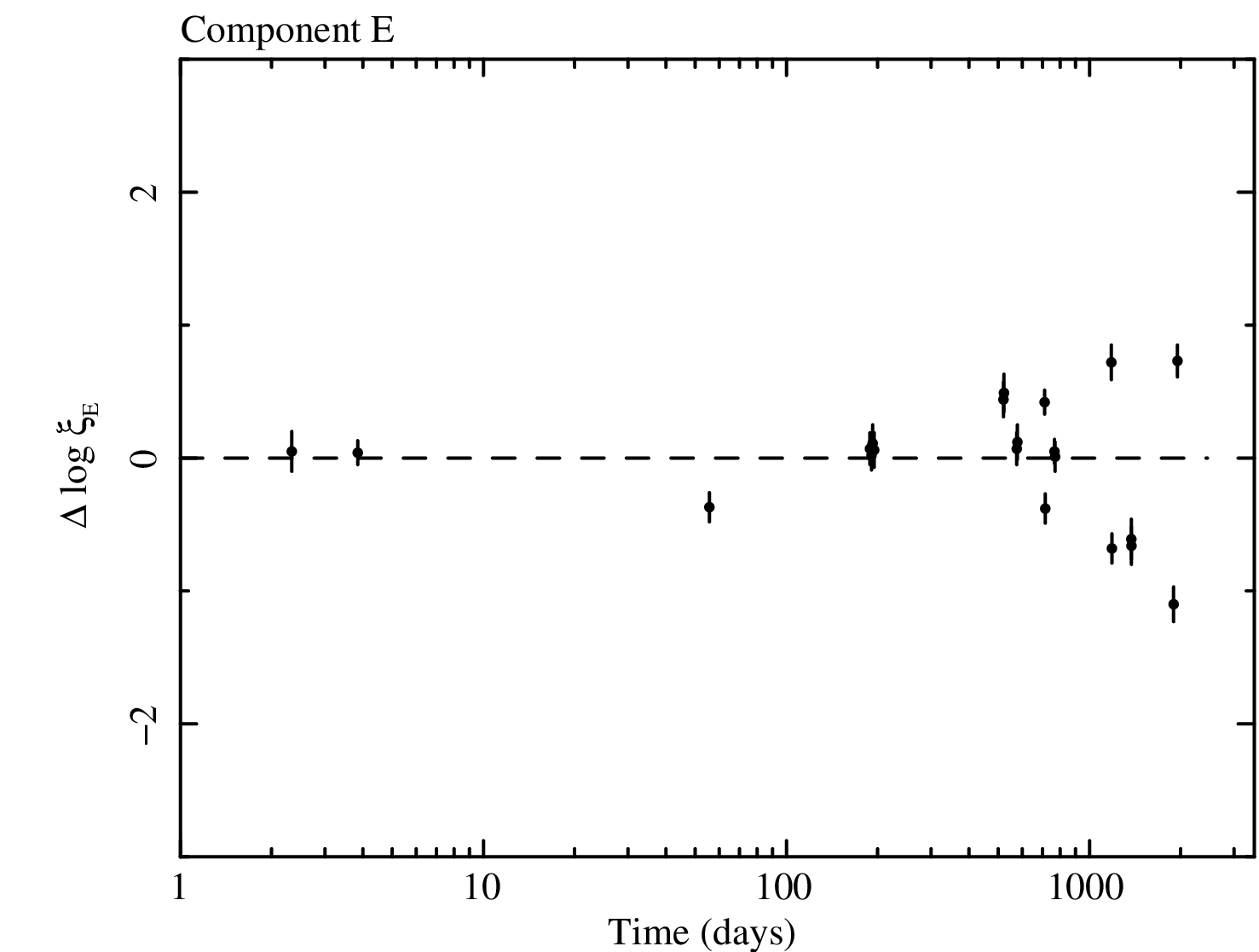}
  \includegraphics[width=6.5cm,angle=-90]{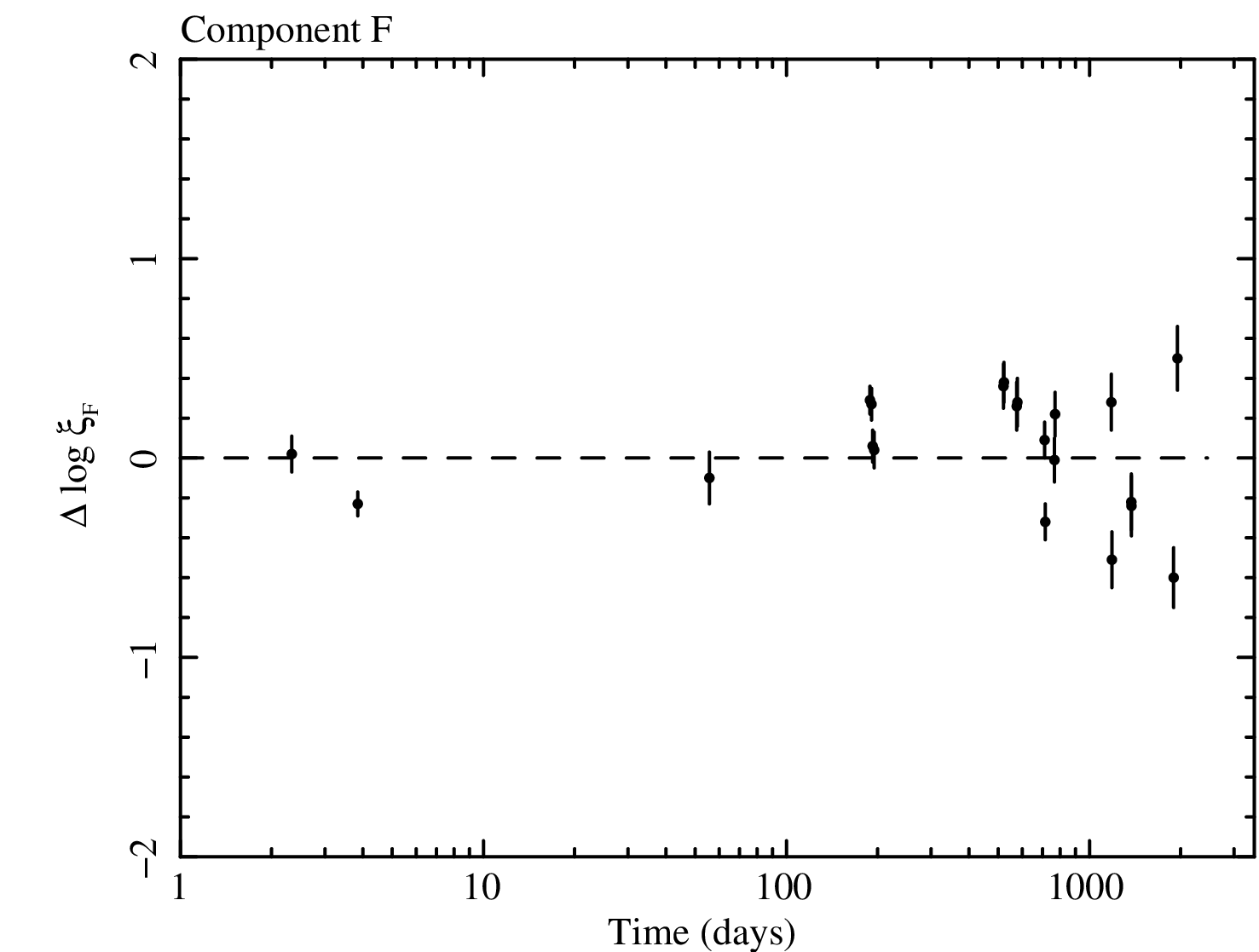}
  }
  \caption{\label{dxitime}Variation in the ionization parameter $\Delta \log \xi$ on different time scales (in days) for components $A$ and $B$ (top panels), $C$ and $D$ (middle panels), and $E$ and $F$ (bottom panels). The dashed line represents no variability.}
\end{figure*}

Motivated by the behavior of the WA components along the stability curves shown in Sect.~\ref{scurves}, we looked for additional signatures of variability that would help to constrain the typical time scales at which each component responds to flux changes. In Fig.~\ref{xilum} we show the values of the ionization parameter $\xi$ plotted against the ionizing luminosity $L$ for each component and archival observation using the L02 parameters as reference values. We chose to use L02 for its quality and because the WA parameters were best determined for this observation. For each component the dotted horizontal line indicates that the WA did not respond to variations in flux with respect to that of L02 (i.e. the ionization parameter $\xi$ is the same as that of L02 for each component). A dashed diagonal line shows $\xi$ proportional to $L$, and marks the trend that the WA would follow if the material responds instantaneously to continuum changes.

The lowest flux observations, L05 and L07, have typically large uncertainties in the ionization parameter $\xi$ for all the WA components. Overall they seem to be consistent with an immediate response for components $A$ to $C$, while components $D$ to $F$ could be consistent with any trend at the $2\sigma$ confidence level. All the WA components of H00 are systematically more ionized that those of L02, even though the ionizing luminosity at the H00 epoch was $\sim$20\% lower than that of L02. Similarly most components are more highly ionized in H02, which has an ionizing flux close to that of L02. The exceptions are components $A$ and $E$, which are inconsistent at only the $2\sigma$ level. A similar behavior is seen for the WA in R00x (ionizing flux comparable to L02) but with systematically lower ionized components, with the exception of components $D$ and $F$. The observations with higher ionizing luminosity, the RGS observations in 2001 and L99, show different trends. In general, both R01y and R01z are closer to the no response trend (i.e. their WA components have a $\xi$ value close to those of L02). However, for components $A$ and $B$ the errors are large enough that they are within $2\sigma$ also consistent with an immediate response trend. On the other hand, the WA components in L99 show a variety of behaviors: $A$ and $B$ are significantly more ionized than in L02, $D$ and $F$ are consistent with the gas responding immediately, while $C$ and $E$ would be closer to a no response scenario.

It is clear that, in spite of the large uncertainties associated with the determination of $\xi$, the WA components show some kind of variability as seen in Figs.~\ref{scurves1}, \ref{scurves2}~and \ref{xilum}, and in the values listed in Table~\ref{archivalbest}. If these changes are due to photoionization or recombination of the ionized gas in response to continuum changes, it is then possible to estimate a lower limit on the density of the absorbing gas via the following equation (\citealt{Bot00}):

\begin{equation}
\label{trec}
t_{\rm rec}(X_i)=\left(\alpha_r(X_i)n\left[\frac{f(X_{i+1})}{f(X_i)}-\frac{\alpha_r(X_{i-1})}{\alpha_r(X_i)}\right]\right)^{-1},
\end{equation}

\noindent where $t_{\rm rec}(X_i)$~is the recombination time scale of the ion $X_i$, $\alpha_r(X_i)$ is the recombination rate from ion $X_{i-1}$ to ion $X_i$, and $f(X_i)$ is the fraction of element $X$ at the ionization level $i$. The recombination rates $\alpha_r$ are known from atomic physics and the fractions $f$ can be determined from the ionization balance of the source. Therefore, for a given ion $X_i$ it is possible to obtain a lower limit on the density $n$ if an upper limit on the recombination time $t_{\rm rec}$ is known, since $t_{\rm rec} \propto 1/n$.

The archival observations of NGC 5548 allow us to probe variability on time scales that range from two days to almost a decade. In Fig.~\ref{dxitime} we show the measured variations in the ionization parameter $\xi$ between each individual observation (e.g., the first observation is compared with the rest, then the second is compared with the rest, and so on). This enables us to estimate the typical time scale at which the different WA components begin to vary, which can be used as a proxy for the recombination time of the gas. In this sense, we will adopt the first point where significant changes can be seen as $t_{\rm rec}$ for a given WA component. For component $A$ there are no signatures of variability in $\xi$ below $\sim 500$~days, and similarly for component $B$ although there are marginal hints of variability at 4 and $\sim 60$~days. Components $C$~and $D$~start showing marginal evidence for variability at 4 days, becoming more significant on time scales of 60 days. Component $E$ also begins to significantly vary at $\sim 60$~days, while component $F$ does so at 4 days.

Another variable in Eq.~\ref{trec} that needs to be determined is the fraction of a given element $X$ in an ionization state $i$. This can be done by means of the ionization balance of NGC 5548 that provides the ionic column densities of all the ions, which can be divided by the total hydrogen column density to obtain $f(X_i)$. We looked for the ions that contribute substantially to the fit for each WA component (i.e. the ones that because of their column density, oscillator strengths, and signal-to-noise of their features in the spectrum, are driving the fit), so that they can be used as a proxy for that component. We used the ionic column densities of these ions to feed the SPEX auxiliary program {\it rec\_time}, which delivers as output the product $nt_{\rm rec}$. The ions used for each component and their corresponding $nt_{\rm rec}$ product are listed in Appendix~\ref{ions}. Using the averaged $<nt_{\rm rec}>$ values, and the variability time scales described above as upper limits for $t_{\rm rec}$, we obtained lower limits to the density of the absorbing gas $n$ for each WA component. We find that the WA components with the lowest density are components $A$ and $B$ (of the order of a few thousands of cm$^{-3}$ or higher), while the higher ionization components $C$ to $F$ have densities of the order of tens of thousands of cm$^{-3}$. These values are listed in Table~\ref{dens}.

\subsection{Location of the warm absorbers}
\label{WAloc}

From the definition of the ionization parameter (Eq.~\ref{xidef}) it follows immediately that we can use the lower limit on the density $n$ calculated in Sect.~\ref{WAvar} to constrain the location of the WA with respect to the central ionizing source:

\begin{equation}
\label{dist}
R \leq \sqrt{L/n\xi}.
\end{equation}

The ionizing luminosity was calculated by integrating the template SED shown in Fig.~\ref{sed_fig} in the $1-1000$~Ryd range, obtaining $L = 1.97 \times 10^{44}$~erg s$^{-1}$. With this value and those of $\xi$ for the L02 dataset, which are representative of the flux state and the ionization state of the WA, we obtained upper limits on the distance of the WA components that are listed in the last column of Table~\ref{dens}. We conclude that the WA in NGC 5548 has a stratified structure, where the lowest ionization components $A$~and $B$~are the farthest, at $\lesssim 50$~and $\lesssim 20$~pc from the central source, respectively. Components $C$~and $D$~lie at the order of $\sim4$~pc or less, whereas components $E$~and $F$ are located at pc and sub-pc distances, respectively.

\begin{table*}
  \centering
  \caption[]{Variability time scales, density, and location of the WA components in NGC 5548.}
  \label{dens}
  \begin{tabular}{l c c c c c c c}
    \hline\hline
    \noalign{\smallskip}
    Component      & $<nt_{\rm rec}>$ (s\,cm$^{-3}$) & $t_{\rm rec}^{\rm lower}$ (days)  &$t_{\rm rec}^{\rm upper}$ (days)  & $n^{\rm lower}$ (cm$^{-3}$) & $n^{\rm upper}$ (cm$^{-3}$) & $R^{\rm lower}$ (pc) & $R^{\rm upper}$ (pc) \\
    \noalign{\smallskip}
    \hline
    \noalign{\smallskip}
    $A$ & $1.1 \times 10^{11}$  & $> 200$ & $< 500$  & $> 2\,600$  & $< 6\,500$ & $> 31$  & $< 50$  \\
    $B$ & $1.2 \times 10^{11}$  & $> 200$ & $< 500$  & $> 2\,850$  & $< 7\,100$ & $> 13$  & $< 20$  \\
    $C$ & $5.9 \times 10^{10}$  & $> 4$   & $< 60$   & $> 11\,600$ & $< 174\,000$ & $> 1.2$ & $< 4.7$  \\
    $D$ & $6.5 \times 10^{10}$  & $> 4$   & $< 60$   & $> 12\,800$ & $< 192\,000$ & $> 0.9$ & $< 3.6$  \\
    $E$ & $4.5 \times 10^{10}$  & $> 4$   & $< 60$   & $> 8\,800$ & $< 132\,000$ & $> 0.6$ & $< 2.2$  \\
    $F$ & $2.6 \times 10^{10}$  & $> 2$   & $< 4$    & $> 77\,300$ & $< 155\,000$ & $> 0.4$  & $< 0.6$ \\
    \noalign{\smallskip}
    \hline
  \end{tabular}
\end{table*}

The determination of the density lower limits described above allows for much better constrained distances compared to other distance estimations. Some of them, such as the one described in \citet{Blu05}, involve assumptions on the conservation of momentum, which are only valid for radiatively-driven winds. Since there is increasing evidence that AGN ionized winds can also be launched by other mechanisms (shocks, thermal instabilities, magnetic fields) or a combination of them, these assumptions need to be handled with caution. The most basic approach to estimate the distance of the intervening gas used in the literature (e.g., \citealt{Blu05}) is by assuming that the bulk of the WA mass is concentrated in a layer with thickness $\Delta r$ that needs to be less than or equal to the distance to the central source $R$ so that $\Delta r /R \leq 1$. Together with the approximation $N_{\rm H} \approx n\Delta r$, it follows that $R \leq L/N_{\rm H}\xi$. Applying this method to NGC 5548 we derive distance constraints that are at least one order of magnitude higher that those obtained from Eq.~\ref{dist}.

Furthermore, we can use a similar argument based on the non-detection of variability on smaller time scales to set upper limits on the density of the absorbing gas, and hence lower limits on their possible location. Following this, we see that components $A$ and $B$ have lower limits on $R$ of the order of $\sim 30$~and $\sim 10$~pc, while the highest ionization components, $C$ to $F$, are at sub-pc scales or further from the black hole (see Table~\ref{dens}).

The scenario depicted by these distance estimations, where high-density highly ionized gas is located much closer to the central SMBH than the less dense lowly ionized absorber, is consistent with the picture sketched in \citet{Kaz12}. In this work, a line of sight at a given inclination angle intersects several layers of a stratified wind, each of them with different $\xi$~and density such as the ones seen in NGC 5548. WA components with a variety of ionization phases are thus originally launched from different parts of the disk, most likely due to the combined effect of magnetic and radiation pressure (\citealt{Fuk10}).

In the context of the unification models of AGN, these ionized winds are much farther from the SMBH than the broad line region (BLR), which is located at a few light-days distance (\citealt{Pan14}; \citealt{Kaa14}). For comparison, the inner edge of the torus can be estimated from the approximate relation for the dust sublimation radius (\citealt{KK01})

\begin{equation}
\label{Rtorus}
R_{\rm torus} \sim L^{1/2}_{\rm ion,44} {\rm (pc)}, 
\end{equation}

\noindent where $L_{\rm ion,44}$ is the ionizing luminosity in units of $10^{44}$~erg s$^{-1}$, obtaining a distance of $\sim 1.4$~pc. The most ionized WA component $F$ is thus generated closer to the disk than the inner edge of the torus, in agreement with the scenario discussed in \citet{Kaz12}. According to their lower limits on $R$, components $C$ to $E$, all of them with $\log \xi > 2$, may also extend down to distances comparable to the torus inner edge. 

With respect to the location of the narrow line region (NLR) in NGC 5548, \citet{Det09} constrained it to a cone extending between 1 and 15 pc from the central black hole, depending on its exact geometry. More recently, \citet{Pet13} used [\OIII{}] variability to estimate the size of the NLR in NGC 5548. They found it to be compact ($R_{\rm NLR} \sim 1 - 3$~pc) although its actual extent is possibly larger if it is elongated toward the line of sight. In this respect, the analysis of the narrow emission lines in the stacked RGS spectrum of NGC 5548 during the 2013-2014 campaign, renders a location for the NLR of $r_{\rm min} = 13.9 \pm 0.6$~pc from the nucleus (\citealt{Whe15}). This analysis also shows the narrow emission lines to be absorbed by WA component $B$, located at $13 < R < 20$~pc, a range consistent with the location of the NLR estimated by \citet{Whe15}, although no conclusions can be firmly drawn on the possible connection between the absorbers along our line of sight and the emitting clouds that extend over a much larger region.

\subsection{Energetics of the warm absorbers}
\label{WAener}

With the distance constraints for the WA derived in Sect.~\ref{WAloc} we can estimate the amount of mass carried by the outflows into the ISM of NGC 5548. For an outflow in the form of a partial thin spherical shell at a distance $R$ from the central source and moving radially at constant speed $v$, the mass outflow rate is:

\begin{equation}
\label{mout}
\dot{M}_{\rm out} = \mu m_{\rm p}N_{\rm H}vR\Omega
\end{equation}

\noindent where $\mu = 1.4$ is the mean atomic mass per proton, $m_{\rm p}$ is the proton mass, and $\Omega$ is the solid angle subtended by the outflow, which ranges between $0$ and $4\pi$~sr. The main source of uncertainty in this equation is introduced by the parameter $\Omega$, which depends on the geometry of the WA outflow and is usually unknown. Typically, $\Omega = \pi/2$~sr is assumed, as inferred from the observed type-1/type-2 ratio in nearby Seyfert galaxies (\citealt{MR95}), and the fact that ionized outflows are detected in approximately half of the observed Seyfert 1 galaxies (\citealt{Dunn07}). This value is only an estimate, and we note that it will not hold if the outflows are bent because of gravitational and/or magnetic processes.

Since $\dot{M}_{\rm out}$ scales with the distance $R$, the farthest components will carry considerably more mass per unit time than the inner ones, even if the former are much less dense than the latter; and similarly for the fastest winds with respect to the slower velocity winds. Indeed we calculate that for components $A$, $B$, and $C$, located at pc-scale distances, their mass outflow rates are of the order of $\dot{M}_{\rm out} \simeq 0.05\Omega$~$\rm M_{\odot}$~yr$^{-1}$. These values are almost a factor of two higher than those of components $E$ and $F$, with mass outflow rates of $\dot{M}_{\rm out} \simeq 0.03\Omega$~$\rm M_{\odot}$~yr$^{-1}$. We have expressed all these $\dot{M}_{\rm out}$ values as a function of $\Omega$ to reflect the uncertainty associated with this parameter.

Assuming that the absorbers subtend a solid angle of $\Omega = \pi/2$~sr, the mass outflow rates for components $A$ to $C$ are in the range of $\dot{M}_{\rm out} \sim 0.06 - 0.09$~$\rm M_{\odot}$~yr$^{-1}$, while those of components $E$ to $F$ are of the order of $\sim 0.04 - 0.05$~$\rm M_{\odot}$~yr$^{-1}$. Component $D$ contributes much less with $\dot{M}_{\rm out} \sim 0.01$~$\rm M_{\odot}$~yr$^{-1}$. The $\dot{M}_{\rm out}$ values for all WA components are reported in Table~\ref{energetics}. In general, these values are similar or slightly higher than the mass accretion rate of the source, $\dot{M}_{\rm acc} \equiv L_{\rm bol}/\eta c^2 = 0.05$~$\rm M_{\odot}$~yr$^{-1}$ (assuming a nominal accretion efficiency of $\eta = 0.1$, and a bolometric luminosity of $L_{\rm bol} = 2.82 \times 10^{44}$~erg cm$^{-1}$). The sum of the mass conveyed per unit time by all of the X-ray WA components is $\sim 0.3$~$\rm M_{\odot}$~yr$^{-1}$, about six times the nominal mass accretion rate in NGC 5548. For comparison, these values are of the same order of magnitude as those measured in other active galaxies (e.g., \citealt{Cos10}; \citealt{CK12}), and they are typically almost one order of magnitude lower than the mass outflow rates measured in powerful starburst galaxies (e.g., \citealt{Vei05}).

The kinetic energy per unit time carried by the outflows is defined as $L_{\rm KE} \equiv \frac{1}{2}\dot{M}_{\rm out}v^2$. The WA components have values of $L_{\rm KE}$ of the order of a few times $10^{39-40}$~erg s$^{-1}$ for the opening angle assumed above (see Table~\ref{energetics}). The total kinetic luminosity injected into the medium by the sum of all the components, $L_{\rm KE} \sim 7.8 \times 10^{40}$~erg s$^{-1}$, constitute a small fraction of the bolometric luminosity of the source, $L_{\rm KE}/L_{\rm bol} \sim 0.03$\%. This value, even if we assume the extreme case of an opening angle of $4\pi$, is well below the range of $0.5-5$\% of the bolometric luminosity typically assumed by feedback models as the minimum amount required to reproduce the observed $M - \sigma$ relation (\citealt{HE10} and references therein).

However, this picture is incomplete if we only focus on the X-ray WA, without taking the effect of the UV absorbers into account. The latter are typically located further away than the X-ray WA, and carry large amounts of matter, which may have a measurable effect in the energy budget of the AGN. In general, a direct comparison between the X-ray and UV absorbers in AGN is difficult because of the higher parameter uncertainties of the former with respect to the latter, especially those regarding spectral resolution. However, in those cases where simultaneous X-ray and UV observations are taken, there is increasing evidence that some of the low ionization X-ray WA components may originate in the same gas responsible for the absorption in the UV band (e.g., Mrk 279, \citealt{Cos07}; Mrk 509, \citealt{Ebr11}; NGC 4593, \citealt{Ebr13}). An extensive discussion on the possible relation between the X-ray and UV absorbers in NGC 5548 can be found in \citet{Arav14}. Here we will assume that the UV absorbers of NGC 5548 contribute to the mass-energy budget as an independent component. \citet{Arav14} provide column densities and distance estimates for three of the six UV components. These parameters are well determined for UV component 1 ($\log N_{\rm H} = 21.5^{+0.4}_{-0.2}$~cm$^{-2}$, and $R = 3.5^{+1.0}_{-1.2}$~pc), but less constrained for the UV components 3 and 5 ($R$ in the $5 - 15$~pc range, and $\log N_{\rm H} < 21.5$ and $< 20.7$~cm$^{-2}$, respectively). These values predict outflow mass rates $\dot{M}_{\rm out}$, assuming also a solid angle of $\Omega = \pi/2$~sr, of $\sim 0.23$~$\rm M_{\odot}$~yr$^{-1}$, $\sim 0.55$~$\rm M_{\odot}$~yr$^{-1}$, and $\sim 0.04$~$\rm M_{\odot}$~yr$^{-1}$, respectively. These values combined translate into a kinetic luminosity of $L_{\rm KE, UV}/L_{\rm bol} \sim 0.17$\%, almost five times higher than that of the X-ray WA. However, even after summing the X-ray and UV contributions, the total amount energy fed back into the medium is $\sim 0.2$\% or less (since part of the contribution of the UV absorbers may overlap with that of the lowest ionization X-ray WA). This is still below the minimum amount required to produce a significant effect in terms of feedback.

If these winds are not transient and are continuously launched from the inner regions of the AGN throughout its activity lifetime ($t \sim 10^8$~yr), the total kinetic energy deposited into the surroundings, $E_{\rm K} = L_{\rm KE}t$, is of the order of a few times $10^{55}$~erg. This is almost two orders of magnitude lower than the energy required to terminate or even regulate star-forming episodes in the host galaxy, which is estimated to be $\sim 10^{57}$~erg (\citealt{Kro10}). It is also negligible compared to the binding energy of massive galactic bulges ($E \sim 10^{60}$~erg), a value required by quasar feedback simulations to effectively regulate structure formation in the inter-galactic medium (e.g., \citealt{SO04}; \citealt{Hop05}).

\begin{table}
  \centering
  \caption[]{Energetics of the X-ray WA components. The solid angle subtended by the WA is assumed to be $\Omega = \pi/2$~sr.}
  \label{energetics}
  \begin{tabular}{l c c c c}
    \hline\hline
    \noalign{\smallskip}
    Component      &  $\dot{M}_{\rm out}$\tablefootmark{a} & $\dot{M}_{\rm out}/\dot{M}_{\rm acc}$ & $L_{\rm KE}$\tablefootmark{b} & $L_{\rm KE}/L_{\rm bol}$ \\
    \noalign{\smallskip}
    \hline
    \noalign{\smallskip}
    $A$  &  $0.061$  & $1.2$  & $0.7$   & $0.002$\% \\
    $B$  &  $0.053$  & $1.7$  & $0.8$   & $0.003$\%  \\
    $C$  &  $0.085$  & $1.7$  & $3.6$   & $0.013$\%   \\
    $D$  &  $0.010$  & $0.2$  & $0.02$  & $7 \times 10^{-5}$\%  \\
    $E$  &  $0.052$  & $1.0$  & $1.0$   & $0.004$\%  \\
    $F$  &  $0.038$  & $0.8$  & $1.8$   & $0.006$\%  \\
    \noalign{\smallskip}
    \hline
    \noalign{\smallskip}
    Total & $0.30$ & $6.6$ & $7.8$ & $0.028$\% \\
    \noalign{\smallskip}
    \hline
  \end{tabular}
  \tablefoot{
    \tablefoottext{a}{Mass outflow rate, in units of $\rm M_{\odot}$~yr$^{-1}$; }
    \tablefoottext{b}{kinetic luminosity, in units of $10^{40}$~erg s$^{-1}$.}
  }
\end{table}


\section{Conclusions}
\label{conclusions}

We have re-analyzed the archival observations of NGC 5548 taken with the grating spectrometers onboard \XMM{}~(RGS) and \Chan{}~(LETGS and HETGS) between 1999 and 2007. The motivations behind this exercise were twofold. In the first place we aimed at characterizing the variability of the ionized absorption features in the spectra of the source through a systematic analysis using up-to-date photoionized codes and atomic physics, since different versions of the codes provide significantly different ionization balances (see Sect.~\ref{motivation}). Secondly, we built a fully characterized and accurate model of the X-ray WA to be used to model the WA in the 2013-2014 RGS data. These data were obtained when the source was heavily obscured in the soft X-rays, imposing severe limitations on our ability to characterize and fit the WA in those spectra (\citealt{Kaa14}; \citealt{Gesu14}).

Using as a baseline the longest archival observation (the combined HETGS+LETGS spectra of 2002), we found that the WA in NGC 5548 is composed of six distinct ionization phases with increasing column densities $N_{\rm H}$ as the ionization parameter $\xi$ increases, moving outwards in four kinematic regimes ($v_{\rm out} \sim -250, \sim -550, \sim -800, \sim -1200$~km s$^{-1}$). Assuming that the changes in the WA between the different archival observations occur only due to ionization/recombination processes (i.e. outflow velocity and column density remain the same), we detect significant variations in ionization of the WA components in response to the changes in the intrinsic flux level of the source. Using the ions that contribute the most to each WA component fit and the variations in their ionic column densities among the observations, we are able to estimate lower limits on the density of the gas, observing that the lowest ionization components are less dense than the higher ionization components.

Lower limits on the density can be used to estimate a stringent upper limit on the location of the various WA components, thus revealing a stratified structure in the absorbing gas. The lower ionization phases are located at distances of the order of pc to tens of pc, whereas the intermediate components lie at pc to sub-pc distances. The highest ionization WA component is located between $\sim 0.4$~and $\sim 0.6$~pc from the central ionizing source. This scenario is consistent with the WA components being launched from different regions close to the accretion disk, possibly due to a combination of magnetic and radiative processes, forming a stratified wind composed of multiple layers that are crossed by our line of sight (\citealt{Kaz12}).

The mass outflow rate summed over all X-ray WA components detected in NGC 5548 is $\dot{M}_{\rm out} \sim 0.3$~$\rm M_{\odot}$~yr$^{-1}$, about six times the nominal accretion rate of the black hole. The kinetic luminosity injected into the surroundings of the central SMBH by the WA is however still a small fraction ($\sim 0.03$\%) of the bolometric luminosity of NGC 5548. Even adding the contribution of the UV absorbers, the total kinetic luminosity ($L_{\rm KE}/L_{\rm bol} \sim 0.2$\%) is not enough to affect the ISM of the host galaxy as invoked by current feedback models, which require that between $0.5$\% and $5$\% of the bolometric luminosity of the AGN must be fed back to the medium. The total kinetic energy deposited into the medium throughout the average lifetime of the AGN ($t \sim 10^8$~yr), assuming a constant level of activity, is two orders of magnitude lower than that required to quench star formation and affect the evolution of the host galaxy.

\begin{acknowledgements}

This work is based on observations obtained by \XMM{}, an ESA science mission with instruments and contributions directly funded by ESA member states and the USA (NASA). The scientific results reported in this article are based on observations made by the \Chan{}~X-ray observatory. We thank the International Space Science Institute (ISSI) in Bern for support and hospitality. SRON is supported financially by NWO, the Netherlands Organization for Scientific Research. This work was supported by NASA through grants for HST program number 13184 from the Space Telescope Science Institute, which is operated by the Association of Universities for Research in Astronomy, Incorporated, under NASA contract NAS5-26555. M.M. acknowledges support from NWO and the UK STFC. M.C. acknowledges financial support from contracts ASI/INAF n.l/037/12/0 and PRIN INAF 2011 and 2012. P-O.P. acknowledges financial support from the CNES and from the CNRS/PICS. S.B. acknowledges INAF/PICS for financial support. G.P. acknowledges support of the Bundesministerium f\"ur Wirtschaft und Technologie/Deutsches Zentrum f\"ur Luft- und Raumfahrt (BMWI/DLR, FKZ 50 OR 1408). H.S. was supported by Grant-in-Aid for JSPS Fellows, 15J02737. M.W. acknowledges the support of a Ph.D. studentship awarded by the UK STFC. We thank the anonymous referee for his/her constructive comments.

\end{acknowledgements}


\begin{appendix}
\section{NGC 5548 archival spectra}
\label{archiveplots}

\begin{figure*}
  \centering
  \hbox{
  \includegraphics[width=6.5cm,angle=-90]{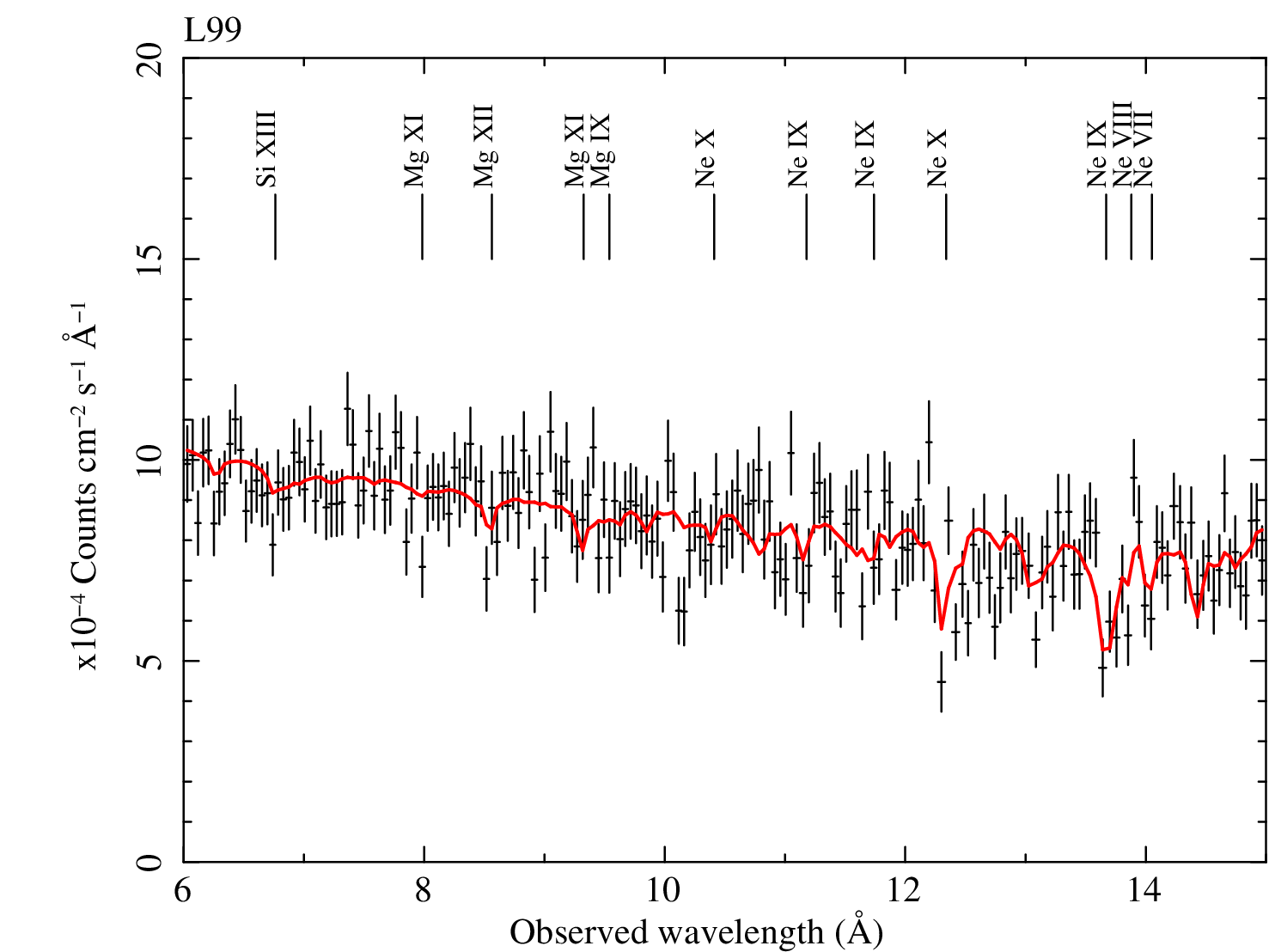}
  \includegraphics[width=6.5cm,angle=-90]{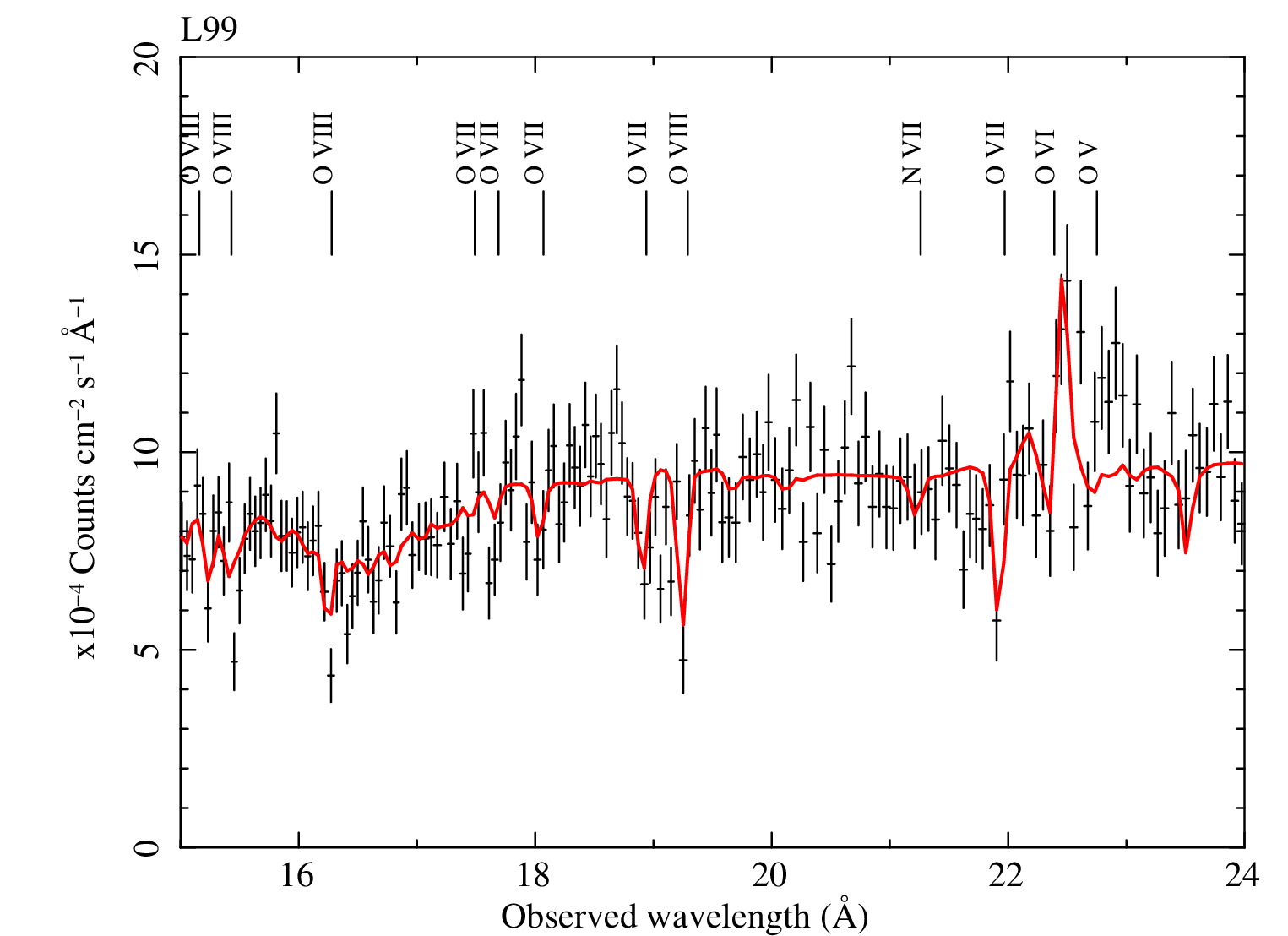}
  }
  \hbox{
  \includegraphics[width=6.5cm,angle=-90]{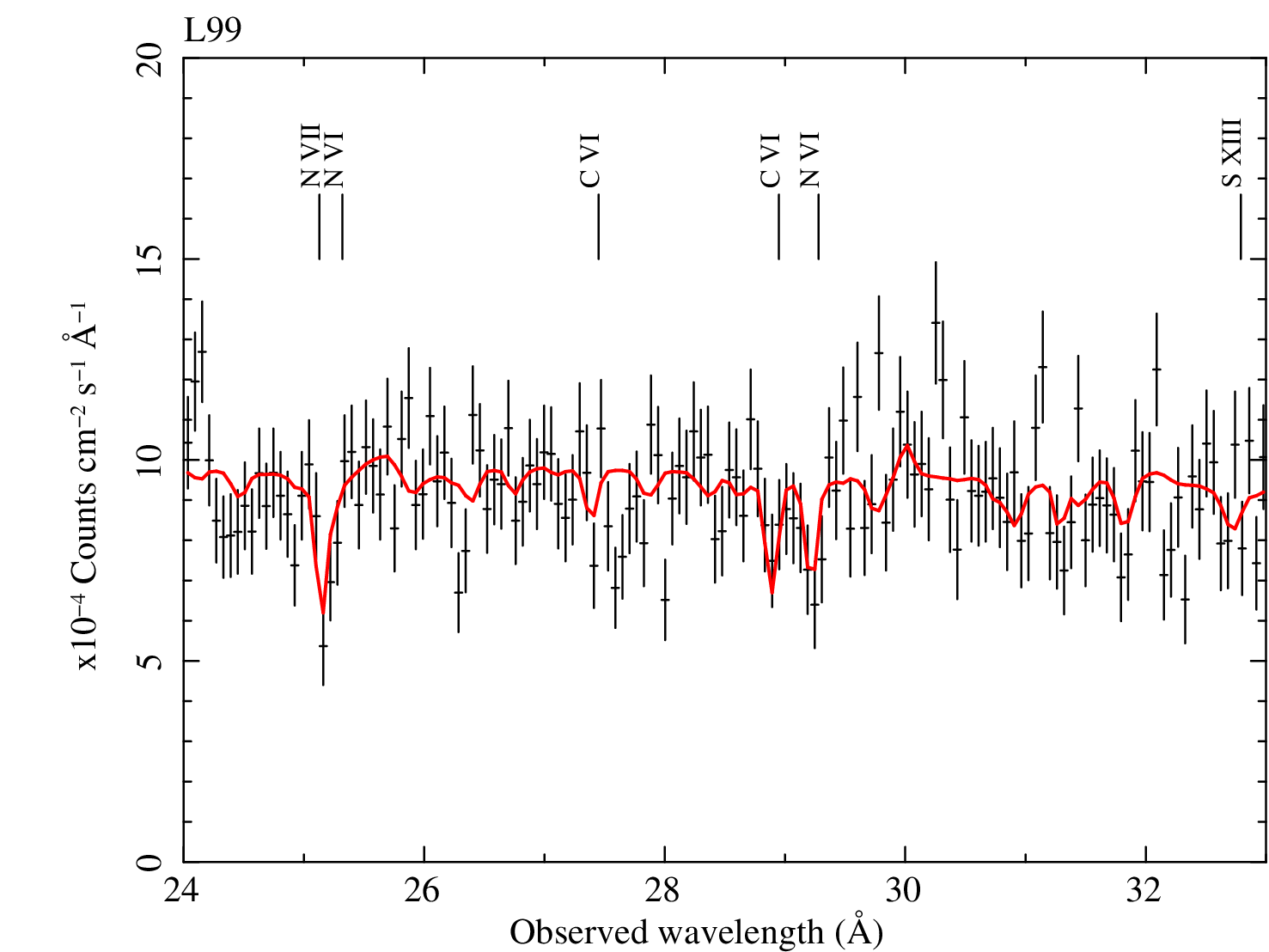}
  \includegraphics[width=6.5cm,angle=-90]{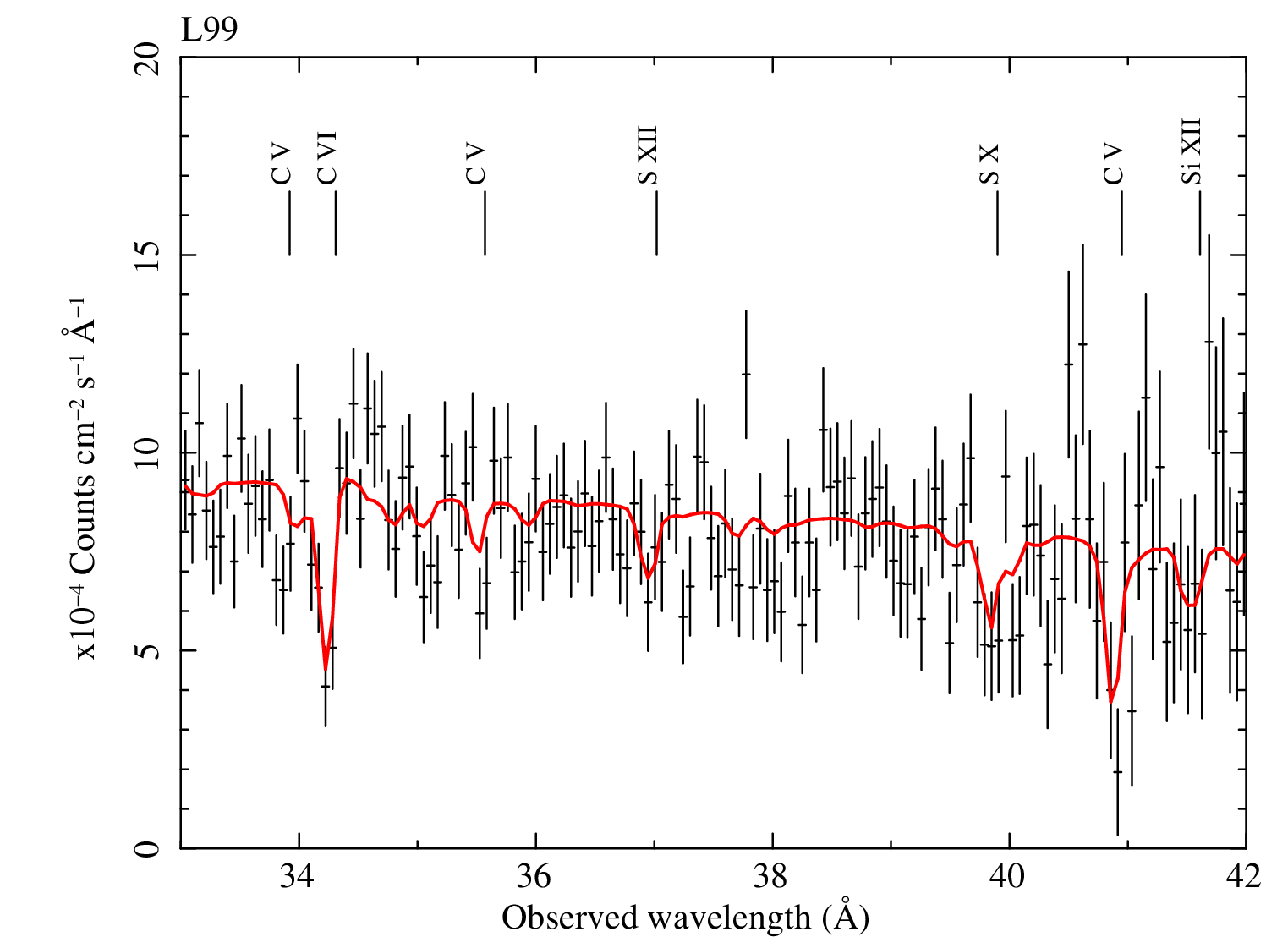}
  }
  \hbox{
  \includegraphics[width=6.5cm,angle=-90]{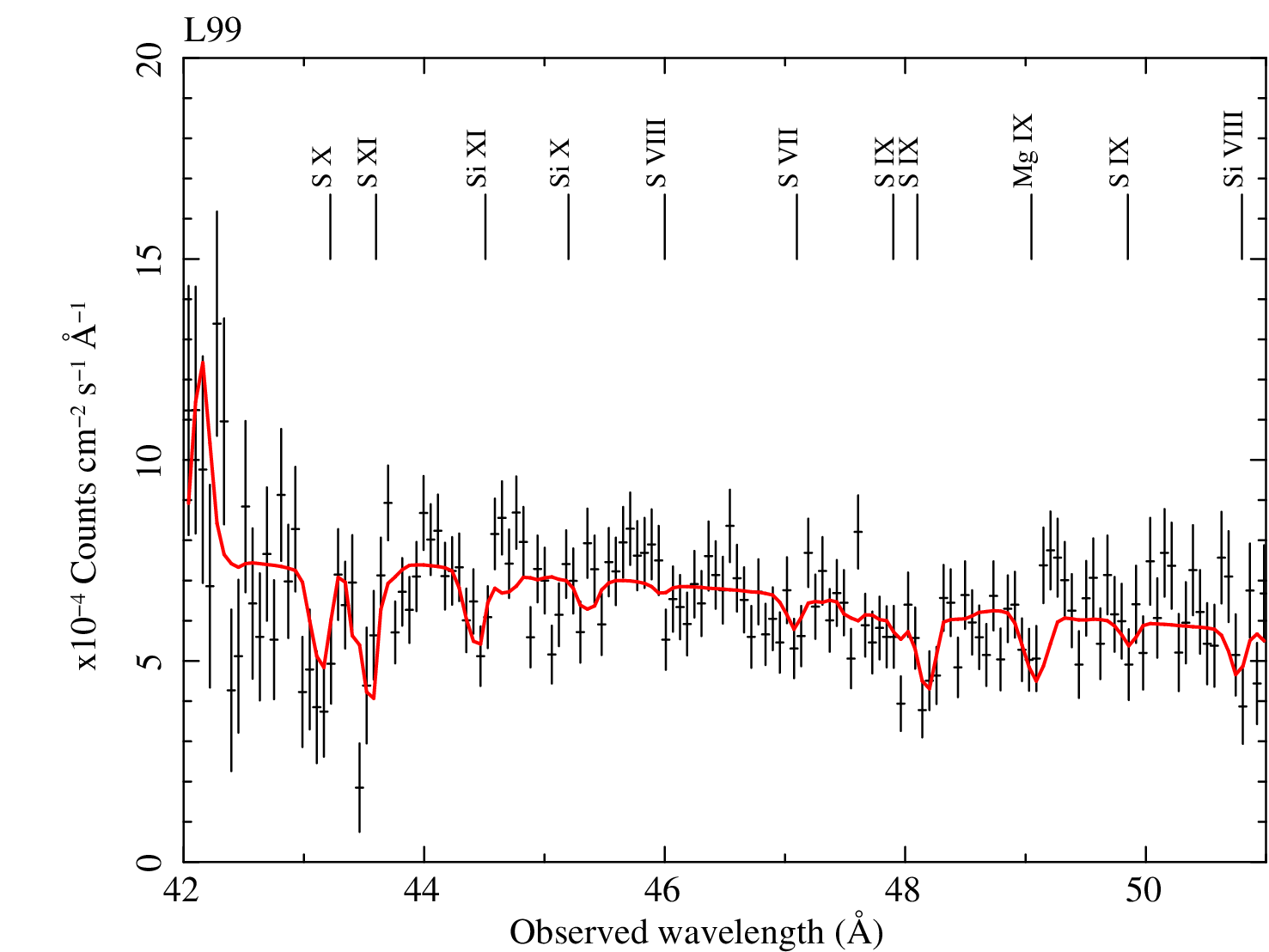}
  \includegraphics[width=6.5cm,angle=-90]{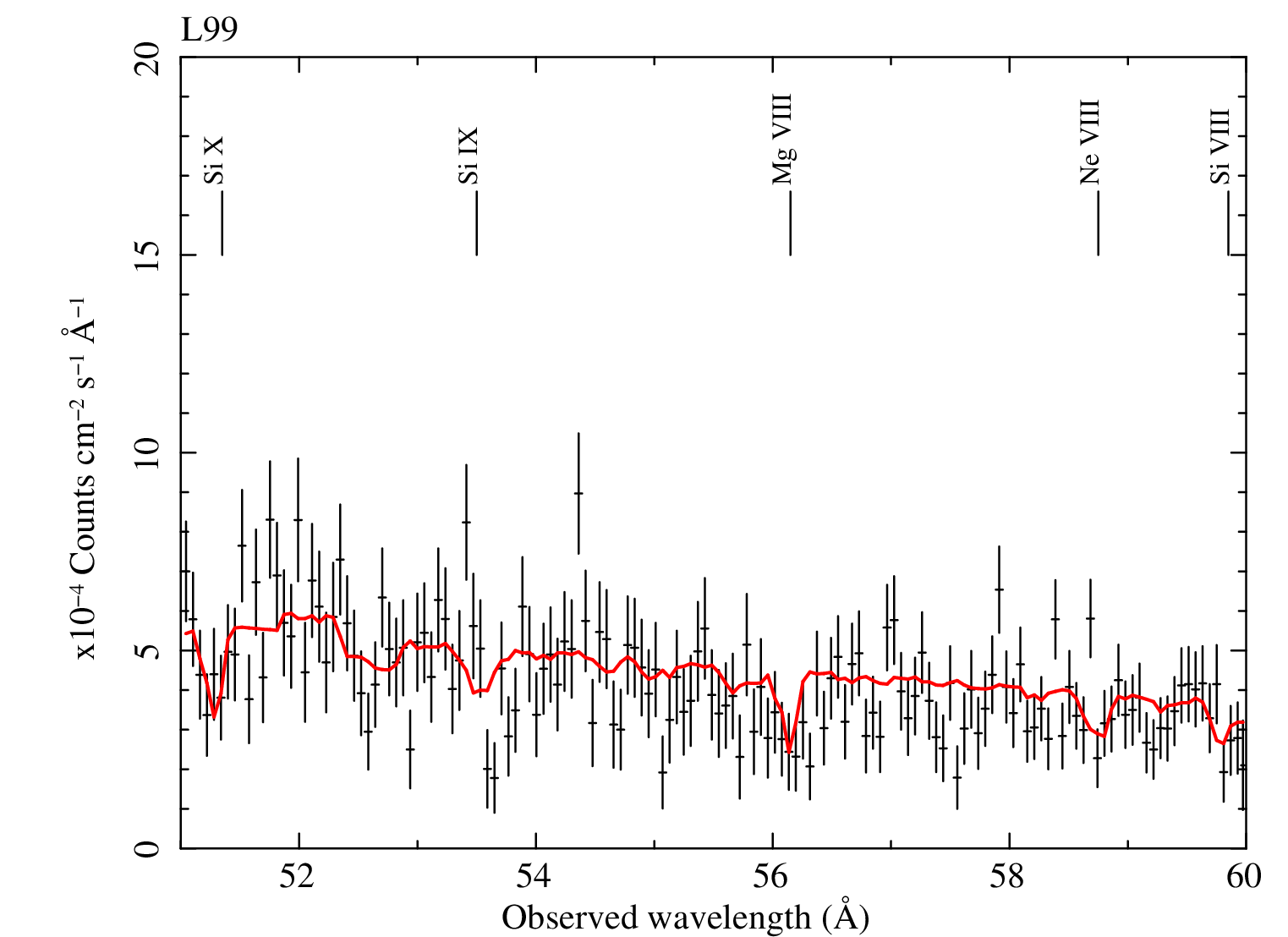}
  }
  \caption{\label{L99fluxed}\Chan{}~LETGS spectrum of NGC 5548 in 1999 (L99). The solid line represents our best-fit model. Some relevant WA absorption features have been labeled.}
\end{figure*}

\begin{figure*}
  \centering
  \hbox{
  \includegraphics[width=6.5cm,angle=-90]{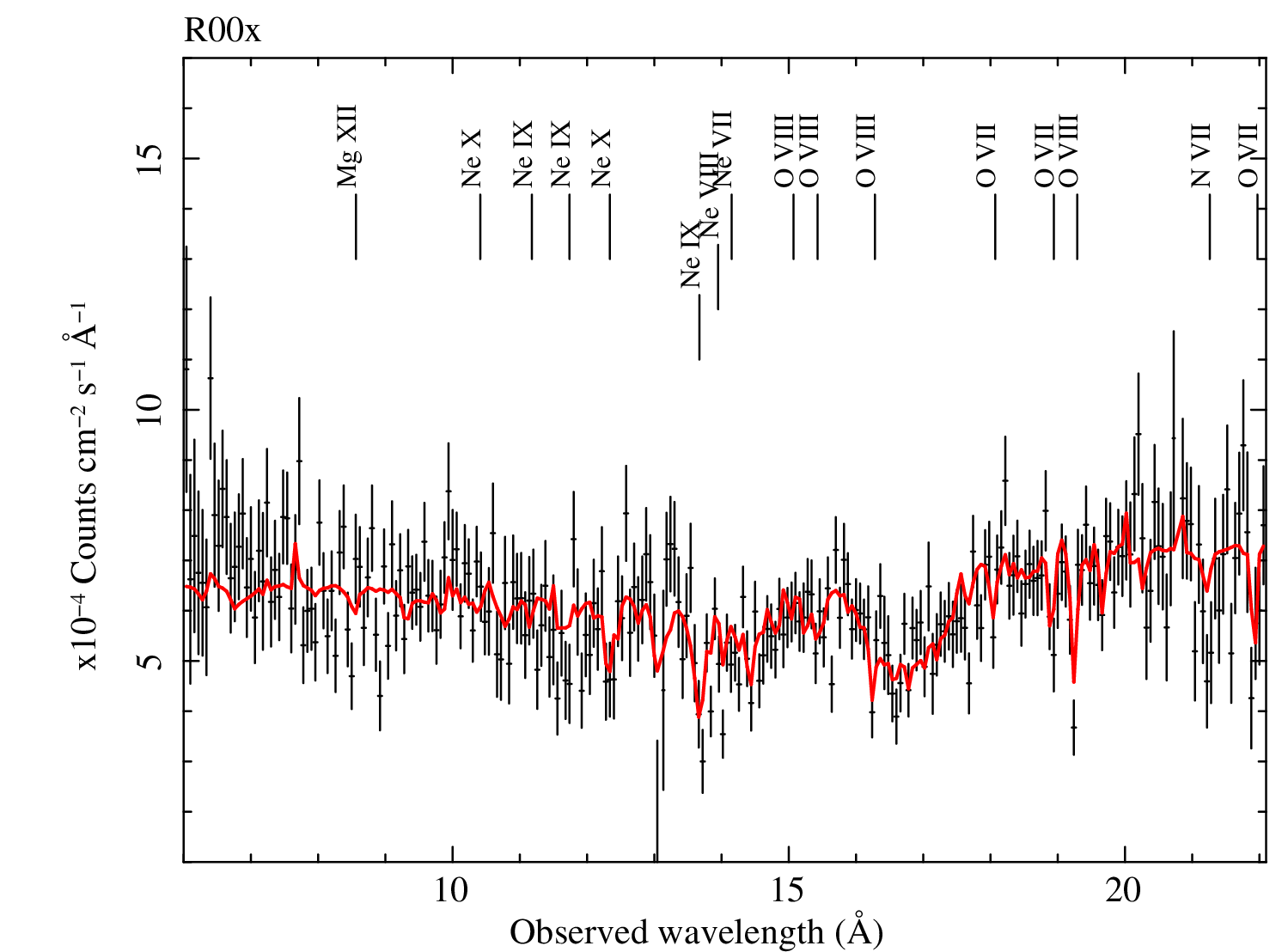}
  \includegraphics[width=6.5cm,angle=-90]{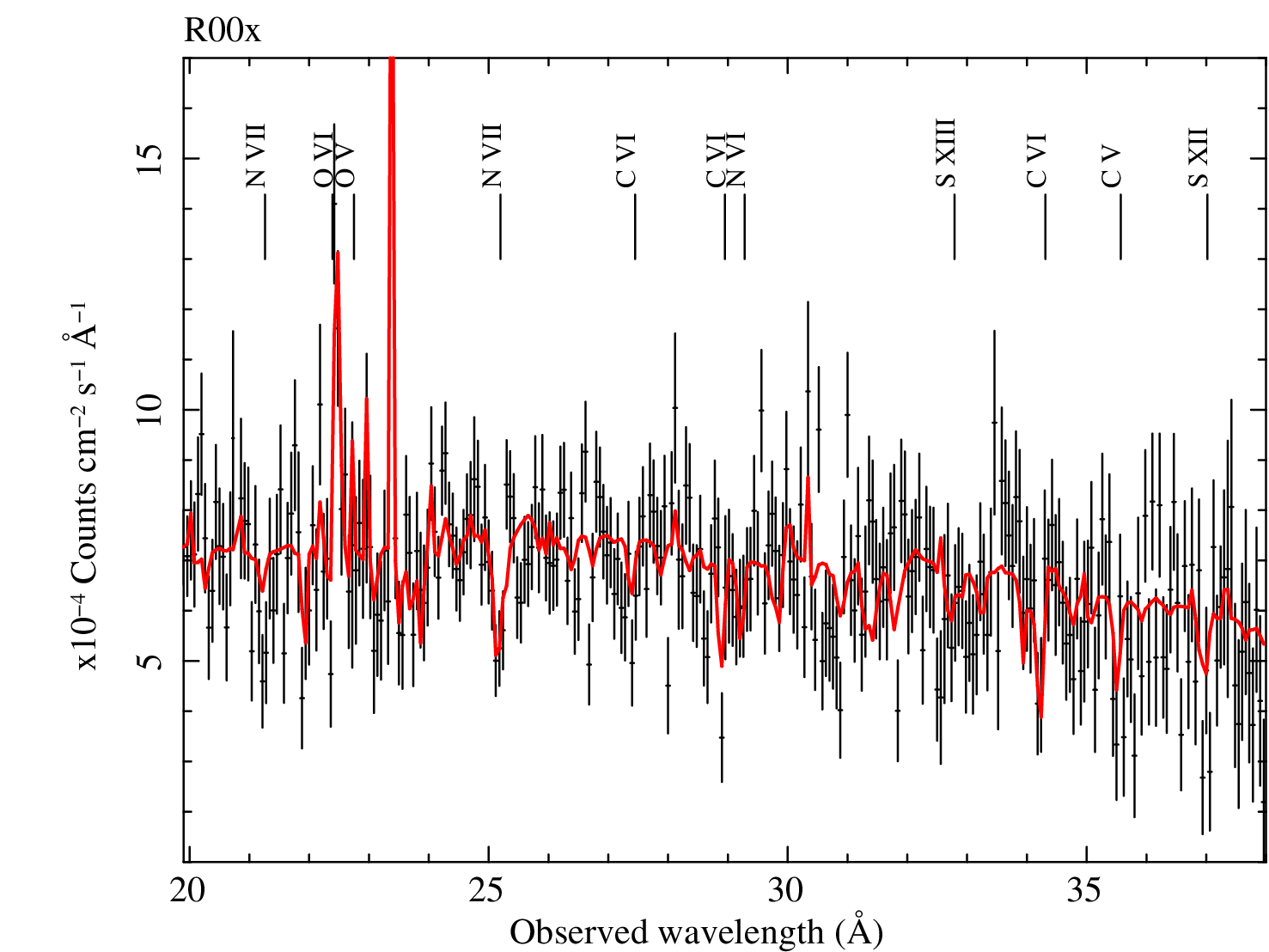}
  }
  \hbox{
  \includegraphics[width=6.5cm,angle=-90]{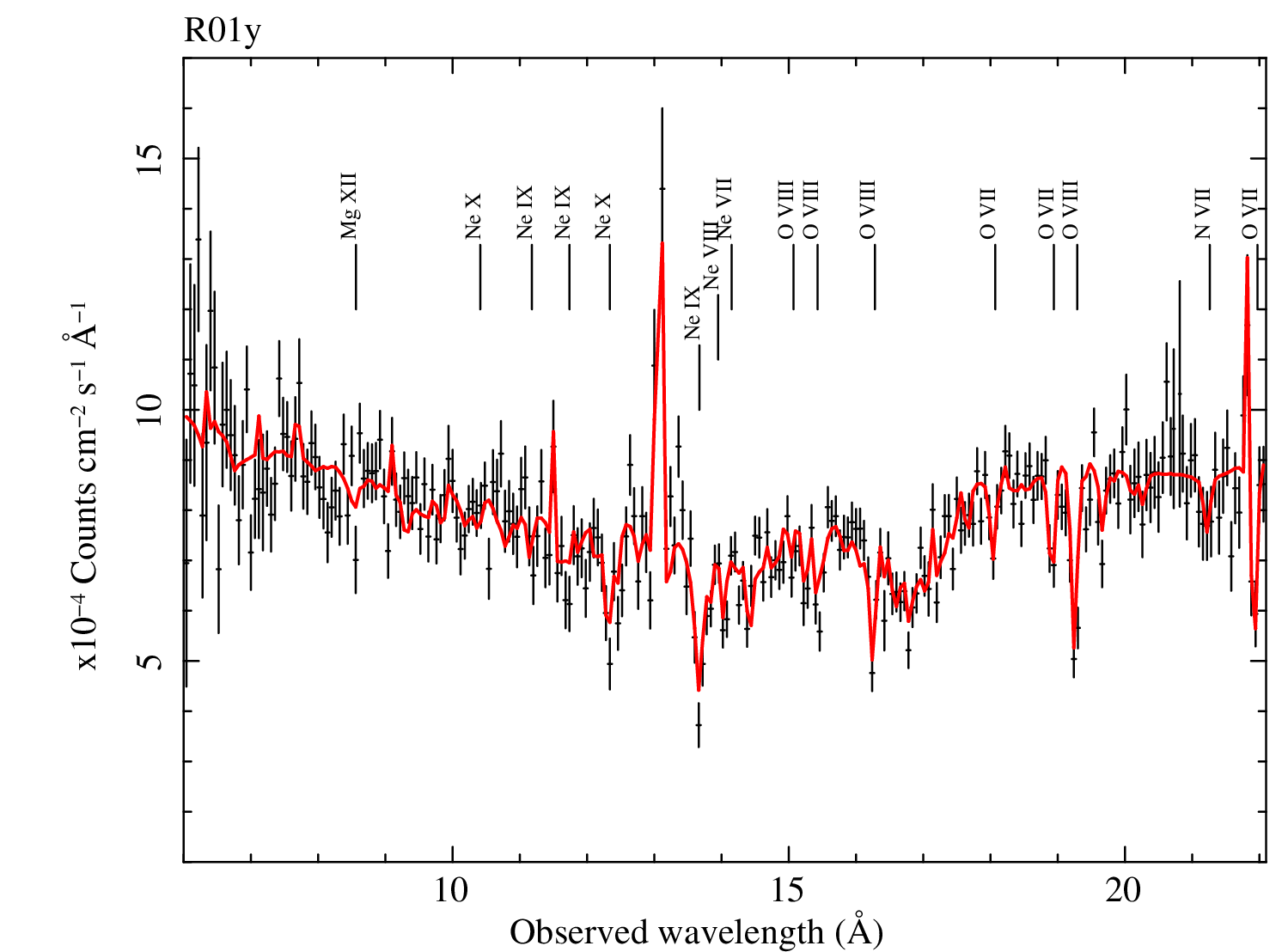}
  \includegraphics[width=6.5cm,angle=-90]{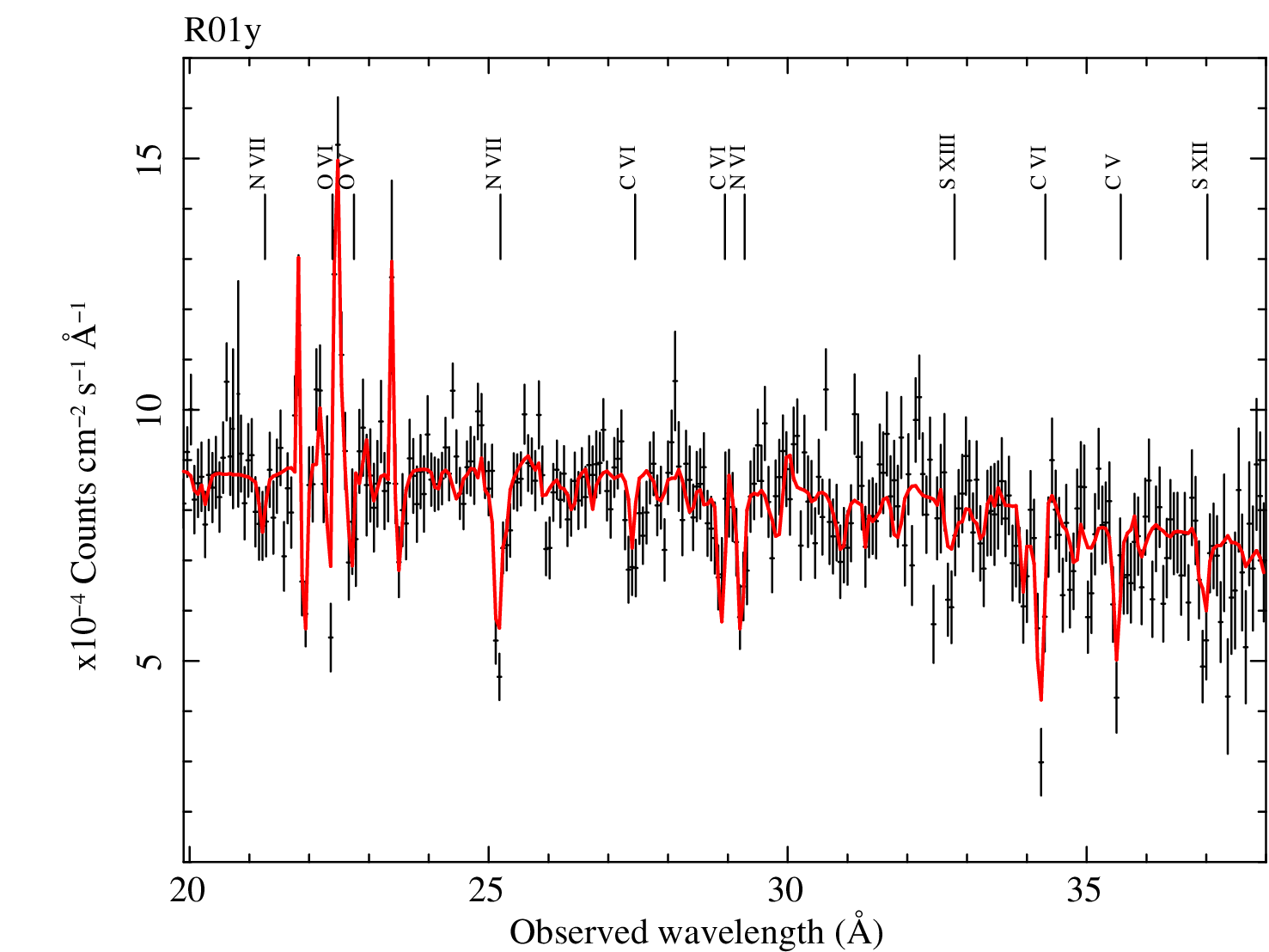}
  }
  \hbox{
  \includegraphics[width=6.5cm,angle=-90]{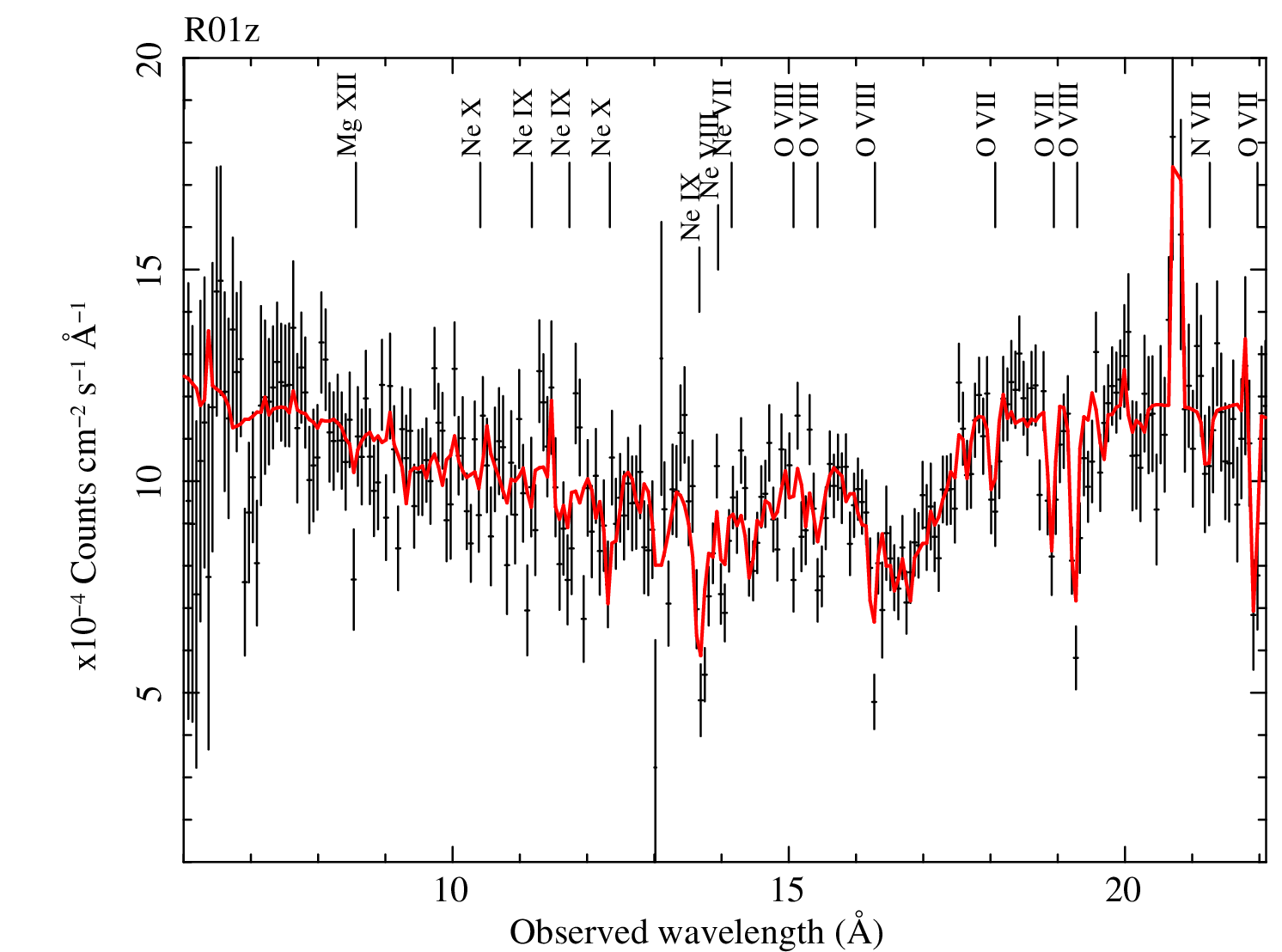}
  \includegraphics[width=6.5cm,angle=-90]{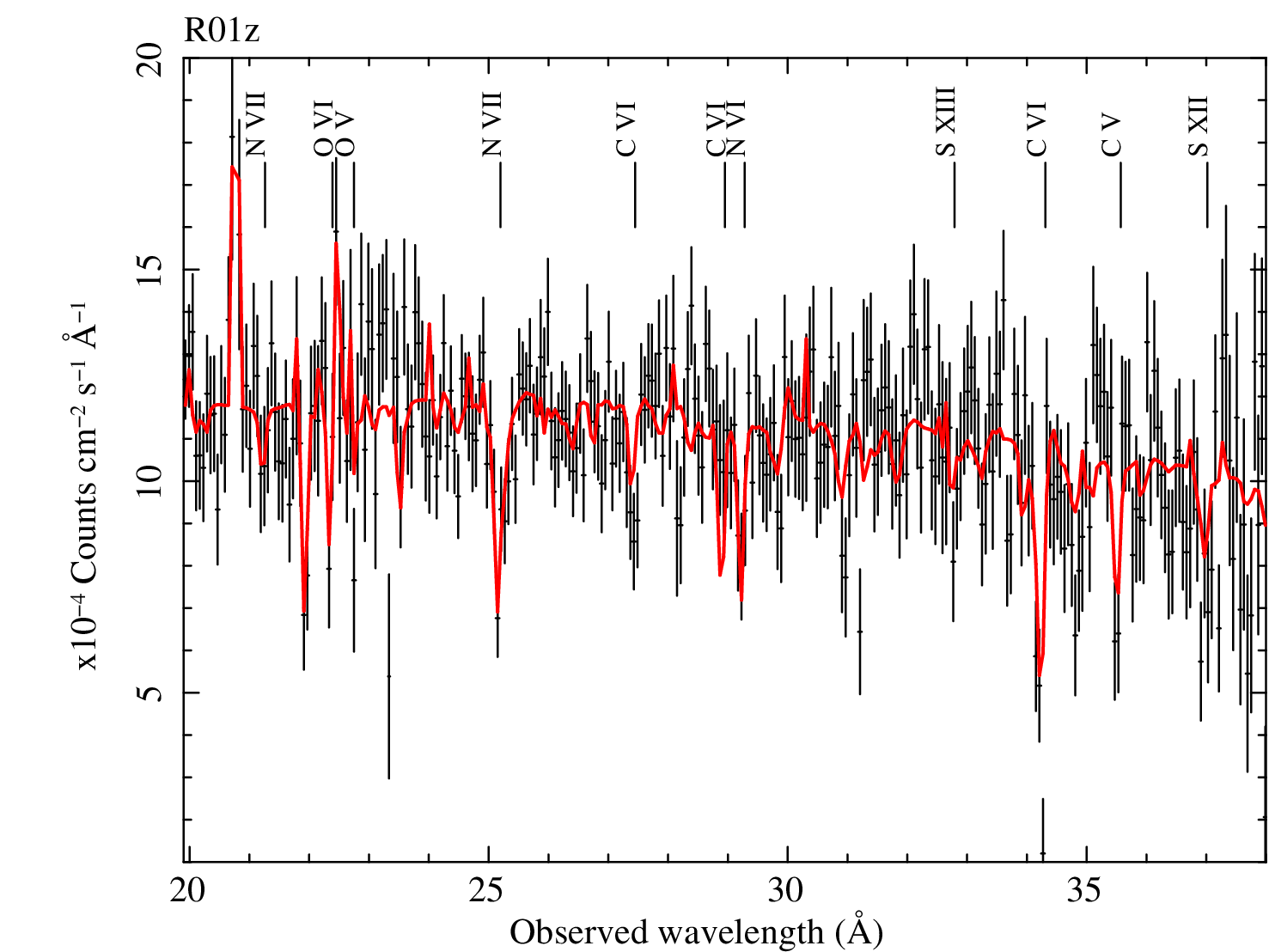}
  }
  \caption{\label{RGSfluxed}\XMM{}~RGS spectra of NGC 5548 in 2000 (R00x; top panels), and in 2001 (R01y, middle panels; R01z, bottom panels). The solid line represents our best-fit model. Some relevant WA absorption features have been labeled.}
\end{figure*}

\begin{figure*}
  \centering
  \hbox{
  \includegraphics[width=6.5cm,angle=-90]{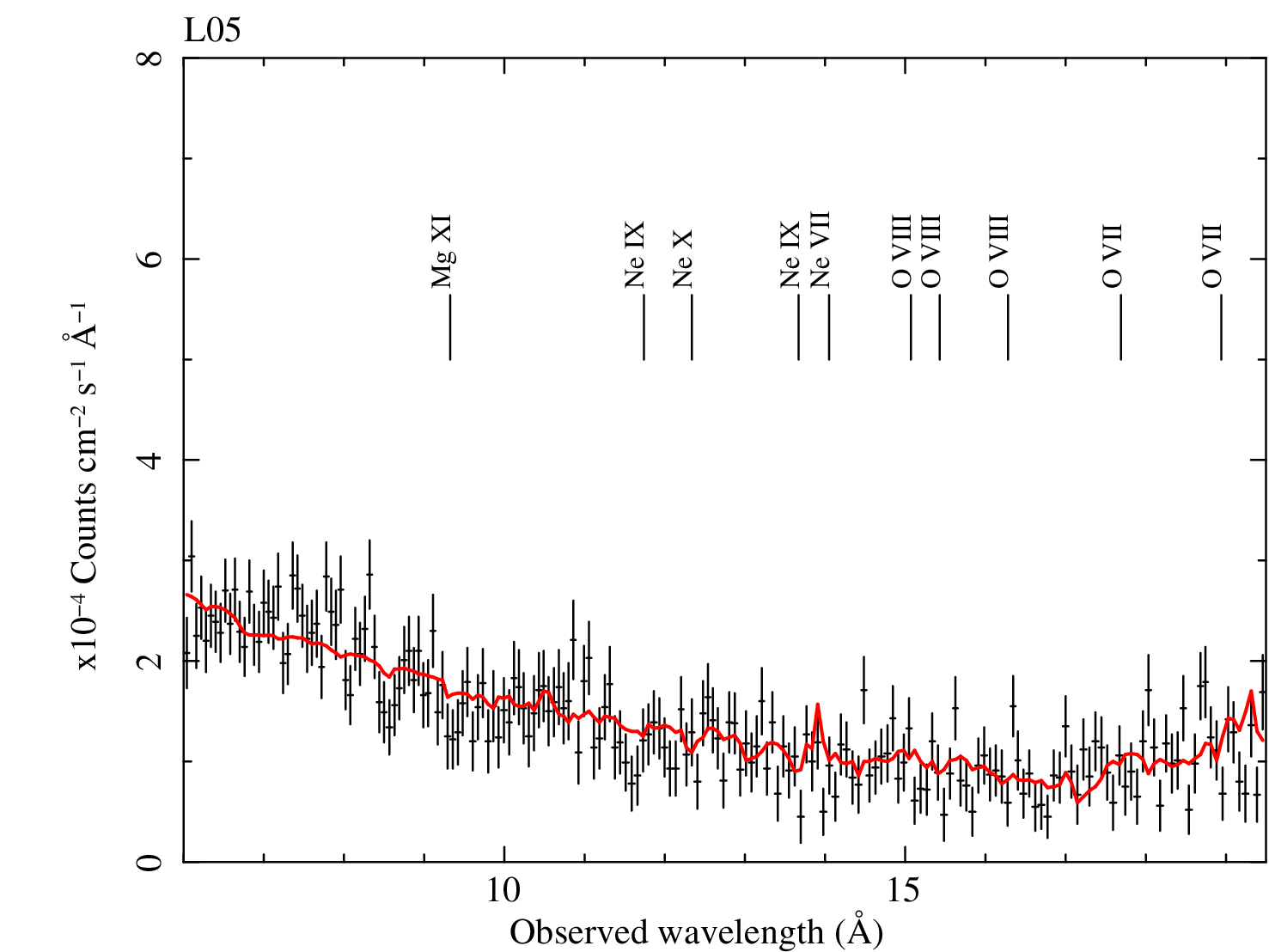}
  \includegraphics[width=6.5cm,angle=-90]{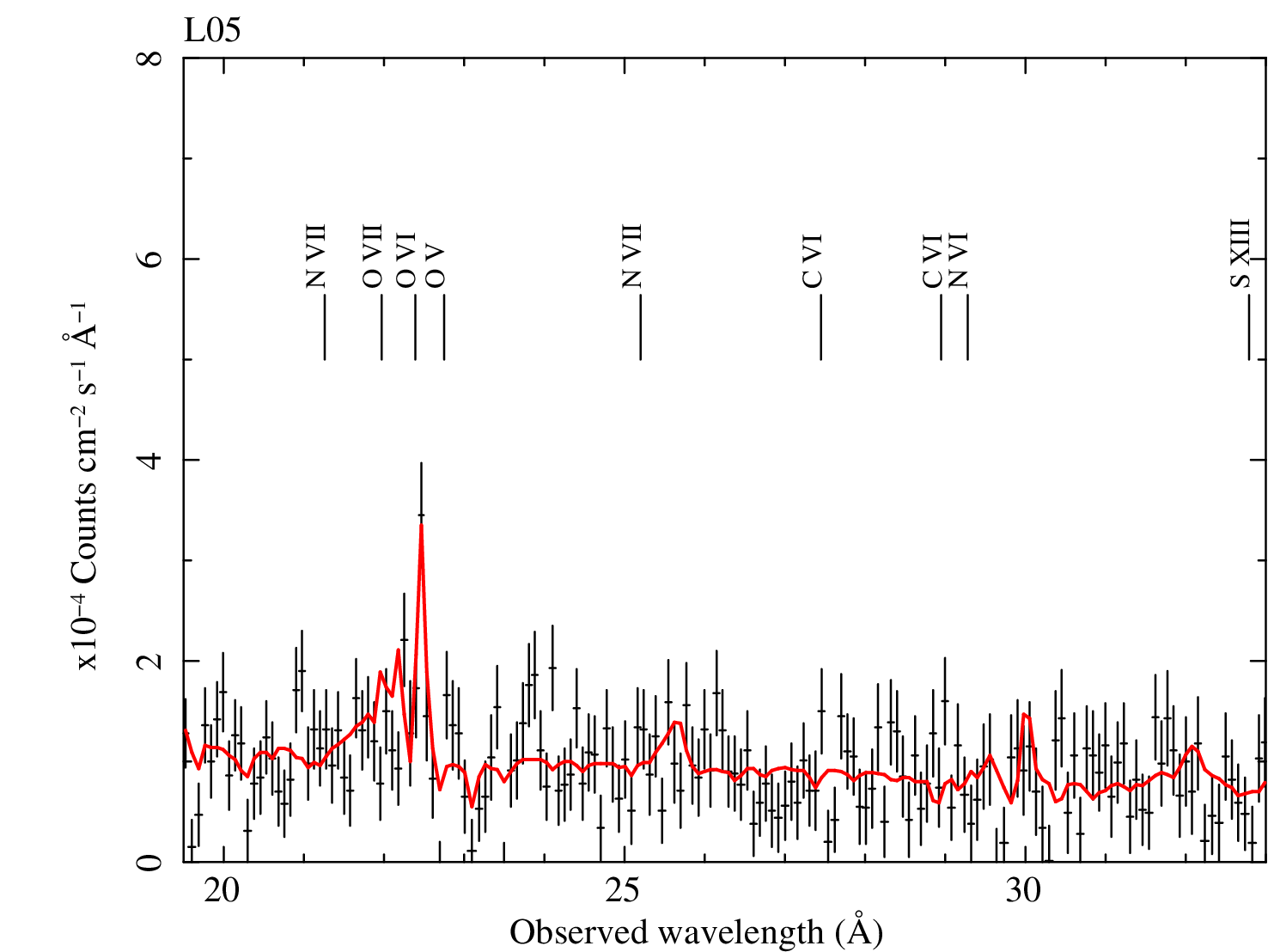}
  }
  \hbox{
  \includegraphics[width=6.5cm,angle=-90]{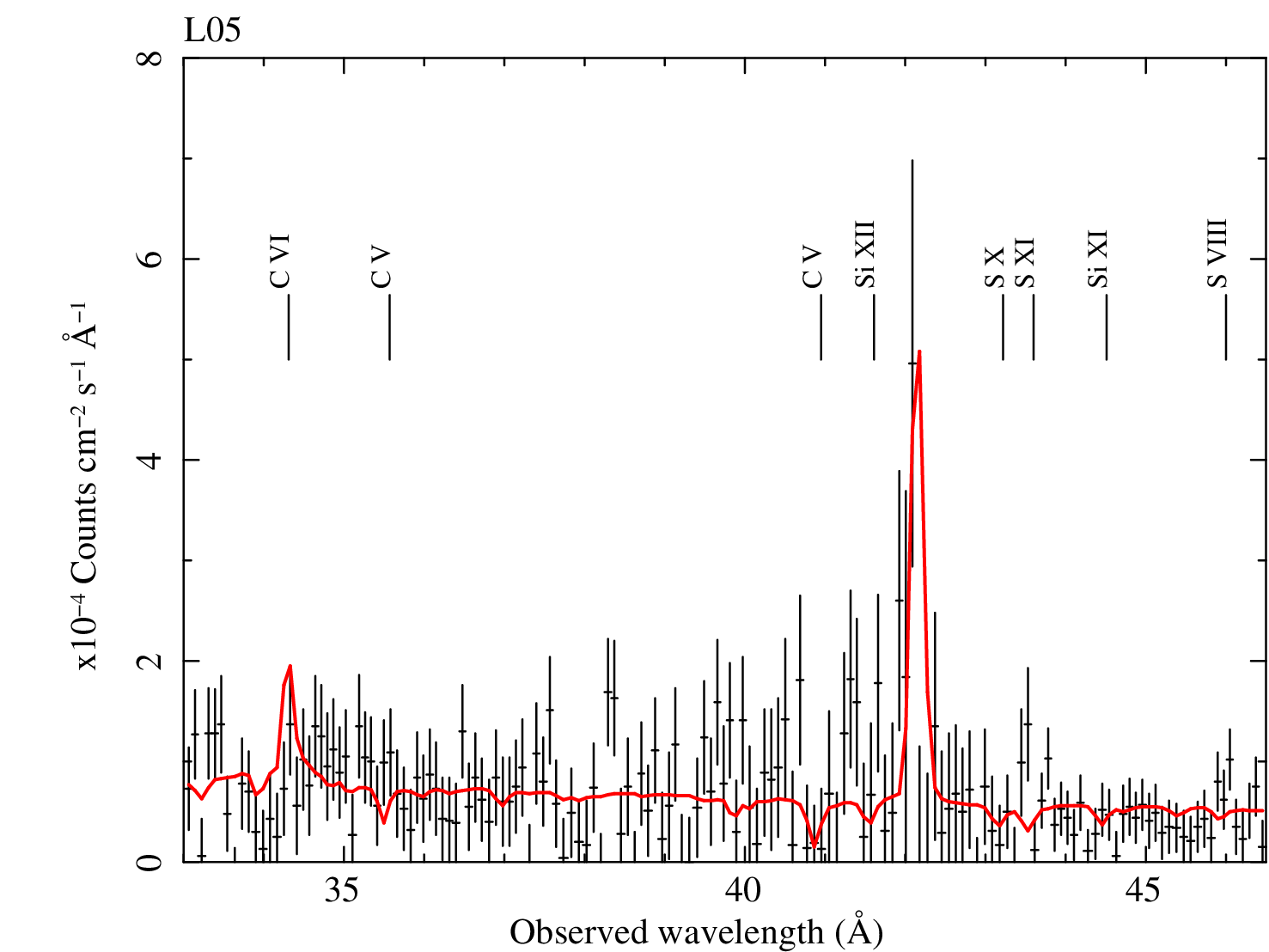}
  \includegraphics[width=6.5cm,angle=-90]{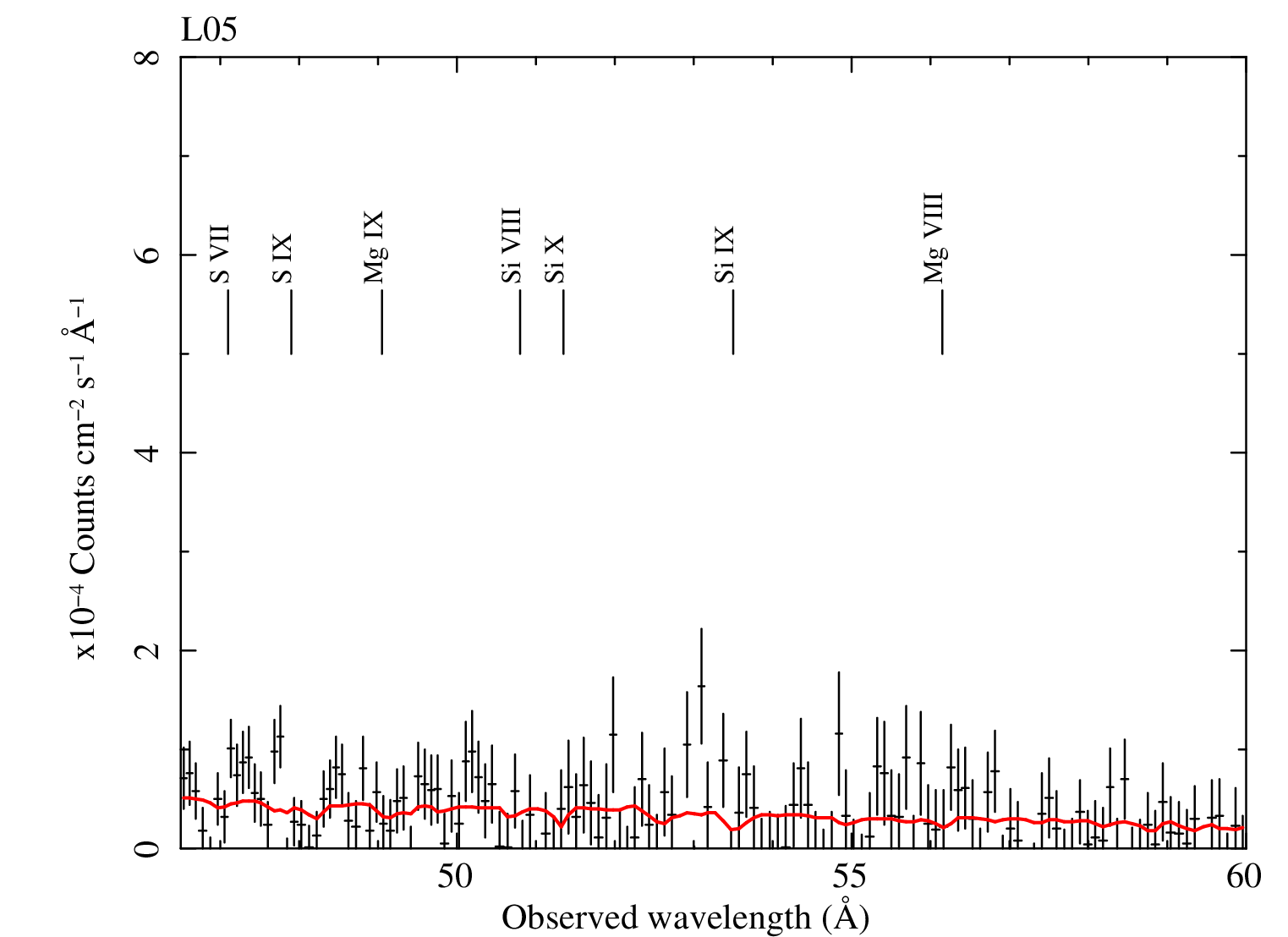}
  }
  \caption{\label{L05fluxed}\Chan{}~LETGS spectra of NGC 5548 in 2005 (L05). The solid line represents our best-fit model. Some relevant WA absorption features have been labeled. The spectrum has been rebinned for clarity.}
\end{figure*}

\begin{figure*}
  \centering
  \hbox{
  \includegraphics[width=6.5cm,angle=-90]{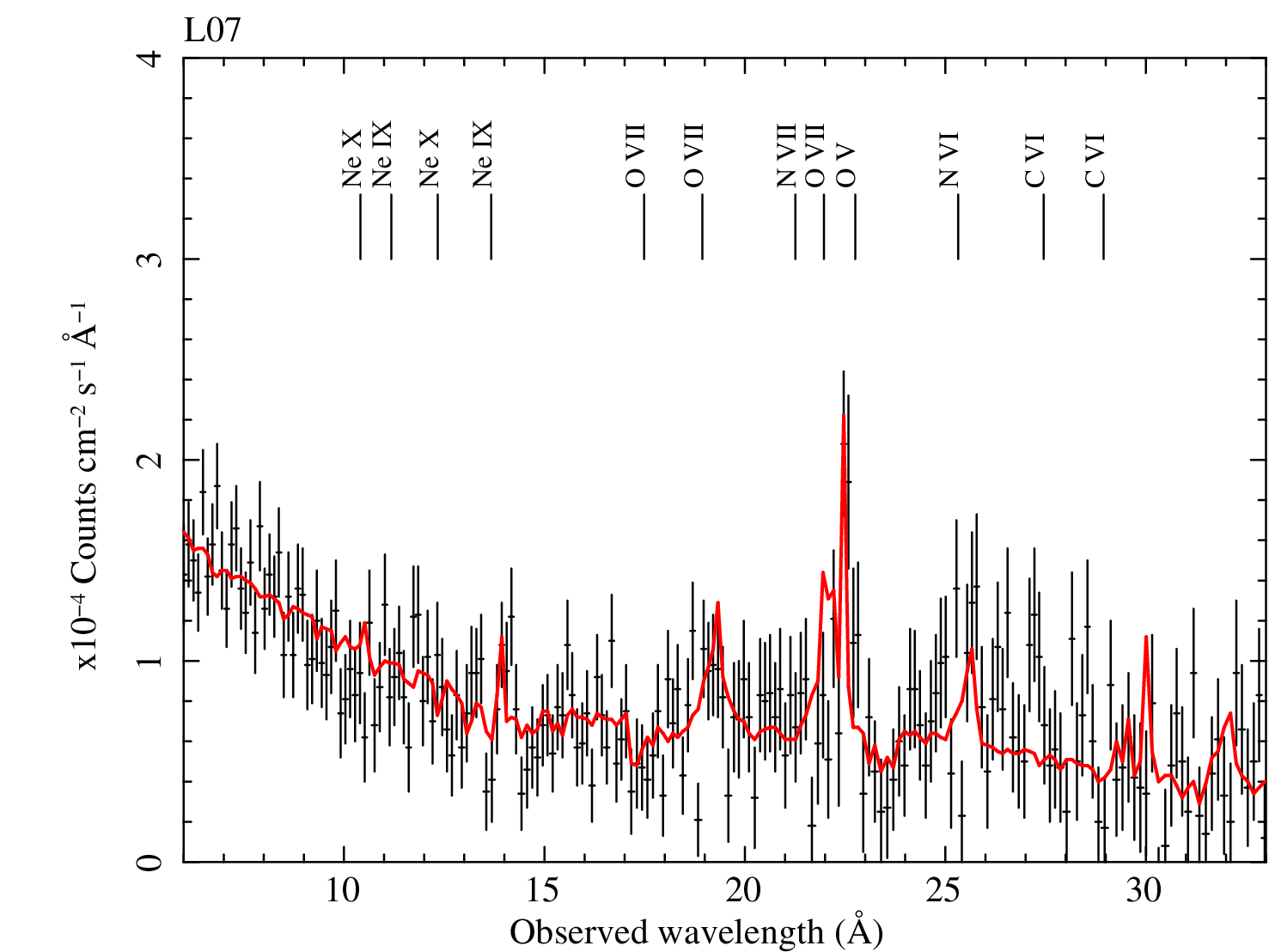}
  \includegraphics[width=6.5cm,angle=-90]{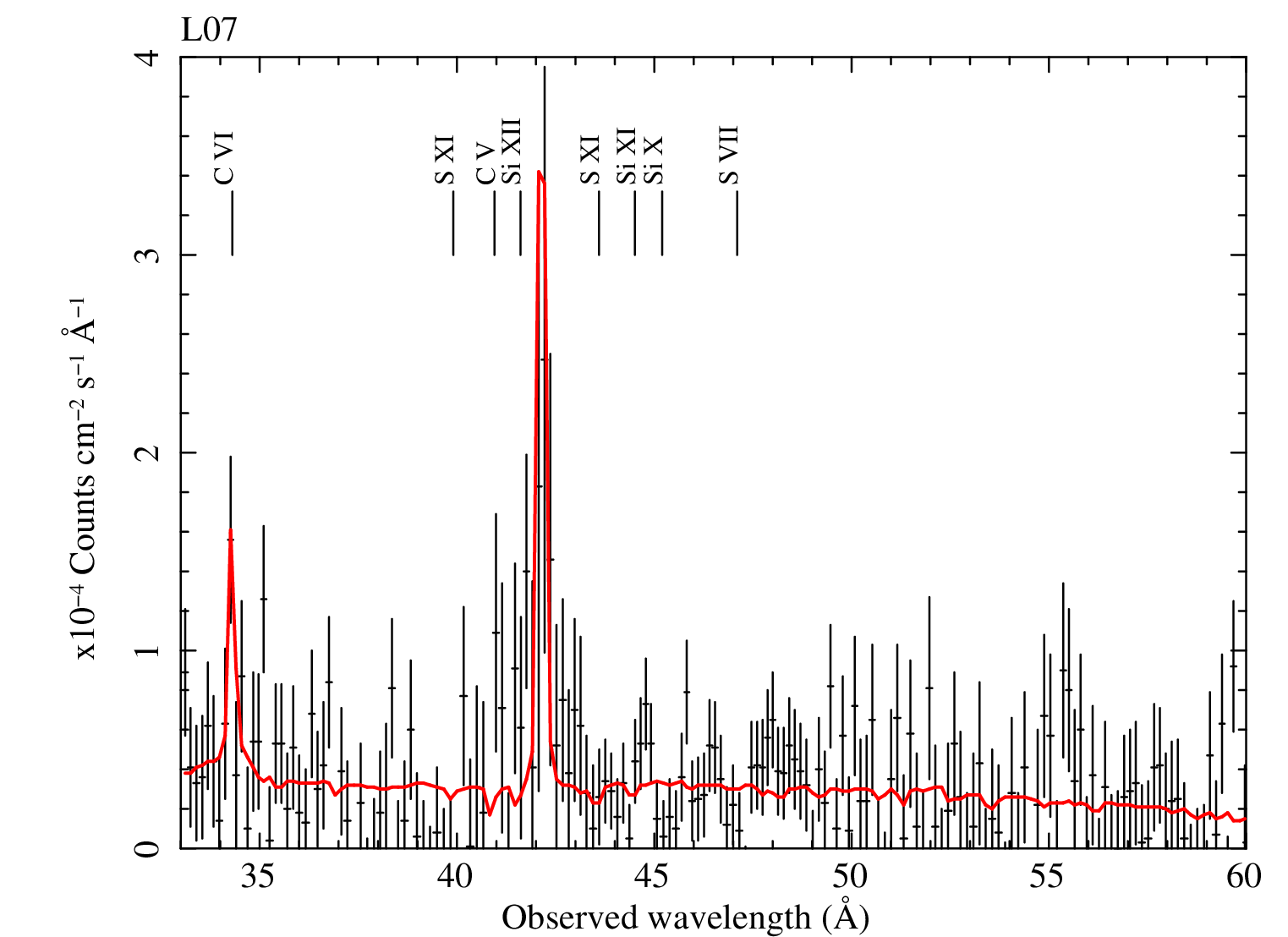}
  }
  \caption{\label{L07fluxed}\Chan{}~LETGS spectra of NGC 5548 in 2007 (L07). The solid line represents our best-fit model. Some relevant WA absorption features have been labeled. The spectrum has been rebinned for clarity.}
\end{figure*}

In this appendix we show some of the NGC 5548 archival spectra used in this work, together with their best fit models: L99 (see Fig.~\ref{L99fluxed}); R00x, R01y, and R01z (see Fig.~\ref{RGSfluxed}); L05 (see Fig.~\ref{L05fluxed}), and L07 (see Fig.~\ref{L07fluxed}).

\section{Ionic species for variability calculations}
\label{ions}

In Tables~\ref{ionlistA} to~\ref{ionlistF} we list the ions that were used in Sect.~\ref{WAvar} as a proxy for each WA component, together with their corresponding product $nt_{\rm rec}$ as provided by the SPEX auxiliary program {\it rec\_time}.

\begin{table}
  \centering
  \caption[]{List of ions used a proxy for the WA component $A$.}
  \label{ionlistA}
  \begin{tabular}{l c}
    \hline\hline
    \noalign{\smallskip}
    Ion   & $nt_{\rm rec}$ (s\,cm$^{-3}$)\\
    \noalign{\smallskip}
    \hline
    \noalign{\smallskip}
    \ion{C}{v} &  $2.5 \times 10^{11}$ \\
    \ion{N}{vi} & $1.7 \times 10^{11}$ \\
    \ion{O}{v} & $1.3 \times 10^{10}$ \\
    \ion{O}{vi} & $6.6 \times 10^{10}$ \\
    \ion{Fe}{viii} & $5.3 \times 10^{10}$ \\
    \noalign{\smallskip}
    \hline
  \end{tabular}
\end{table}

\begin{table}
  \centering
  \caption[]{List of ions used a proxy for the WA component $B$.}
  \label{ionlistB}
  \begin{tabular}{l c}
    \hline\hline
    \noalign{\smallskip}
    Ion   &   $nt_{\rm rec}$ (s\,cm$^{-3}$)  \\
    \noalign{\smallskip}
    \hline
    \noalign{\smallskip}
    \ion{C}{vi} &  $3.8 \times 10^{11}$ \\
    \ion{N}{vi} &  $1.0 \times 10^{11}$ \\
    \ion{N}{vii} &  $2.3 \times 10^{11}$ \\
    \ion{O}{vii} &  $2.1 \times 10^{11}$ \\
    \ion{Ne}{vii} &  $3.9 \times 10^{10}$ \\
    \ion{Ne}{viii} &  $1.8 \times 10^{11}$ \\
    \ion{Mg}{viii} &  $1.0 \times 10^{11}$ \\
    \ion{Mg}{ix} &  $9.7 \times 10^{10}$ \\
    \ion{Si}{viii} &  $6.0 \times 10^{10}$ \\
    \ion{Si}{ix} &  $9.0 \times 10^{10}$ \\
    \ion{S}{ix} &  $5.9 \times 10^{10}$ \\
    \ion{S}{x} &  $1.2 \times 10^{11}$ \\
    \ion{Fe}{viii}  & $1.2 \times 10^{11}$ \\
    \ion{Fe}{ix} &  $8.9 \times 10^{9}$ \\
    \ion{Fe}{x} &  $5.8 \times 10^{9}$ \\
    \noalign{\smallskip}
    \hline
  \end{tabular}
\end{table}

\begin{table}
  \centering
  \caption[]{List of ions used a proxy for the WA component $C$.}
  \label{ionlistC}
  \begin{tabular}{l c}
    \hline\hline
    \noalign{\smallskip}
    Ion   &   $nt_{\rm rec}$ (s\,cm$^{-3}$)  \\
    \noalign{\smallskip}
    \hline
    \noalign{\smallskip}
    \ion{N}{vii} &  $6.5 \times 10^{10}$ \\
    \ion{O}{viii} &  $2.8 \times 10^{11}$ \\
    \ion{Ne}{ix} &  $1.4 \times 10^{11}$ \\
    \ion{Ne}{x} &  $1.6 \times 10^{11}$ \\
    \ion{Mg}{ix} &  $2.0 \times 10^{10}$ \\
    \ion{Mg}{x} &  $3.1 \times 10^{10}$ \\
    \ion{Mg}{xi} &  $9.1 \times 10^{10}$ \\
    \ion{Si}{ix} &  $5.6 \times 10^{9}$ \\
    \ion{Si}{x} &  $1.6 \times 10^{10}$ \\
    \ion{Si}{xi} &  $1.5 \times 10^{10}$ \\
    \ion{S}{x} &  $1.6 \times 10^{10}$ \\
    \ion{S}{xi} &  $3.7 \times 10^{10}$ \\
    \ion{S}{xii}  & $6.0 \times 10^{10}$ \\
    \ion{Fe}{xi} &  $3.3 \times 10^{9}$ \\
    \ion{Fe}{xii} &  $2.4 \times 10^{10}$ \\
    \ion{Fe}{xiii} &  $4.0 \times 10^{9}$ \\
    \ion{Fe}{xviii} &  $4.0 \times 10^{10}$ \\
    \noalign{\smallskip}
    \hline
  \end{tabular}
\end{table}

\begin{table}
  \centering
  \caption[]{List of ions used a proxy for the WA component $D$.}
  \label{ionlistD}
  \begin{tabular}{l c}
    \hline\hline
    \noalign{\smallskip}
    Ion   &   $nt_{\rm rec}$ (s\,cm$^{-3}$)  \\
    \noalign{\smallskip}
    \hline
    \noalign{\smallskip}
    \ion{Ne}{x} &  $1.0 \times 10^{11}$ \\
    \ion{Mg}{xi} &  $2.0 \times 10^{11}$ \\
    \ion{Si}{xi} &  $2.3 \times 10^{10}$ \\
    \ion{S}{xii} &  $2.7 \times 10^{10}$ \\
    \ion{Fe}{xviii} &  $2.2 \times 10^{10}$ \\
    \ion{Fe}{xix} &  $1.9 \times 10^{10}$ \\
    \noalign{\smallskip}
    \hline
  \end{tabular}
\end{table}

\begin{table}
  \centering
  \caption[]{List of ions used a proxy for the WA component $E$.}
  \label{ionlistE}
  \begin{tabular}{l c}
    \hline\hline
    \noalign{\smallskip}
    Ion   &   $nt_{\rm rec}$ (s\,cm$^{-3}$)  \\
    \noalign{\smallskip}
    \hline
    \noalign{\smallskip}
    \ion{Ne}{x} &  $3.5 \times 10^{10}$ \\
    \ion{Mg}{xii} &  $9.4 \times 10^{10}$ \\
    \ion{Fe}{xix} &  $1.6 \times 10^{10}$ \\
    \ion{Fe}{xx} &  $3.4 \times 10^{10}$ \\
    \noalign{\smallskip}
    \hline
  \end{tabular}
\end{table}

\begin{table}
  \centering
  \caption[]{List of ions used a proxy for the WA component $F$.}
  \label{ionlistF}
  \begin{tabular}{l c}
    \hline\hline
    \noalign{\smallskip}
    Ion   &   $nt_{\rm rec}$ (s\,cm$^{-3}$)  \\
    \noalign{\smallskip}
    \hline
    \noalign{\smallskip}
    \ion{Ne}{x} &  $2.1 \times 10^{10}$ \\
    \ion{Mg}{xii} &  $4.6 \times 10^{10}$ \\
    \ion{Fe}{xx} &  $1.1 \times 10^{10}$ \\
    \ion{Fe}{xxiv} &  $2.7 \times 10^{10}$ \\
    \noalign{\smallskip}
    \hline
  \end{tabular}
\end{table}

\end{appendix}

\end{document}